\documentclass[reprint,superscriptaddress,amsmath,amssymb,
 aps,nofootinbib,prx]{revtex4-2}
\usepackage{siunitx}
\usepackage{floatpag}
\usepackage{adjustbox}
\usepackage{cancel}
\usepackage{wrapfig}
\usepackage[nointegrals]{wasysym}
\usepackage{subcaption}
\usepackage{lineno,hyperref}
\usepackage{graphicx}
\usepackage{dcolumn}
\usepackage{bm}
\usepackage{array}
\usepackage{amssymb}
\usepackage{amsmath}
\usepackage{textcomp}
\usepackage[euler]{textgreek}
\usepackage{nicefrac}
\usepackage[]{acronym} 
\usepackage{array}
\usepackage{booktabs}
\usepackage[margin=1.25in]{geometry}
\usepackage{xcolor}
\usepackage{cancel}
\usepackage[nottoc,numbib]{tocbibind}

\modulolinenumbers[5]

\acrodefplural{BD}[BDs]{beam dumps}
\acrodefplural{ALPS}[\ac{ALPS}\,I]{Any Light Particle Search I}
\acrodefplural{FSR}[FSRs]{free spectral ranges}
\acrodefplural{HWP}[HWPs]{half-wave-plates}
\acrodefplural{PLL}[PLLs]{phase-locked loops}

\begin{document}


\preprint{APS}

\title{Any Light Particle Searches with ALPS II: Description of the first science campaign}

\author{Aaron~D.~Spector}\email{aaron.spector@desy.de}
\affiliation{Deutsches Elektronen-Synchrotron DESY, 22603 Hamburg, Germany}

\author{Daniel~C.~Brotherton}
\affiliation{Department of Physics, University of Florida, 32611 Gainesville, Florida, USA}

\author{Ayman~Hallal}
\altaffiliation[Present address: ]{https://optiwave.com/}
\affiliation{Max-Planck-Institut f\"ur Gravitationsphysik (Albert-Einstein-Institut) and Leibniz Universit\"at Hannover, 30167 Hannover, Germany}

\author{Henry~Fr\"adrich}
\affiliation{Deutsches Elektronen-Synchrotron DESY, 22603 Hamburg, Germany}

\author{Jacob Egge}
\affiliation{Deutsches Elektronen-Synchrotron DESY, 22603 Hamburg, Germany}

\author{Li-Wei~Wei}
\altaffiliation[Present address:  ]{Max-Planck-Institut f\"ur Gravitationsphysik (Albert-Einstein-Institut) and Leibniz Universit\"at Hannover, 30167 Hannover, Germany}
\affiliation{Deutsches Elektronen-Synchrotron DESY, 22603 Hamburg, Germany}

\author{Todd~Kozlowski}\altaffiliation[Present address:  ]{Helmut-Schmidt-Universität, 22043 Hamburg, Germany}
\affiliation{Deutsches Elektronen-Synchrotron DESY, 22603 Hamburg, Germany}

\author{Kanioar~Karan}\altaffiliation[Present address:  ]{Max-Planck-Institut f\"ur Gravitationsphysik (Albert-Einstein-Institut) and Leibniz Universit\"at Hannover, 30167 Hannover, Germany}
\affiliation{School of Physics and Astronomy, Cardiff University, Cardiff CF24 3AA, United Kingdom}

\author{Zachary~R.~Bush}\altaffiliation[Present address:  ]{https://www.ionq.com/}
\affiliation{Department of Physics, University of Florida, 32611 Gainesville, Florida, USA}

\author{Mauricio~Diaz-Ortiz~Jr.}
\altaffiliation[Present address: ]{Donders Institute - Biophysics, Radboud University, 6525 AJ, Nijmegen, Nederland}
\affiliation{Department of Physics, University of Florida, 32611 Gainesville, Florida, USA}

\author{Aldo~Ejlli}
\affiliation{Max-Planck-Institut f\"ur Gravitationsphysik (Albert-Einstein-Institut) and Leibniz Universit\"at Hannover, 30167 Hannover, Germany}

\author{Joe Gleason}
\affiliation{Department of Physics, University of Florida, 32611 Gainesville, Florida, USA}

\author{Hartmut~Grote}\affiliation{School of Physics and Astronomy, Cardiff University, Cardiff CF24 3AA, United Kingdom}

\author{Michael T. Hartman}
\altaffiliation{Now at: Institute Fresnel, 13397 Marseille, France}
\affiliation{Deutsches Elektronen-Synchrotron DESY, Notkestr. 85, 22607 Hamburg, Germany}

\author{Harold~Hollis}
\affiliation{Department of Physics, University of Florida, 32611 Gainesville, Florida, USA}

\author{Katharina-Sophie Isleif}
\affiliation{Helmut-Schmidt-Universität, 22043 Hamburg, Germany}

\author{Alasdair L. James}
\altaffiliation[Present address: ]{LIGO Laboratory, California Institute of Technology, Pasadena, CA 91125, USA}
\affiliation{School of Physics and Astronomy, Cardiff University, Cardiff CF24 3AA, United
Kingdom}

\author{Giuseppe~Messineo}
\altaffiliation[Present address:  ]{Sezione di Padova $\cdot$ Istituto Nazionale di Fisica Nucleare, 35131 Padua, Italy}
\affiliation{Department of Physics, University of Florida, 32611 Gainesville, Florida, USA}

\author{Guido~Mueller}
\affiliation{Department of Physics, University of Florida, 32611 Gainesville, Florida, USA}
\affiliation{Max-Planck-Institut f\"ur Gravitationsphysik (Albert-Einstein-Institut) and Leibniz Universit\"at Hannover, 30167 Hannover, Germany}

\author{Ryan~Netrval}
\affiliation{Max-Planck-Institut f\"ur Gravitationsphysik (Albert-Einstein-Institut) and Leibniz Universit\"at Hannover, 30167 Hannover, Germany}

\author{Isabella~Oceano}
\altaffiliation[Present address:  ]{Universit\"at Hamburg, 22761 Hamburg, Germany}
\affiliation{Deutsches Elektronen-Synchrotron DESY, 22603 Hamburg, Germany}

\author{Jan~H.~P\~old}
\altaffiliation[Present address:  ]{Leuze electronic, 73277 Owen, Germany}
\affiliation{Deutsches Elektronen-Synchrotron DESY, 22603 Hamburg, Germany}

\author{Richard~C.~G.~Smith}
\altaffiliation[Present address:  ]{https://aerospace.org/}
\affiliation{Max-Planck-Institut f\"ur Gravitationsphysik (Albert-Einstein-Institut) and Leibniz Universit\"at Hannover, 30167 Hannover, Germany} 

\author{David~B.~Tanner}
\affiliation{Department of Physics, University of Florida, 32611 Gainesville, Florida, USA}

\author{Benno~Willke}
\affiliation{Max-Planck-Institut f\"ur Gravitationsphysik (Albert-Einstein-Institut) and Leibniz Universit\"at Hannover, 30167 Hannover, Germany}

\author{Axel~Lindner}
\affiliation{Deutsches Elektronen-Synchrotron DESY, 22603 Hamburg, Germany}


%
\begin{abstract}
From February to May of 2024 the Any Light Particle Search II (ALPS\,II) conducted its first science campaign using the `light-shining-through-a-wall' technique to search for pseudo-Goldstone bosons that lie beyond the Standard Model of particle physics and which are inaccessible by accelerator-based experiments. The experimental setup consists of two strings of superconducting dipole magnets, each more than 100\,m long, that are separated by a wall. Laser light is directed through the first magnet string and a heterodyne detection system is used to measure the electromagnetic power that traverses a wall via the conversion to and then from a bosonic field. After the wall, a high-finesse optical cavity resonantly enhances the signal power. Two searches were carried out, one with the laser polarized perpendicular to the magnetic field direction and another with its polarization state aligned parallel to the magnetic field. No evidence for the existence of new bosons was found. In its first science campaign, ALPS\,II reached photon-boson conversion probability sensitivities of a few $10^{-13}$. The ongoing upgrade of the optical system aims to increase this sensitivity by about four orders of magnitude.

\end{abstract}

\maketitle

\tableofcontents


\newcommand{\murm}{%
  \ifmmode
    \mathchoice
        {\hbox{\normalsize\textmu}}
        {\hbox{\normalsize\textmu}}
        {\hbox{\scriptsize\textmu}}
        {\hbox{\tiny\textmu}}%
  \else
    \textmu
  \fi
}

\onecolumngrid
    \vspace{1.7in}
    \section*{List of Acronyms}
\twocolumngrid
\begin{acronym}[DESY] 
\acro{ADC}{analog-to-digital converter}
\acro{AL}{auxiliary laser}
\acro{ALPS}{Any Light Particle Search}
\acro{BD}{beam dump}
\acro{BSM}{beyond the Standard Model}
\acro{CH}{central hall}
\acro{COB}{central optical bench}
\acro{DESY}{Deutsches Elektronen-Synchrotron}
\acro{EOM}{electro-optic modulator}
\acro{FIR}{finite impulse response}
\acro{FSR}{free spectral range}
\acro{HPL}{high-power laser}
\acro{HR}{highly reflective}
\acro{HWP}{half-wave-plate}
\acro{LO}{local oscillator laser}
\acro{LSW}{Light-shining-through-a-wall}
\acro{MZ}{Mach Zehnder}
\acro{NL}{north left}
\acro{NR}{north right}
\acro{PC}{production cavity}
\acro{PD}{photodetector}
\acro{PDH}{Pound-Drever-Hall}
\acro{PLL}{phase-locked loop}
\acro{PSD}{power spectral density}
\acro{RC}{regeneration cavity}
\acro{RMS}{root-mean-square}
\acro{PDF}{probability density function}
\acro{RL}{reference laser}
\acro{ULE}{ultra-low expansion}
\end{acronym}

\setlength{\tabcolsep}{6pt} 
\renewcommand{\arraystretch}{1.4} 

\newpage
\section{Introduction}

\label{Sec:intro}


While the necessity of physics \ac{BSM} is increasingly apparent, experimental evidence that can explain phenomena like dark matter, dark energy, or the small mass of the Higgs boson remains scarce. Many theories for \ac{BSM} physics predict the existence of very lightweight and extremely weakly interacting pseudo-Goldstone bosons like the axion, which are out of reach at accelerator-based experiments (see \cite{ALPSII_science} and references therein).
This calls for complementary approaches.

\ac{LSW} experiments at optical frequencies offer unique opportunities to search for new physics~\cite{okun1982limits,Anselm:1985obz,VanBibber1987759}. Here, a laser is directed through a conversion region (typically a beam tube with vacuum or well-controlled gas-pressure, which might be permeated by a magnetic dipole field), generating (potentially) a beam of bosons capable of propagating through a wall that blocks the laser light. The bosonic field then travels through another (re-)conversion region beyond the wall, where some of its energy is reconverted to an electromagnetic field. By measuring the rate at which energy converts between the electromagnetic and bosonic fields, it is possible to deduce their coupling.

Optical cavities can be used to enhance significantly this process. A cavity before the wall will increase the intensity of the electromagnetic field traveling through the conversion region, while a cavity after the wall will resonantly enhance the electromagnetic field reconverted from the interaction between the bosons field~\cite{Hoogeveen19913,PhysRevLett.98.172002,PhysRevD.80.072004}. \ac{LSW} experiments in the optical regime generate an enormous photon flux while also being able to accommodate a long string of magnets due to a small beam divergence, making them compelling tools in the hunt for \ac{BSM} physics. A laser operating at a wavelength of 1064\,nm with a high-finesse cavity before the wall can direct up to $\sim10^{24}$ photons per second through a string of dipole magnets before the wall.
Behind the wall, a cavity might boost the re-conversion probability by more than a factor 10,000 and photon rates of $10^{-5}\,\mathrm{s}^{-1}$ can be sensed~\cite{bahre2013any}. 
Therefore, photon-boson-photon conversion probabilities below $10^{-33}$ at O(100\,m) long conversion and re-conversion regions can be probed by such optical LSW-experiments.

The \ac{ALPS}, an \ac{LSW} experiment conducted at \ac{DESY} from 2007 to 2010, utilized an optical cavity before the wall with a single 5\,T HERA dipole magnet to set, at the time, new exclusion limits for purely laboratory based searches for scalar, pseudoscalar bosons, hidden photons and other hypothetical bosons~\cite{ehret2010new}. This was followed by the OSQAR experiment based at CERN, which utilized two 9\,T LHC magnets to become the most sensitive \ac{LSW} experiment for scalar and pseudoscalar bosons of its time \cite{PhysRevD.92.092002}. 

In 2012, a new \ac{LSW} experiment, \ac{ALPS}\,II, was started at \ac{DESY} and in 2024 \ac{ALPS}\,II launched its first science campaign. With two strings of 12 HERA dipole magnets providing 5.3\,T magnet fields, a \ac{HPL} generating the bosonic field before the wall, and a high-finesse optical cavity after the wall implemented for its first science campaign, \ac{ALPS}\,II is now the most sensitive \ac{LSW} experiment ever built. Over the course of four months, from February to May 2024, two data runs were conducted searching for beyond the Standard Model lightweight bosons. 

We would like to emphasize that these searches are sensitive to a rich variety of bosons including vector and tensor bosons in addition to the typically cited scalar and pseudoscalars \cite{ALPSII_science,PhysRevD.82.115018,garciacely2025stellarboundslightspin2}. Some of the electromagnetic couplings of these fields, such as vector bosons, do not even require the magnets. The results in this manuscript are therefore reported generally in terms of the conversion rate between electromagnetic and \ac{BSM} fields during a single pass through one of the magnet strings, denoted as $\mathcal{P}_{\gamma\leftrightarrow \phi}$. 
An interpretation of these results in terms of the coupling between electromagnetic and scalar, pseudoscalar, vector, and tensor bosons is reported in Ref.~\cite{ALPSII_science}. Additionally,  throughout this paper we refer to lightweight bosons, bosonic fields, and \ac{BSM} fields interchangeably as potential sources of a signal. A discovery of any such field would represent a fundamental breakthrough in our understanding of the universe as the first evidence of a new energy scale between the electroweak and Planck scales as well as the first direct measurement of a particle residing in the \ac{BSM} `dark sector'.

This document provides an overview of the optical system, calibration procedure, results, and performance of the instrument during the first science campaign, and is organized as follows. The remainder of Section~\ref{Sec:intro} discusses the design of the \ac{ALPS}\,II optical system as well as the optical system that was implemented during the first science campaign, and then explains the fundamentals of the heterodyne detection system, and the central optical bench. Section~\ref{Sec:Cal} explains how the results are calibrated and the various sources of systematic error in the calibration. The results, expected background rates, and derived exclusion limits are then presented in  Section~\ref{Sec:Results_full}. The performance of the optical system during the first science campaign is examined in Section~\ref{Sec:Perf} and our conclusions are presented in Section~\ref{Sec:Conc}.

\subsection{ALPS II Design}
\label{Sec:ALPSII_des}

\begin{figure*}
    \centering
    \includegraphics[width=0.99\textwidth]{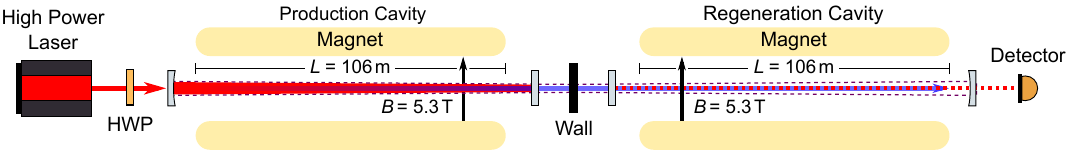}
    \caption{Side view of the design of the full \ac{ALPS}\,II experimental system. The \ac{HPL} beam and \ac{PC} spatial eigenmode is shown in red, while the \ac{BSM} field traveling to the right is shown in blue. The dashed purple lines show the projection of the \ac{RC} spatial eigenmode. A \ac{HWP} before the \ac{PC} can be used to configure the polarization state of the \ac{HPL} with respect to the direction of the magnetic field (shown as the black arrow).}
    \label{fig:ALPS_Des_IIc}
\end{figure*}

As Figure~\ref{fig:ALPS_Des_IIc} shows, the \ac{ALPS}\,II optical system enhances the observable photon flux from photon-boson-photon conversion via the resonant power buildup of optical cavities before and after the wall \cite{bahre2013any,ortiz2020design}. A \ac{HPL} \cite{Frede:07} is injected into the optical cavity before the wall, referred to as the \ac{PC}, which increases the laser power propagating through the initial magnetic field. The cavity after the wall, referred to as the \ac{RC}, resonantly enhances the signal power measured at the detection system. The electromagnetic field generated by the reconversion of the \ac{BSM} field in the \ac{RC} will be referred to as the regenerated, reconverted, or signal field. 
 
The magnetic fields for \ac{ALPS}\,II are generated by superconducting dipole magnets formerly used by the Hadron Electron Ring Accelerator (HERA). With 12 magnets in each of the strings before and after the wall, the products of the magnetic flux density times the lengths of the magnetic fields are $BL = 563.2\pm0.1\rm\,T\cdot m$ for each string \cite{meinke1991superconducting,albrecht2021straightening}. The magnets are oriented such that their magnetic fields are pointing in the vertical direction with respect to the ground. The polarization state of the \ac{HPL} can be controlled with a \ac{HWP} in the injection path to the \ac{PC}. This allows us to search for different types of \ac{BSM} fields based on whether the \ac{HPL} is polarized parallel or perpendicular to the magnetic field.

At sufficiently low masses ($<1\rm\,meV$) the \ac{BSM} field will remain coherent over the length of the experiment with the electromagnetic field that generates it. Therefore, for the experiment to work properly, it is essential that the light circulating in the \ac{PC} is resonant with the \ac{RC} and for the cavities to share the same spatial eigenmode. These conditions are quantified by a parameter that we refer to as the field overlap. The wall is also equipped with a shutter that can be opened to allow light to couple directly from the \ac{PC} to the \ac{RC} to check the field overlap by measuring the power in transmission of the \ac{RC}, referred to as $P_{\rm open}$, and comparing it to the \ac{HPL}  power incident on the \ac{RC}. As we explain in more detail in Section~\ref{Sec:Cal}, the open shutter measurements can be used to directly calibrate the results in terms of the conversion rate between electromagnetic and \ac{BSM} fields $\mathcal{P}_{\gamma\leftrightarrow \phi}$.

To provide specific examples, we will examine the interaction of scalar fields and a laser polarized perpendicular to the external magnetic field, as well as pseudoscalar fields and a laser polarized parallel to the magnetic field. As Ref.~\cite{PhysRevD.82.115018} explains, with the shutter closed, the power measured after the wall, $P_\phi$, generated by the reconversion of the \ac{BSM} field, is the same in both of these cases and can be approximated with the equation 
\begin{align}
    P_\phi & = \eta^2 \beta_{\!_{\rm RC}}  P_{\!_{\rm PC}}  \,\mathcal{P}_{\gamma\leftrightarrow \phi}^2  \\
    &\simeq \eta^2  \beta_{\!_{\rm RC}}  P_{\!_{\rm PC}} \left(\frac{g_{\phi\gamma\gamma}BL}{2}\right)^4,
\end{align}
for scalar or pseudoscalar fields with a rest mass energy below 0.1\,meV.
In this equation $\beta_{\!_{\rm RC}}$ is the resonant enhancement factor of the \ac{RC} and $P_{\!_{\rm PC}} $ is the laser power circulating in the \ac{PC} \cite{ortiz2020design}. 
To maximize the total field overlap between the light circulating in the \ac{PC} and the \ac{RC} eigenmode $\eta$, the frequency of the \ac{HPL} must be held on a resonance of the \ac{RC} and coupled to its spatial eigenmode. The field overlap is the product of the spatial, spectral, and polarization components, 
\begin{equation}
    \eta = \eta_{xy} \eta_{\rm pol} \eta_{\hat z} .
\end{equation}
Here, $\eta_{xy}$ is the spatial component field overlap between the light circulating in the \ac{PC} and the \ac{RC} eigenmode, while  $\eta_{\rm pol}$ is the polarization overlap between the field generating the \ac{BSM} field and the direction of the magnetic field, and finally $\eta_{\hat z}$, is the spectral component of the total overlap and expresses the degree to which the \ac{HPL} is resonant with the \ac{RC}. Additional inefficiencies in the detection systems are neglected for now and discussed later. 

In this manuscript the direction of the magnetic fields will be referred to as the $\hat y$ direction, while the optical axis of the system (and hence both cavities) will be defined as the $\hat z$ direction, and the horizontal component with respect to the ground that is also orthogonal to the optical axis will be the $\hat x$ direction. These directions are illustrated in the diagram of the optical system in Figure~\ref{fig:Opt_sys}, discussed later in Section~\ref{Sec:Opt_sys}.
The wavefronts of fields propagating along the optical axis of the system will therefore be parallel to the $xy$ plane (ignoring their curvature).
With this in mind, the magnitude of the spatial component of the field overlap will be denoted as $\eta_{xy}$. For two fields that share the same polarization state and have complex spatial field distributions $E_1(x,y,z)$ and $E_2(x,y,z)$, $\eta_{xy}$ can be defined with the following overlap integral  \cite{paschotta2007mod},
\begin{align}
    \eta_{xy} &= \frac{|\int E_1^* E_2 \,dx\,dy|}{\sqrt{\int |E_1|^2 \, dx\,dy \int |E_2|^2     \,dx\,dy}}\\ 
    &= \frac{|\int E_1^* E_2 \,dx\,dy |}{\sqrt{P_1 P_2}}.
    \label{eq:eta_xy}   
\end{align}
In this case $P_1$ and $P_2$ give the powers of the fields; however, this equation can also be applied using any complex field distribution including that of a cavity eigenmode. In the case of \ac{ALPS}\,II, $E_1$ refers to the spatial mode of the \ac{HPL} light propagating though the production magnet string and $E_2$ refers to the spatial mode of the \ac{RC}.

The polarization component of $\eta$, denoted by $\eta_{\rm pol}$, is equal to the ratio of the circulating field in the \ac{PC} that is in the ideal polarization state,
\begin{equation}
    \eta_{\rm pol,\perp} = \frac{|E_x|}{|\vec E|} 
    \end{equation}
    and
    \begin{equation}
    \eta_{\rm pol,\parallel} = \frac{|E_y|}{|\vec E|},
\end{equation}
either perpendicular or parallel to the magnetic field for a scalar or pseudoscalar search, respectively. 

The spectral component of the field overlap $\eta_{\hat z}$, also referred to as the longitudinal field overlap, is discussed in more detail Section~\ref{Sec:Long_over}. It is dependent on the resonance condition of the cavity and can be approximated by the cavity Lorentzian: 
\begin{equation}
\eta_{\hat z}
\simeq \frac{1}{\sqrt{1+\left(\frac{\Delta\nu}{\nu_0}\right)^2}}.
\end{equation}
Here $\Delta \nu$ is the difference between the steady state frequency of the \ac{HPL} light circulating in the \ac{PC} and the nearest resonance of the \ac{RC}, $\nu_0$ is the frequency half-width-half-maximum of the cavity. In principle, the longitudinal field overlap also contains a component related to the frequency noise of the \ac{HPL} light with respect to the resonance of the \ac{RC} \cite{ortiz2020design}, however, as discussed later in this section, the contribution of this term is insignificant due to the configuration of the optical system for the first science campaign.


While \ac{ALPS}\,II was designed to utilize both a \ac{PC} and an \ac{RC}, the first science campaign was performed without the \ac{PC} in place for a number of reasons that will be discussed later in this section. Without the \ac{PC}, $P_{\!_{\rm PC}}$ is equivalent to the input power of the \ac{HPL} $P_{\rm i}$ in the calculation of the power measured at the output of the \ac{RC}:
\begin{equation}
    P_\phi = \eta^2 \beta_{\!_{\rm RC}}  P_{\rm i} \,\mathcal{P}_{\gamma\leftrightarrow \phi}^2 .
    \label{Eq:P_phi_intro}
\end{equation}
A magnetic field length product of $563.2\rm\,T\cdot m$ and a coupling of $4\times10^{-10}\rm\,GeV^{-1}$ (a factor of 20 above the \ac{ALPS}\,II design sensitive with two cavities) would result in $\mathcal{P}_{\gamma\leftrightarrow \phi} \approx 1\times10^{-14}$.
Therefore, using a laser wavelength of 1064\,nm, with an input power of 30\,W, a resonant enhancement factor in the \ac{RC} of 7000, and $\eta^2 = 100\%$ would lead to roughly 1 photon per day\footnote{Both Watts and `Photons/time' will be used as a unit of power throughout this paper. To convert power between each set of units the equation $P = h\nu\times({\rm photons/s})$ can be used, with $P$ representing power in Watts, $h$ being Planck's constant, and $\nu$ the frequency of the light.} at the output of the regeneration cavity.

With the shutter open the \ac{HPL} power directly after the \ac{RC} is given by
\begin{equation}
    P_{\rm open} = \eta^2\beta_{\!_{\rm RC}}  P_{\rm i} (T_1T_2...).
\label{Eq:P_open_ALPS_Intro}
\end{equation}
In this equation the factor $(T_1T_2...)$ refers to the combined transmissivity of all components mounted along the optical axis of the system in between the magnet strings and and is analogous to $\mathcal{P}_{\gamma\leftrightarrow \phi}$ in the closed shutter case. In fact, under the assumption that the parameters of the optical system $\eta$, $\beta_{\!_{\rm RC}}$, and $P_{\rm i}$ do not change between open and closed shutter measurements, the conversion rate $\mathcal{P}_{\gamma\leftrightarrow \phi}$ can be expressed entirely in terms of $(T_1T_2...)$ and the ratio $P_\phi/P_{\rm open}$:
\begin{equation}
   \mathcal{P}_{\gamma\leftrightarrow \phi} =  \sqrt{(T_1T_2...)\frac{P_\phi}{P_{\rm open}}}  .
\end{equation}
As Section~\ref{Sec:COB} explains, the optics mounted between the magnet strings not only generate the interference beatnotes that are used to maintain the resonance condition of the \ac{HPL} with respect to the \ac{RC}, but also play a role in the containment of the stray-light background. Therefore the transmissivity of these optics is very low with a measured combined transmissivity of the optics used in \ac{ALPS}\,II of $T_1T_2...=(9.7\pm1.2)\times10^{-23}$ (see Section~\ref{Sec:COB_in} for a detailed description of these measurements). Again assuming an input power $P_{\rm i} = 30$\,W, a power build of  $\beta_{\!_{\rm RC}}=7000$, and a field overlap of $\eta=100\%$, this would produce an open shutter power just after the cavity of $P_{\rm open}\sim2\times10^{-17}$\,W or roughly 100 photons per second.

To sense such ultra-weak signals, a heterodyne detection system was used \cite{PhysRevD.99.022001}. This technique works by optically mixing the weak signal field with an additional local oscillator laser to generate an interference beatnote that can then be measured with a photodiode, whereas the signal field alone would be much smaller than the photodiode's dark noise. This approach is extremely practical, as even with off-the-shelf components sensing the beatnote, heterodyne interferometry is capable of reaching the shot noise limit corresponding to a background of roughly one photon per measurement time \cite{hallal2022heterodyne}.

Searching for signals on the order of one photon per day when the shutter is closed means that the suppression of the stray-light background is critical to the sensitivity of the experiment. Any stray light from the \ac{HPL} that couples to the detection system can create a background that is indistinguishable from the science signal. This necessitates strict requirements on the `light tightness' of the optical system. Therefore, the goal is to allow less than a single photon of the 30\,W of input \ac{HPL} power at the exact frequency of the \ac{HPL} to couple spuriously to the \ac{RC} over the duration of the measurement (an equivalent power of $2\times10^{-25}$\,W for a measurement of one million seconds). Random backgrounds with a mean value $\gamma_{_{\rm BR}}$ significantly higher than 1 photon per measurement time will cause a $\sqrt{\gamma_{_{\rm BR}}}$ loss in sensitivity in terms of $\mathcal{P}_{\gamma\leftrightarrow \phi}$ and a $\sqrt[4]{\gamma_{_{\rm BR}}}$ loss in sensitivity in terms of $g_{\phi\gamma\gamma}$.

The exceptionally fine frequency resolution of the heterodyne detection system has the advantage of being able to exclude spurious signals from scattered light that experience more than a cycle of phase drift over the period of data taking. Therefore, for the measurement to be shot noise limited, only the power spectral density of the background integrated over the frequency resolution of the detection system must be below one photon per measurement time. Achieving such a sensitivity does not come without its own set of challenges. One of these challenges is that the phase noise between the \ac{HPL} and the laser used as the local oscillator laser must be held to a fraction of a cycle over the duration of the measurement. This is explained in more detail in Section~\ref{Sec:Het}.

While the \ac{PC} considerably boosts the sensitivity of the experiment, there are several advantages to characterizing the system without a \ac{PC} initially in place. Because the \ac{PC} has a nominal transmissivity\footnote{In this document transmissivity (and reflectivity) will refer to the ratio of power transmitted (and reflected) by an optic (or an optical system like a cavity) with respect to the input power.} of only $2.5\%$, operating the optical system without it also allows for a factor of $40\times$ more laser power to be incident on the wall, reducing the integration time needed to find the `light leaks'. It is also important to note that the design of the system, even without the PC in place, significantly attenuates the power of the \ac{HPL} incident on the \ac{RC} in order to contain the stray light, making it more challenging to measure the field overlap. Its characterization is therefore much easier without the \ac{PC} in place due to the increase in \ac{HPL} power at the detection system. Finally, in the two-cavity design, the field overlap is maintained by a control system designed to change the length of \ac{PC} to track the length changes of the free-running \ac{RC}. Not using a \ac{PC} significantly simplifies the control of the optical system. In this case, the \ac{RC} length can be left free-running, but the feedback control loop can actuate on the frequency of  the \ac{HPL} instead of the position of one of the \ac{PC} mirrors. Because the resonances in the piezoelectric actuator attached to the laser crystal are at much higher frequencies than the resonances of the mount used to perform the length actuation of one of the \ac{PC} mirrors it is possible to actuate much faster with this setup. The faster actuation is therefore able to suppress the residual \ac{RMS} frequency noise of the \ac{HPL} with respect to the \ac{RC} resonance to a level in which its contribution to the loss in field overlap is less that 1\%. This noise is therefore not considered in this analysis because it is responsible for significantly lower losses in the field overlap than the other sources and their corresponding uncertainties.

\subsection{Optical System}
\label{Sec:Opt_sys}

\begin{figure*}
    \centering
    \includegraphics[width=0.99\textwidth]{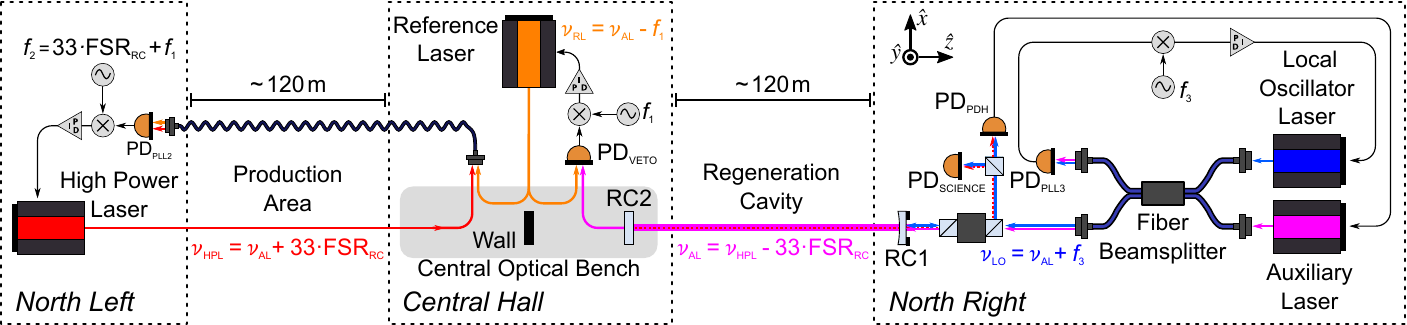}
    \caption{Top down view of the optical system used during the \ac{ALPS}\,II first science campaign.}
    \label{fig:Opt_sys}
\end{figure*}

The optical system for the first science campaign was designed to achieve five goals: to maximize the (1) \ac{HPL} power, (2) resonant enhancement in the \ac{RC}, and (3) coupling between the \ac{HPL} and \ac{RC} as much as possible while suppressing (4) the background signals due \ac{HPL} stray light as well as (5) the relative phase noise between the \ac{HPL} and the laser used as the local oscillator. In the following sections we explain how this is done.

A diagram of the optical system (top down view) used for the first science campaign is shown in Figure~\ref{fig:Opt_sys}. The optical components are located in one of three clean rooms, \ac{NL}, \ac{CH}, and \ac{NR}, that appear in the diagram in order from left to right. 
These cleanrooms are connected by the magnet strings, each of which are 116.8\,m in physical length (not shown in this diagram). The magnet string on the left side of the diagram will be referred to as the production magnet string, while the regeneration magnet string is on the right side of the figure. The \ac{RC} is shown on the right side of the figure, its eigenmode propagating through the bore of the regeneration magnet string. The  curved mirror of the \ac{RC}, housed in \ac{NR}, will be referred to as RC1, while the flat \ac{RC} mirror mounted to the \ac{COB} (discussed later in this section) will be referred to as RC2. This configuration will produce a cavity eigenmode with a waist position at RC2. A detailed description of the \ac{RC} and its characterization can be found in Ref.\,\cite{kozlowski2024designperformancealpsii}. There, a measurement of the cavity mirror $g$ parameters is used to infer (assuming that RC2 is flat) a spatial eigenmode with waist position at RC2 having a radius of 7.2\,mm along an axis oriented $11^\circ\pm5^\circ$ with respect to the horizontal and 6.6\,mm along an axis oriented $150^\circ\pm5^\circ$ with respect to the horizontal. Likewise, the results of calculating the beam radius at RC1 for the same orientations are 9.1\,mm and 9.2\,mm respectively. The minimum free aperture of the magnets varied between 47\,mm and 51\,mm and the magnets with the widest aperture were positioned near the end stations of the experiment to ensure that the losses from the cavity circulating fields clipping on the beam tube were as low as possible \cite{albrecht2021straightening}. In doing this, the free aperture of each of the individual magnets should be at least a factor 2.7 times the projected beam diameter at their position. Scans of the excess optical losses as a function of the \ac{RC} eigenmode position on its end mirrors presented in Ref.\,\cite{kozlowski2024designperformancealpsii} suggest the free aperture of the entire regeneration magnet string may be smaller than this due to an error in the positioning of one of the magnets. Nevertheless, this error is believed to only contribute roughly 16\,ppm of extra clipping losses, well below other sources of optical losses such as scattering  due to the surface roughness of the cavity mirrors. A summary of the \ac{RC} parameters during the closed shutter periods of the two science runs is also shown in Table~\ref{tab:RC_param}. A more detailed explanation of these parameters can be found in Section~\ref{Sec:RC}.

The \ac{HPL} supplies the power that drives the production of the \ac{BSM} field. Its frequency must therefore be held on a resonance of the \ac{RC}. This is accomplished with a series of \acp{PLL} using two other lasers, referred to as the \ac{AL} and \ac{RL}. Here, the frequency of the \ac{AL} is stabilized to a resonance of the \ac{RC} using the \ac{PDH} technique \cite{black2001introduction,pound1946electronic,drever1983laser}. 
The \ac{RL} is then interfered with the transmitted \ac{AL} on the \ac{COB} and phase locked to it. Finally, the \ac{HPL} is phase locked to the \ac{RL}. With this configuration the length changes of the \ac{RC} will be encoded in the phase of the \ac{AL} in transmission of the cavity and will transfer to the phase of the \ac{RL} and then the \ac{HPL} via the \acp{PLL} allowing the \ac{HPL} to be frequency stabilized to the \ac{RC} while maintaining some degree of isolation between them. This isolation is critical because the presence of stray \ac{HPL} light that couples to the \ac{RC} will create a background that is indistinguishable from a signal created from the regeneration of \ac{BSM} fields. As Section~\ref{Sec:COB} describes, the \ac{COB} utilizes two `light-tight' boxes (not shown in Figure~\ref{fig:Opt_sys}) to help block possible paths that scattered light from the \ac{HPL} could take to the \ac{PC}.
 
For all frequency stabilization loops in the optical system, whether they are \acp{PLL} or \ac{PDH} loops, Piezoelectric devices mounted to the laser crystals provide the fast actuation at frequencies from 100\,mHz to 40\,kHz, while a Peltier element also mounted to the laser crystal changes the laser crystal temperature, providing the slow, long-range actuation at frequencies below 100\,mHz.

A fourth laser on the \ac{NR} table, referred to as the \ac{LO}, is incident on the \ac{RC}, but held off resonance. It is used to form the interference beatnote with the regenerated signal for the heterodyne detection system. With all of this, the \ac{HPL} is held on a resonance of the \ac{RC} at a frequency 33 \acp{FSR} away from the \ac{AL}.\footnote{Here the \ac{FSR} of the cavity is defined as $c/2L_{_{\rm RC}}$ with $c$ being the speed of light in vacuum and $L_{_{\rm RC}}$ being the length of the cavity} The system also maintains the phase stability of the beatnote between the \ac{HPL} and the \ac{LO}.
This is critical as the regenerated signal is measured via this beatnote, by a heterodyne detection system (see Section~\ref{Sec:Het}) at the science \ac{PD} in \ac{NR}. A \ac{PD} located in \ac{CH}, referred to as the veto \ac{PD}, is used to monitor the stray light contribution to the light reflected from the \ac{RC} along with other parameters critical to evaluating the state of the optical system.

\begin{table*}
    \centering
    \makebox[\textwidth][c]{
    \begin{tabular}{c|cc}
Regeneration cavity parameter            &  Value $\left(\rm S_{\perp}\right)$ & Value $\left(\rm S_{\parallel}\right)$ \\ \hline
Free spectral range ($f_0$) & $\left(1222632.34^{\,+0.18}_{\,-0.46}\right)$\,Hz & $\left(1222632.41^{\,+0.79}_{\,-0.55}\right)$\,Hz        \\
Length ($L_{_{\rm RC}}$)    & $\left(122.601230^{\,+0.000030}_{\,-0.000017}\right)$\,m & $\left(122.601223^{\,+0.000055}_{\,-0.000079}\right)$\,m          \\
Storage time ($\tau$)       & $\left(6.850^{\,+0.079}_{\,-0.129}\right)$\,ms & $\left(6.753^{\,+0.076}_{\,-0.117}\right)$\,ms         \\
Total round-trip attenuation ($A$) & $\left(238.8^{+2.7}_{-4.6}\right)$\,ppm  & $\left(242.2^{+2.3}_{-4.7}\right)$\,ppm \\
Resonant enhancement ($\beta_{_{\rm RC}}$) & $7010^{\,+190}_{\,-300}$ & $6820^{\,+190}_{\,-270}$ \\
Power overlap ($\eta^2$) &  $0.47^{+0.18}_{-0.14}$  &  $0.49^{+0.23}_{-0.22}$  \\
    \end{tabular}}
    \caption{Summary of the measured parameters of the \ac{RC} during $\rm S_{_\perp}$ and  $\rm S_{_\parallel}$. All parameters in the table are dynamic that fluctuated during the science campaign. These fluctuations were observed between measurements performed during maintenance periods in the run. To assess the performance of the optical system during the closed shutter periods, the nearest set of maintenance parameters was assigned to  each point of valid closed shutter data. The mean of these assigned values over the full set of valid closed shutter data is what is reported here, while the uncertainties give the combination of the peak to peak range in combination with the systematic uncertainty for each measurement. For the \ac{FSR}, length, and storage time of the \ac{RC}, the dynamic drifts were the dominant source of uncertainty, while for the excess optical losses and the resonant enhancement the uncertainty had roughly equal contributions from the dynamic drifts of the parameters and the systematic uncertainty of the RC mirrors transmissivities. A detailed description of the \ac{RC} and its characterization can be found in Ref.\,\cite{kozlowski2024designperformancealpsii} as well as Section~\ref{Sec:RC}.
        \label{tab:RC_param}}
\end{table*}

The first science campaigns consisted of two measurement runs. In the first measurement run, referred to as $\rm S_{_\perp}$, the polarization states of the lasers traveling through the magnetic fields were aligned in the $\hat x$ direction, perpendicular to the magnetic fields $(\vec{E}\perp\vec B_{\rm ext})$, while for the second measurement run, $\rm S_{_\parallel}$, the laser polarization states were aligned in the $\hat y$ direction, parallel to the magnetic fields $(\vec{E}\parallel\vec B_{\rm ext})$. The laser polarization states on either end of both magnetic fields could be adjusted using \acp{HWP} (not shown in Figure~\ref{fig:Opt_sys}).

\begin{figure*}
    \centering
    \includegraphics[width=0.75\textwidth]{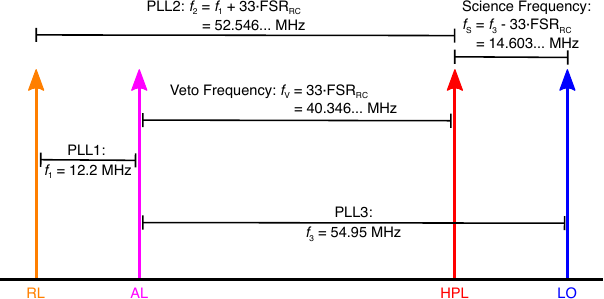}
    \caption{Frequencies of the lasers and PLLs.}
    \label{fig:freq_plan}
\end{figure*}

Figure~\ref{fig:freq_plan} shows how the frequencies of the various lasers were configured for the first science campaign. The optical frequencies of the lasers are roughly 282\,THz and their frequencies are all referenced to the length of the \ac{RC} via the frequency of the auxiliary laser which is frequency stabilized to a resonance of the cavity. The \ac{LO} is then phase locked to the \ac{AL} with an offset frequency of $f_3=54.95$\,MHz. This \ac{PLL} will be referred to as \ac{PLL}3. As Figure~\ref{fig:Opt_sys} shows, both lasers are coupled to optical fibers and interfered using a fiber beam splitter and therefore share the same spatial eigenmode as they propagate toward the cavity.

The \ac{COB}, located at the center of the experiment is a critical component of the optical system. It is designed to help sense the relative phase changes between \ac{HPL} and \ac{AL}, while also suppressing the stray-light background and housing the wall/shutter system. This is discussed in more detail in Section~\ref{Sec:COB}. On the \ac{COB}, the \ac{RL} is interfered with the \ac{AL} light in transmission of the \ac{RC}. This beatnote is then detected with the veto PD and is used to phase lock \ac{RL} via \ac{PLL}1 to \ac{AL} with an offset frequency of $f_1 = 12.2$\,MHz to the \ac{AL} light in transmission of the \ac{RC}.  The \ac{RL} light is also interfered with the \ac{HPL} light exiting the production magnet string on the \ac{COB}. The combine laser fields are then coupled to a 140\,m polarization-maintaining single mode optical fiber and sent to \ac{NL} where it is used to phase lock the \ac{HPL} to the \ac{RL} at \ac{PLL}2. Here the \ac{PLL}2 offset frequency is set to 33 times the \ac{FSR} of the cavity (the \ac{FSR} is roughly 1.22263\,MHz) plus an additional 12.2\,MHz offset to account for the \ac{PLL}1 frequency. This ensures that the \ac{HPL} is held at a frequency 33 \ac{FSR}s away from the \ac{AL} and thus on resonance with the \ac{RC}.

The transmissivities of the cavity mirrors are configured such that the majority of the regenerated field will exit the cavity via the mirror in \ac{NR} and be sensed by the science \ac{PD}. This light will form an interference beatnote with the \ac{LO} as it exits the \ac{RC}, at the heterodyne frequency $f_{\rm s}=54.95\,{\rm MHz}-33 {\rm \,FSR}\sim14.603{\rm \,MHz}$. Some of the regenerated light will also exit the cavity via the mirror in \ac{CH} and be incident on the veto detector. The regenerated signal will interfere with \ac{AL} inside the \ac{RC} to form an interference beatnote at the veto heterodyne frequency of $f_{\rm v}=-33$\,\ac{FSR}.

The phase stability of all of the interference beatnotes is critical to the \ac{ALPS}\,II heterodyne detection scheme. Therefore, the oscillators used to demodulate the science and veto signals, as well as those used for the \ac{PLL}s, are all synchronized to a global clock and left to run uninterrupted for the duration of a science run. To set the frequencies relative to the \ac{RC} \ac{FSR} (referred to from here on as $f_0$), $f_0$ is measured and used to fix the beatnote frequencies before the run. Changes to the length of the \ac{RC} that push its \ac{FSR} away from the initial value of $f_0$ can introduce phase offsets to the measurement. This is discussed in more detail in Section~\ref{Sec:Long_over}. Another point  discussed in more detail in Appendix~\ref{APP:Sec_DFR}, is that because the oscillators are driven by digital clocks, they cannot use any arbitrary frequency, but must instead use a frequency that can be represented by an integer fraction of the clock frequency.

The \ac{HPL} must remain coupled to the spatial eigenmode of the \ac{RC} with a frequency corresponding to one of its resonances. This can be checked by opening the shutter in the wall and allowing a small portion of the \ac{HPL} light to be injected to the \ac{RC} and measuring this power in transmission of the \ac{RC}. 
Due to the highly reflective optics used to contain the stray light on the\ac{COB}, the open-shutter HPL power ($P_{\rm open}$) is too weak to be observed directly using a power meter or photodetector. Instead, the amplitude of the \ac{HPL}-\ac{LO} beatnote must be measured at the science detector using the heterodyne detection system. The demodulation and calibration procedure is then applied to the data to give the total coupling of the \ac{HPL} to the \ac{RC}, along with information on the \ac{HPL}'s phase coherence with respect to the heterodyne detection system. During these open shutter periods the veto detector can also be used to check the spatial coupling of the \ac{HPL} to the \ac{RC}, as well as its phase coherence.

\subsection{Heterodyne Sensing}
\label{Sec:Het}
\ac{ALPS}\,II utilizes a heterodyne detection system in which the \ac{LO} is used as the local oscillator laser and is interfered with the signal field exiting the \ac{RC} at RC1. The power of the resulting field traveling away from the curved mirror of the cavity toward the science \ac{PD} can be expressed by
\begin{widetext}
\begin{equation}
    P_{\!_{\rm out}}(t) = P_{\!_{\rm LO}} + P_{\!_{\rm S}} +2\eta_{\!_{xy,{\rm LO}}}\sqrt{P_{\!_{\rm LO}}P_{\!_{\rm S}}}\cos{\{2\pi(\nu_{\!_{\rm LO}}-
    \nu_{\!_{\rm S}})t+\theta(t)}\} + ... {\hspace{2mm}.}
    \label{Eq:het_P_tot}
\end{equation}    
\end{widetext}
In this equation, the terms that include the power of \ac{AL} are not shown. $P_{\!_{\rm LO}}$ is the power of the \ac{LO} in reflection of the \ac{RC}, while $P_{\!_{\rm S}}$ is the power of the field at the signal frequency. It should be noted here that $P_{\!_{\rm S}}$ can refer to either the power of the regenerated field with the shutter closed ($P_\phi$, discussed in Section~\ref{Sec:ALPSII_des}), or the power in transmission of the cavity when the shutter is open ($P_{\rm open}$, discussed in more detail later in Section~\ref{Sec:Cal}). Because the \ac{LO} is not resonant with the \ac{RC} nearly all of its power is reflected from the cavity, while the signal field is generated inside the cavity itself and is transmitted through the \ac{RC} curved mirror. In reality, there is also some residual power of \ac{AL} reflected from the cavity; however due to the \ac{RC} impedance matching, this power is much lower than the \ac{LO} power and is ignored in this analysis. The spatial overlap between the signal field and \ac{LO} is given by $\eta_{\!_{xy,{\rm LO}}}$. This is also equivalent to the spatial overlap between the \ac{LO} and the \ac{RC} as the signal field is entirely in the spatial mode of the \ac{RC}. The frequencies of the fields are $\nu_{\!_{\rm LO}}$ for \ac{LO} and $\nu_{\!_{\rm S}}$ for the signal field which is identical to the frequency of the \ac{HPL}. $\theta(t)$ represents the time dependent relative phase offset between the interference beatnote at $\nu_{\!_{\rm LO}} - \nu_{\!_{\rm S}}$ and the phase of the oscillator used to demodulate the data at the heterodyne frequency. It can be expressed as a static phase plus a dynamic term $\theta(t)=\theta_0+\delta\theta(t)$. The static phase offset $\theta_0$ is arbitrary; however, the dynamic term $\delta\theta(t)$ must be suppressed as much as possible. Here, a \ac{RMS} of  0.1\,rad would lead to a loss of 1\,\% in the power measured by the heterodyne detection system \cite{hallal2022heterodyne}.

Due to several diagnostic pick-off mirrors, as well as the imperfect reflectivity and aperture of the mirrors in the path to the cavity, there are some optical losses as the light travels from the \ac{RC} to the science \ac{PD}. The power transmissivity on this path $T_{\!_{\rm sci}}$, was measured to be between 80\% and 90\% throughout the first science campaign. The power incident on the photodetector results in a voltage output by the photodetector that is proportional to its conversion gain $G$. The static voltage is removed with an analog high-pass filter, suppressing the terms $P_{\!_{\rm LO}}$ and $P_{\!_{\rm S}}$ in Equation~\ref{Eq:het_P_tot}, and the resulting voltage, approximated by
\begin{widetext}
\begin{equation}
    V_{\!_{\rm ADC}}(t) \simeq 2\eta_{\!_{xy,{\rm LO}}} T_{\!_{\rm sci}} G_{\!_{\rm AC}} \sqrt{P_{\!_{\rm LO}}P_{\!_{\rm S}}}\cos{\{2\pi(\nu_{\!_{\rm LO}}-
    \nu_{\!_{\rm S}})t+\theta(t)}\},
\end{equation}
\end{widetext}
is digitized by the \ac{ADC} with a sampling frequency $f_{m_1}=500{\rm \,MSa/s}$. Here $G_{\!_{\rm AC}}$ refers to the total conversion gain of the electronic chain of the science \ac{PD} and its electronic filters at the heterodyne demodulation frequency ($\sim14\rm\,MHz$). The amplitude of the voltage signal will be referred to as $V_{\!_{\rm AC}} = 2\eta_{\!_{xy,{\rm LO}}} T_{\!_{\rm sci}} G_{\!_{\rm AC}} \sqrt{P_{\!_{\rm LO}}P_{\!_{\rm S}}}$, while its frequency will be denoted by $f_{\!_{\rm S}} = \nu_{\!_{\rm LO}}-\nu_{\!_{\rm S}}$. While $V_{\!_{\rm AC}}$ in this equation refers to the amplitude of the deterministic signal measured at the science \ac{PD}, from here on the measured value of $V_{\!_{\rm AC}}$ is assumed to include some measurement noise.

The amplitude of this beatnote is the primary measurable for the experiment as it can be used to deduce the signal power $P_{\!_{\rm S}}$. The measurement of $V_{\!_{\rm AC}}$ is complicated by the fact that the phase of the beatnote (which is assumed to be static) is unknown. Therefore, the time series of the beatnote must be projected onto both cosine and sine components. Here, the `in-phase' term $I_{\!_{\rm DM1}}$ is used to refer to demodulation via a cosine function, while the `quadrature' term $Q_{\!_{\rm DM1}}$ refers to demodulation with a sine function. To avoid offsets being introduced by asymmetries in the demodulation process, a technique referred to as double demodulation is used \cite{hallal2022heterodyne}.
\begin{figure*}
    \centering
    \includegraphics[width=\textwidth]{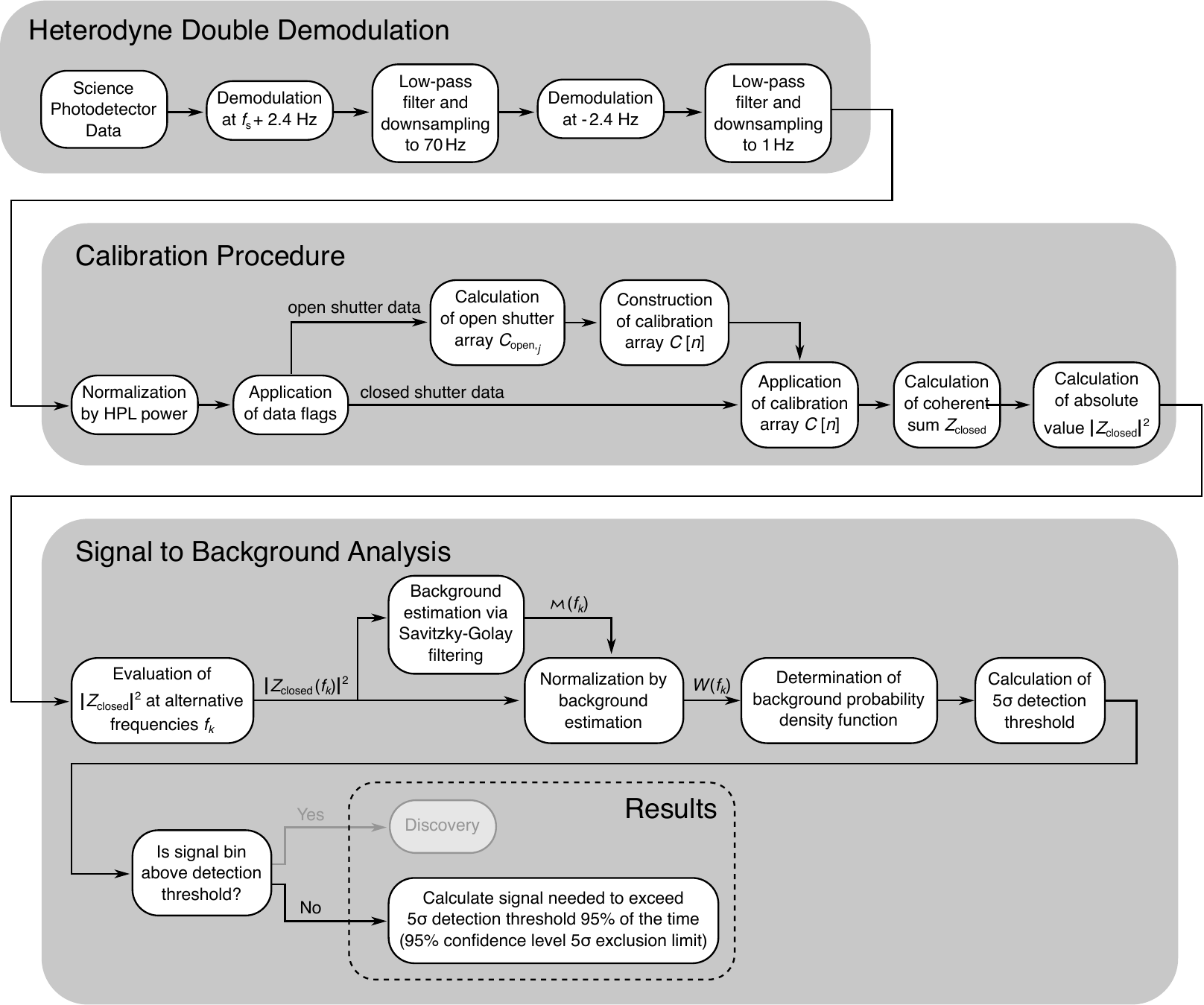}
    \caption{Flow chart of the analysis for the science \ac{PD} data. The data first undergo the process of double demodulation by the heterodyne detection system. The calibration procedure is then applied to the resulting data series to calculate the final result at the heterodyne signal frequency. The background is then estimated by performing the same analysis at alternative frequencies and a detection threshold is calculated based on the statistics of the background at these other frequencies. A discovery can be claimed if the signal at the expected frequency is above the detection threshold. If the signal is below the detection threshold, as was the case in the case in both runs of the first science campaign, exclusion limits can then be calculated.}
    \label{fig:flow}
\end{figure*}
This process is shown in the upper gray box of the flow chart in Figure~\ref{fig:flow}. Here, the first stage of demodulation is performed digitally at the nominal frequency difference between the two lasers plus a 2.4\,Hz offset: $f_{\!_{\rm S}}+2.4\,\rm{Hz}$. By expressing the data series resulting after the first demodulation $H_{\!_{\rm DM1}}[m]$, as the complex number $I_{\!_{\rm DM1}}+iQ_{\!_{\rm DM1}}$ the result of the first demodulation can be treated as a series of points on the complex plane:
\begin{widetext}
\begin{align}
    H_{\!_{\rm DM1}}[m] & = \, \overbrace{ V_{\!_{\rm ADC}}[m] \,\cos\left\{2\pi(f_{\!_{\rm S}}+2.4\,{\rm Hz})\frac{m}{f_{m_1}}\right\}}^{I_{_{\rm DM1}}} + i \,\overbrace{V_{\!_{\rm ADC}}[m] \sin\left\{2\pi(f_{\!_{\rm S}}+2.4\,{\rm Hz})\frac{m}{f_{m_1}}\right\}}^{Q_{_{\rm DM1}}}   +... \label{Eq:HetDM0}\\
    & = V_{\!_{\rm AC}}\cos{\left\{2\pi m \frac{f_{\!_{\rm S}}}{f_{m_1}}+\theta[m] \right\}} \, e^{i2\pi(f_{\!_{\rm S}}+2.4\,{\rm Hz})\frac{m}{f_{_{m_1}}}}    +... \label{Eq:HetDM1}\\
    & = \frac{V_{\!_{\rm AC}}}{2}\left(e^{i\left(2\pi m \frac{f_{\!_{\rm S}}}{f_{_{m_1}}}+\theta[m]\right) } + e^{-i\left(2\pi m \frac{f_{\!_{\rm S}}}{f_{_{m_1}}}-\theta[m]\right) }\right) \, e^{i2\pi(f_{\!_{\rm S}}+2.4\,{\rm Hz})\frac{m}{f_{_{m_1}}}}  +...    \label{Eq:HetDM2}\\
    & = \frac{V_{\!_{\rm AC}}}{2} e^{i\left(2\pi m \frac{2.4\,{\rm Hz}}{f_{_{m_1}}}-\theta[m]\right) } +... \hspace{0.3cm}.
    \label{Eq:HetDM3}
\end{align}   
\end{widetext}
An \ac{FIR} filter is then used to down-sample $I_{_{\rm DM1}}$ and $Q_{_{\rm DM1}}$ to the desired data rate of $f_m=70$ samples per second. Terms at higher frequencies, such as those at the doubled frequency $2f_{\!_{\rm S}}$, or at the \ac{PLL}3 frequency, are not relevant as they are eliminated by this process. Also, the result will be the same whether the demodulation is expressed in terms of the arrays $I_{\!_{\rm DM1}}[n]$ and $Q_{\!_{\rm DM1}}[n]$ or using the complex amplitude and phase $\frac{V_{\!_{\rm AC}}}{2} e^{i2\pi(f_{\!_{\rm S}}+2.4\,{\rm Hz})\frac{n}{f_{m}}}$. To reduce the number of terms to keep track of in the second demodulation we used the formulation on the complex plane. The science signal is thus a complex 2.4\,Hz oscillation in this data.\interfootnotelinepenalty=10000\footnote{ 2.4\,Hz is chosen as the offset frequency as it is well within the Nyquist frequency of the downsampled sampling frequency at 70\,Hz and also shares a greatest common divisor with this frequency of less than one. This is meant to reduce potential effects from aliasing when the signal is down sampled \cite{hallal2022heterodyne}.} 

The second demodulation is then applied in post-processing by multiplying the result of the first demodulation with $e^{i2\pi(-2.4\,{\rm Hz})\frac{n}{f_{m}}}$. Here the sign of the frequency matters because the data are now complex. Performing the additional demodulation in post-processing helps mitigate the effects of static offsets in the result of the first demodulation stage. More details on this process can be found in Ref.\,\cite{hallal2022heterodyne}. After the data series is demodulated for the second time the data is down-sampled to the final data rate of 1 sample per second using another \ac{FIR} filter. An investigation into the down-sampling showed no appreciable difference for measurements derived with the down-sampled data using this process versus the data at 70 samples per second.

The full process of two stages of demodulation, then filtering and down-sampling produces what we will refer to as the `$H$-function' or,
\begin{equation}
    H[n] = \frac{V_{\!_{\rm AC}}}{2}e^{-i\theta[n]}
    \label{eq:H_n}.
\end{equation}
In this equation $H[n]$ is a data series with an index $n$ that also happens to be $n$ seconds into the run due to the data rate of 1\,Hz. While it is not explicitly shown in the equation above, each point in the series will also have a random value drawn from a probability distribution on the complex plane to account for the measurement noise in the system. 

\begin{figure}[t]
    \centering
 \begin{subfigure}[b]{0.3\textwidth}
    \centering
    \includegraphics[width=\textwidth]{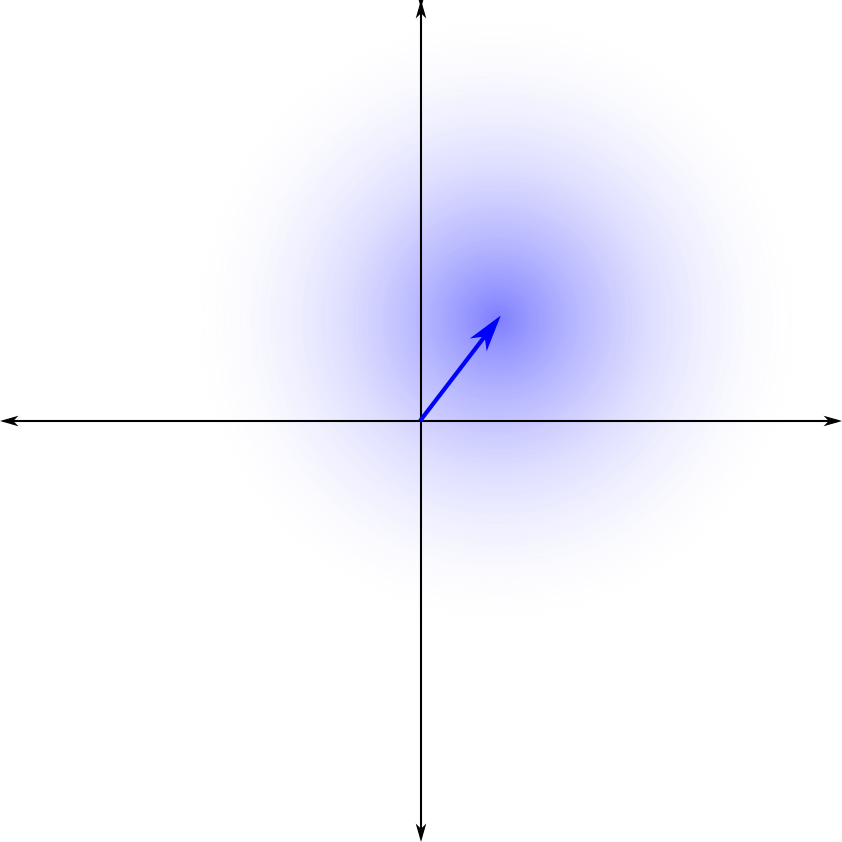}
    \caption{
    \label{fig:H_n}}
    \end{subfigure}\hspace{2cm}
     \begin{subfigure}[b]{0.3\textwidth}
    \centering
    \includegraphics[width=\textwidth]{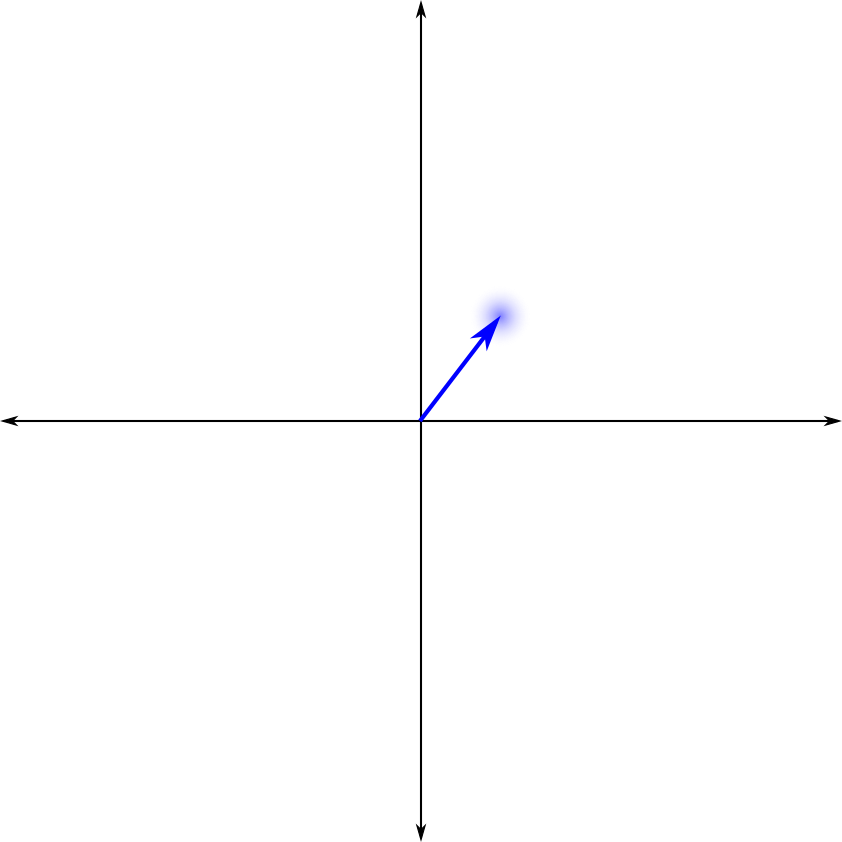}
    \caption{
    \label{fig:Z_n}}
    \end{subfigure}
    \caption{(a) $H[n]$ represented on the complex plane. (b) $Z$-function represented on the complex plane after summing over 100 coherent points of $H[n]$.}
\end{figure}

The expectation value of $H[n]$ for a given sample is a complex number with an amplitude given by half the amplitude in voltage of the beatnote between the \ac{LO} and the signal field at that time and a phase equal to the phase difference between the beatnote and the oscillators used to perform both demodulations:
\begin{align}
    \langle H[n]\rangle &=  \frac{V_{\!_{\rm AC}}}{2}e^{-i\theta_0} \\
    & = \eta_{\!_{xy,{\rm LO}}} T_{\!_{\rm sci}} G_{\!_{\rm AC}} \sqrt{P_{\!_{\rm LO}} P_{\!_{\rm S}}} e^{-i\theta_0} .
    \label{eq:h_func}
\end{align}
Figure~\ref{fig:H_n} shows a representation of  $H[n]$ on the complex plane. Here the `cloud' is the  probability distribution representing the measurement noise. As the figure shows, the sum of the signal and the cloud is offset from the origin by a blue vector which for $H[n]$ has an amplitude of $\frac{V_{\!_{\rm AC}}}{2}$ and a phase of $-\theta_0$. As mentioned above, the actual values of $\theta_0$ from the runs are unknown and do not affect the results; however, it is extremely important to understand the dynamic component of the phase $\delta\theta$, as it could lead to decoherence of the appearance of the signal in the data.

The values of $H[n]$ can then be integrated to produce what is referred to as the $Z$-function as the equation
\begin{equation}
     Z = \frac{1}{N}\sum^N_{n = 1} H[n]
     \label{Eq:Z_orig}
\end{equation}
shows. The $Z$-function is equivalent to calculating the average value of $H[n]$. Thus, as $H[n]$ is summed, the expectation value of $Z$ remains constant. If the measurement noise present in $V_{_{\rm AC}}$ (and hence $H[n]$) is uncorrelated, the variance of $Z$ decreases with $1/N$ (this is shown experimentally later on in Figure~\ref{fig:exp_op_cl}). The reason is that even though the variance in the sum $H[n]$ increases linearly with $N$, when calculating the variance of $Z$, the variance of the sum is multiplied by the square of the prefactor, or $1/N^2$. Figure~\ref{fig:Z_n} shows the result of calculating $Z$ from a series of 100 values of $H[n]$ with an expectation value and uncertainty equivalent to Figure~\ref{fig:H_n}. Here, the area on the complex plane that contains some fraction of the probability distribution in $Z$ is a factor of 100 less than the area that contains the same fraction of the probability distribution for individual values of $H[n]$.

The power of the signal leaving the \ac{RC} can then be expressed as the absolute square of the $Z$-function using the equation
\begin{equation}
    P_{\!_{\rm S}} =  \frac{|Z|^2}{\eta_{\!_{xy,{\rm LO}}}^2 T_{\!_{\rm sci}}^2 G_{\!_{\rm AC}}^2 P_{\!_{\rm LO}}},
\end{equation} 
assuming that the relative phase $\theta(t)$ is constant, and ignoring the effects of photon counting statistics and technical noise on $Z$.
If the measurement is limited by photon counting statistics, the expected measured photon rate will be one photon per measurement time. Therefore the sensitivity of the system in terms of power will improve linearly with integration time. If a signal is present above quantum noise,\footnote{Throughout this manuscript the phrases quantum noise and shot noise will refer to the noise due to photon counting statistics.} the signal-to-noise ratio of the measurement of $|Z|^2$ and thus $P_{\!_{\rm S}}$ will be equal to the expected number of photons measured over that period of time in the signal field \cite{PhysRevD.99.022001}. 

For the heterodyne detection system to work properly, the phase difference $\theta$ between the \ac{LO}, the signal field, and the oscillators used in the double demodulation scheme must remain stable during the period of time in which the summation takes place. This is one of the most significant tasks of the optical system. As mentioned earlier, the phase coherence between these fields can be directly monitored during the open shutter periods by calculating $Z$ for these periods and examining the stationarity of its phase. When the shutter is open, the signal field is the \ac{HPL} light incident on the \ac{RC} after passing through the \ac{COB} optics. 

With a total run time of more than a month for $\rm S_{_\parallel}$ the frequencies of the \ac{PLL}s and demodulation stages must be known relative to each other with nHz accuracy to ensure that the error in these frequencies does not contribute to significant phase noise in the measurement.\footnote{While roughly 1,000,000 seconds of valid data were acquired during $\rm S_{_\parallel}$, the experiment ran for over 4,000,000 seconds during this time frame. Therefore, 0.1\,rad would translate to roughly a 4\,nHz error in the frequencies used.} 
It is therefore important to understand how these signals are actually represented by the digital oscillators producing them. In all cases, the signals are supplied by function generators with a clock rate of 1\,GHz and 48 bit precision.\footnote{It should also be noted here that all devices used in the experiment are synchronized to the same global clock.} This means that the actual frequency of any signal output by these devices must be some integer multiple of $1\rm\,GHz/2^{48}$ ($\sim$3.5527\,{\textmu}Hz). To be clear, this does not prevent the  frequencies from being \textit{known} with nHz precision, only that they cannot be set with that level of precision, which is unimportant. This difference is discussed in more detail in Appendix~\ref{APP:Sec_DFR}.

\subsection{Central Optical Bench}
\label{Sec:COB}

\begin{figure*}
    \centering
    \includegraphics[width=\textwidth]{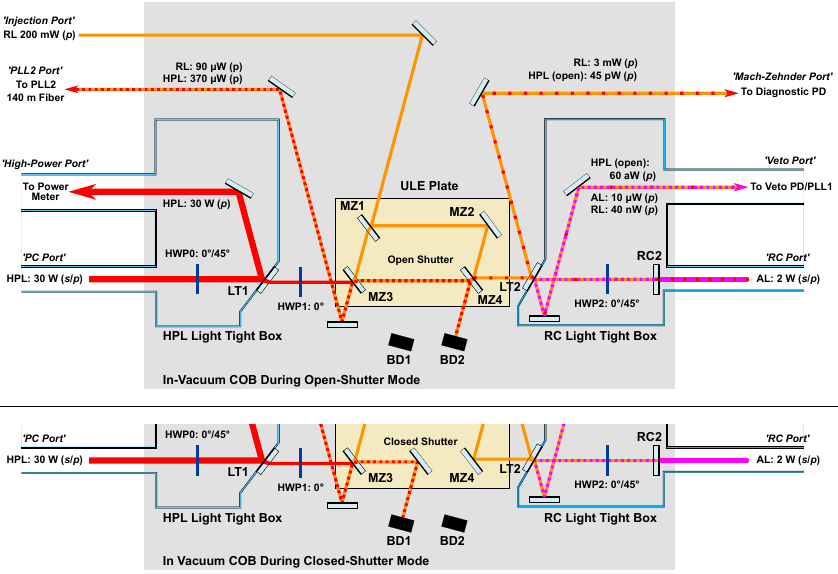}
    \caption{The \ac{COB} layout for the first science campaign during open-shutter mode is shown in the upper diagram with the lower diagram showing the optical axis of the \ac{COB} during closed shutter mode. Here, the lasers powers at different points on the \ac{COB} are inexact, but represent typical values measured during the science runs.}
    \label{fig:COB}
\end{figure*}

The \ac{COB} has three primary functions in the experiment. It houses the wall and shutter system, suppresses the spurious coupling of \ac{HPL} light to the \ac{RC}, and supports the transfer of the phase information of the \ac{AL} light transmitted by the \ac{RC} to the \ac{HPL} even when the shutter is closed. Figure~\ref{fig:COB} shows the design of the \ac{COB} for the first science campaign. In this diagram the thicker black edge of the mirrors indicate their reflective surface. The main practical differences between this layout and the \ac{COB} design for the full optical system with a \ac{PC} is that this layout does not have a \ac{PC} mirror on it. Thus, up to 30\,W of \ac{HPL} power is available on the \ac{COB}, when only 1\,W would be possible if the \ac{PC} were in place.

The shutter system is a remotely operable wheel in which different materials can be used to fill slots and, depending on the position of the wheel, act as the wall. One of the slots must be left unoccupied for the open shutter periods. The upper diagram of the \ac{COB}in Figure~\ref{fig:COB} shows the system when the shutter is open, while the lower diagram shows the optical axis of the \ac{COB} when the shutter is closed. Here, the \acp{BD} \ac{BD}1 and \ac{BD}2 contain the \ac{HPL} stray light when the shutter is closed and open respectively. For $\rm S_{_\perp}$, a silver mirror that directed the residual \ac{HPL} light to a beam dump was used as the wall. Between the $\rm S_{_\perp}$ and $\rm S_{_\parallel}$ run, two other shutter materials were implemented, including a piece made from solid aluminum, and another solid aluminum piece covered in foil coated with a black ultra-highly-absorptive layer. Subsequent testing indicated that the aluminum piece with the black foil provided the best suppression of stray light over short test phases and therefore it was used as the wall for $\rm S_{_\parallel}$.\footnote{This may also have been related to the fact that only one of the five open slots on the filter wheel was occupied by a shutter for $\rm S_{_\perp}$ and all but one was occupied for $\rm S_{_\parallel}$. This is under further investigation.} No matter what material was used for the wall no light is expected to directly transmit it as the silver mirror should provide over 25 \textit{orders of magnitude} suppression of the directly transmitted light (in addition to the suppression of the \ac{COB} optics), while the solid aluminum pieces should provide more than 100,000 \textit{orders} of suppression.

The \ac{COB} has two light-tight boxes that sub-divide it into three areas. The \ac{HPL} area on the left, the \ac{RC} area on the right, and the intermediate \ac{RL} area in between them. The purpose of this design is to contain the stray light from th \ac{HPL} as well as possible and try to prevent it from coupling to the \ac{RC}, while also allowing the \ac{HPL} to be phase locked to the \ac{AL}. The light-tight boxes therefore cannot be completely sealed in order to allow for the stabilization of the \ac{HPL} frequency as well as open shutter measurements. To accommodate this they are equipped with the \ac{HR} mirrors LT1 and LT2. Because there are also light-tight boxes outside the vacuum system connected to the exit flanges of the light-tight boxes inside the vacuum system, the only way in and out of them is through the magnet bore or LT1 and LT2. Each has a transmissivity  of $12.4\pm0.4$\,ppm for $p$-polarized ($\hat x$ direction) light with a wavelength of 1064\,nm at room temperature using an angle of incidence of $35^\circ$. The low transmissivity was chosen to help suppress the stray light signals that pass between the areas of the experiment.

To keep the \ac{HPL} and the \ac{RC} as separated as possible, \ac{RL} is split into two paths to perform the transfer of phase information from the \ac{AL} to the \ac{HPL}. This splitting is done by a \ac{MZ} interferometer. The beam is initially split by the mirror MZ1. Any changes in the differential path-length  between the two paths from MZ1 to MZ4 would lead to a phase offset between the heterodyne local oscillator field and the regenerated signal, as well as the open shutter signal. For this reason the \ac{MZ} mirrors are mounted to an \ac{ULE} glass plate. The \ac{MZ} alignment was performed with the \ac{COB} installed in the central vacuum tank while it was vented. This was done using a CCD outside the vacuum system to observe the position of the beam in transmission of one of the \ac{MZ} paths, while blocking the other path. With this process, it was possible to align the paths to within $\pm10$\,{\textmu}rad of each other.

The \ac{RL} light for the \ac{RL}-\ac{AL} interference is reflected by MZ1 ($R_{\!_{\rm MZ1}}=1.51\pm0.02\%$) and then passes through LT2 via reflections from two \ac{HR} mirrors, MZ2 and MZ4. With 200\,mW of \ac{RL} power injected to the \ac{COB} at the `Injection Port', this leaves roughly 40\,nW of power incident on the \ac{RC} at the mirror RC2, where most of it is reflected. With 10\,{\textmu}W of power in \ac{AL} in transmission of \ac{RC}, the interference beatnote formed here has an amplitude on the order of 1\,{\textmu}W.{\interfootnotelinepenalty=10000\footnote{Assuming unity spatial overlap, the interference beatnote amplitude in power $P_{_{\rm AC}}$, for two fields with powers $P_1$ and $P_2$ can be found using $P_{_{\rm AC}}=2\sqrt{P_1P_2}$.}} This light is then directed to the out-of-vacuum veto \ac{PD} which is used to perform the sensing for \ac{PLL}1.

To form the interference beatnote with \ac{HPL}, \ac{RL} passes through MZ1 and MZ3 ($T_{\!_{\rm MZ3}}=452\pm10$\,ppm) where it interferes with the \ac{HPL} field transmitted by LT1. At this point the \ac{RL} power is 90\,{\textmu}W while the \ac{HPL} power here with 30\,W injected to the `\ac{PC} port' is 370\,{\textmu}W, leading to an interference beatnote with an amplitude on the order of 360\,{\textmu}W. This light is sent out of the vacuum system at the `\ac{PLL}2 port' and coupled to the 140\,m fiber that transfers the light to the \ac{PLL}2 photodetector in \ac{NL}.

Because tests of the backgrounds before the first science campaign did not show any dependence on the \ac{HPL} polarization, the \ac{HPL} light was set to be $p$-polarized upon incidence on LT1 to provide the maximum signals for the interference beatnotes and open shutter periods, as all on axis \ac{COB} mirrors were more transmissive for $p$-polarized light. All of the lasers in the setup are therefore tuned to nominally be $p$-polarized when incident on the 35$^\circ$ angle of incidence mirrors on the \ac{COB}. For this reason, the transmissivity values quoted in this manuscript are all for $p$-polarized light unless stated otherwise.

For $\rm S_{_\perp}$, the \ac{HPL} light injected to the production magnet string was already $p$-polarized so no other optics were necessary to achieve this and $\rm HWP_0$ was not installed. For $\rm S_{_\parallel}$, the light injected to the production magnet string was $s$-polarized and $\rm HWP_0$ was oriented to rotate the polarization of the \ac{HPL} light by 90 degrees. For $\rm S_{_\perp}$ $\rm HWP_2$ was tuned to have no effect on the \ac{AL} light in transmission of the cavity as it was already $p$-polarized (oriented in the $\hat x$ direction). For $\rm S_{_\parallel}$, because \ac{AL} was $s$-polarized (oriented in the $\hat y$ direction), $\rm HWP_2$ was adjusted to rotate the \ac{AL} light in transmission of the \ac{RC} to be $p$-polarized. $\rm HWP_1$ and $\rm HWP_2$ were mounted to remotely rotatable stages in order to periodically invert the sign of specific stray light paths and break their coherence, however tests of this procedure showed its impact on the stray light was only marginal and therefore it was not used during the science runs to avoid the risk of introducing additional systematic effects.

When the shutter is open, the \ac{HPL} must pass through mirrors LT1, MZ3, MZ4 ($T_{\!_{\rm MZ4}}=275\pm10$\,ppm), LT2 before being incident on the \ac{RC}.
The combined transmissivity of these optics, and hence $T_{\!_{\rm COB}}$, was measured to be $T_{\!_{\rm COB}} = 1.9\pm0.1\times10^{-17}$. These measurements are discussed in more detail in Section~\ref{Sec:COB_in}.  With the shutter open, the \ac{HPL} incident on the \ac{RC} was roughly $6\times10^{-16}$\,W when 30\,W of \ac{HPL} power was used. Including mirror RC2, the on-axis optical attenuation between the two magnet strings was $T_{\!_{\rm COB}} \cdot T_{\!_{\rm RC2}}= 9.7\pm1.2\times10^{-23}$ when the shutter was open. When considering the transmissivity of the \ac{RC} and the coupling of the \ac{HPL} to its eigenmode, on the order of 10\,aW of \ac{HPL} power is expected at the science detector when the shutter is open.

With \ac{RL} aligned to \ac{AL} in transmission of the cavity, \ac{RL} serves as a reference of the position of the cavity eigenmode along the optical axis after MZ4. Therefore, a diagnostic \ac{PD} at the output of the `Mach-Zehnder Port' in reflection of LT2 was used to help align the \ac{HPL} to \ac{RL} where the power of the \ac{HPL} was much higher than after LT2. 

The \ac{HPL} light in reflection of LT1 is directed out of vacuum at the `High-Power Port', to a power meter that measured the \ac{HPL} power continuously during the science runs. Tests over short time periods showed no apparent difference in the background levels when this light was incident on the power meter versus a number of other beam dumps. The power meter is housed in an out-of-vacuum light-tight box mentioned earlier in this section, which connects to the vacuum flange at the High-Power port to try to limit any stray light from scattering into the central cleanroom.

\section{Closed-Shutter Calibration}
\label{Sec:Cal}

As explained in Section~\ref{Sec:ALPSII_des}, the conversion rate of an electromagnetic field passing once through one of the magnet strings and converting to a \ac{BSM} field (and vice versa) can be expressed using measured parameters as\footnote{Here the double pass conversion rate, $\mathcal{P}_{\gamma\rightarrow a\rightarrow\gamma}$, of an electromagnetic field converting to a \ac{BSM} field in the magnet string before the wall and reconverting to an electromagnetic field in the magnet string after the wall is simply $\mathcal{P}_{\gamma\rightarrow a\rightarrow\gamma} = \mathcal{P}_{\gamma\leftrightarrow \phi}^2$.}
\begin{equation}
\label{Eq:Prob_meas}
\mathcal{P}_{\gamma\leftrightarrow \phi} = \sqrt{\frac{P_\phi}{\eta^2 \beta_{\!_{\rm RC}}  P_{\rm i}}}.
\end{equation}
Assuming there are no effects that alter the power or pointing of the \ac{HPL} outside of the transmissivity on the optics on the \ac{COB}, the equation
\begin{equation}
    P_{\rm open} = \eta^2 \beta_{\!_{\rm RC}}  T_{\!_{\rm RC2}} T_{\!_{\rm COB}}  P_{\rm i} ,
    \label{Eq:P_open}
\end{equation}
gives the \ac{HPL} power in transmission of the \ac{RC} when the shutter is open. Here, Equation~\ref{Eq:P_open_ALPS_Intro} from Section~\ref{Sec:ALPSII_des} was modified to include the specific transmissivities for the factor $(T_1T_2...)$. In this case $T_{\!_{\rm COB}}$ is the combined transmissivity of the four mirrors oriented at 35 degrees on the \ac{COB} (for $p$-polarized light), while $T_{\!_{\rm RC2}}$ is the transmissivity of the flat cavity mirror on the \ac{COB} (RC2 in Figure~\ref{fig:Opt_sys}). The factor $\beta_{\!_{\rm RC}} T_{\!_{\rm RC2}}$ gives the nominal transmissivity of the \ac{RC} for  a laser on resonance with a perfectly spatially coupled beam. The open shutter power is also dependent on the total field overlap $\eta$. The spatial and spectral components of $\eta$ will determine the coupling of the \ac{HPL} to the \ac{RC}. Because the \ac{COB} is designed to be significantly more transmissive for the properly configured polarization state, light in the incorrect polarization state will not pass through the \ac{COB} optics. Thus, the open shutter power will depend on the polarization component of $\eta$.



With this the equation for $\mathcal{P}_{\gamma\leftrightarrow \phi}$  can be further simplified using the measured open shutter power:
\begin{equation}
    \mathcal{P}_{\gamma\leftrightarrow \phi} = \sqrt{T_{\!_{\rm COB}}T_{\!_{\rm RC2}} \frac{P_\phi}{P_{\rm open}}}
    \label{Eq:g_a_rat}.
\end{equation}
This shows that if the open shutter power could be predicted for a given period of closed shutter data, $\mathcal{P}_{\gamma\leftrightarrow \phi}$ could be calculated in terms of the transmissivity of the optics between the two magnetic field regions and the mean of the ratio of the measured power when the shutter is closed versus the predicted open shutter power. This is critical as the transmissivity of the optics on the \ac{COB} and the flat cavity mirror are essentially static parameters that do not need to be measured during data taking.  
Therefore the only dynamic variables which must be measured during the run are $P_\phi$ and $P_{\rm open}$.\footnote{Other dynamic variables such as the \ac{HPL} power injected to the production string were measured during the run, but under these assumptions they are not strictly necessary to calibrate the results in terms of $\mathcal{P}_{\gamma\leftrightarrow \phi}$.}

In using the ratio $P_\phi/P_{\rm open}$ to calibrate the results, we rely on the assumption that the \ac{COB} optics will only attenuate the power of the \ac{HPL} and not perturb the spatial eigenmode with respect to the eigenmode of the \ac{BSM} field. There are a number of systematic effects that could invalidate this assumption and, as Section~\ref{Sec:Sys_eff_op} explores, several checks were in place specifically to ensure that this was not the case.

The open and closed shutter powers cannot be measured simultaneously. However, because the misalignment of the system due to drifts typically only became significant over several days, it was possible to calibrate the closed shutter data using the nearest neighboring open shutter periods. Open shutter measurements were performed daily during $\rm S_{_\perp}$ and several times per day for $\rm S_{_\parallel}$, to minimize the drifts between open and closed shutter periods. For $\rm S_{_\perp}$ this corresponded to 19 open shutter periods during the 15 day run, while for  $\rm S_{_\parallel}$ 78 open shutter periods were performed over 41 days. The average time between the closed shutter data and its nearest open shutter period was roughly 8\,hours for both runs. A critical component of this is ensuring that the system was in a stable state during the periods in which data was acquired for both the times when the shutter was open or closed. The process of identifying these times and flagging the data based on the status of the system is explained in Section~\ref{Sedc:flags}. 

Because it was possible to continuously measure $P_{_{\rm HPL}}$, the closed and open shutter data was also normalized by the \ac{HPL} power to further reduce the systematic uncertainty of the measurement, and this is described in Section~\ref{Sec:HPL_norm}. The rest of the calibration procedure to calculate the result from the heterodyne detection system in terms of $Z$ and then calibrate this in $\mathcal{P}_{\gamma\leftrightarrow \phi}$ is explained in Section~\ref{Sec:Op_cl_cal}. 
A second \ac{PD} called the veto detector, was implemented on the opposite side of the \ac{RC} from the science detector, to verify potential signals and we examine this in Section~\ref{Sec:veto_det}. The uncertainty in the calibration of the ratio $P_\phi/P_{\rm open}$ and the results in terms of $\mathcal{P}_{\gamma\leftrightarrow \phi}$ is then scrutinized in Sections~\ref{Sec:C_n_uncert} and \ref{Sec:results_uncert} respectively.

It should be noted that the purpose of the calibration procedure is to provide the results in terms of the electromagnetic field to \ac{BSM} field conversion rate $\mathcal{P}_{\gamma\leftrightarrow \phi}$ and therefore this manuscript does not consider the effect of the magnet strings on the coupling of the \ac{BSM} fields. Exclusion limits, derived from the results presented in Section~\ref{Sec:results}, on the coupling between electromagnetic fields and pseudoscalar, scalar, vector, and tensor bosons can instead be found in Ref.~\cite{ALPSII_science}. 
 
\subsection{Potential Systematic Effects}

\label{Sec:Sys_eff_op}

There are several effects that could spoil the calibration of the regenerated signal via the open shutter signal. For example, any wedge angle in the optics between the two magnetic fields will deflect the light that transmits from the production area to the regeneration cavity during the open shutter periods. The \ac{BSM} fields however, will be unaffected by these wedge angles during the closed shutter periods. Therefore, if the system is aligned by optimizing the open shutter signal, compensating for the wedge angles would lead to a misalignment of the \ac{BSM} field with respect to the cavity eigenmode. The wedge angles of the \ac{COB} substrates were measured by the manufacturer to be between 1.7 to 5.5\,{\textmu}rad, producing a deflection angle on the beams between 0.8 to 2.8\,{\textmu}rad. 

The total deflection angle of the \ac{COB} due to the wedge angles of the on-axis optics was measured by two different methods. First, a green He-Ne laser was directed through the beam tube from \ac{NR} to \ac{NL} through the \ac{COB} and it position was recorded on a camera. Then, the \ac{COB} was removed and the change in the position of the green beams on the camera was measured. This gave a deflection angle of $10\pm5$\,{\textmu}rad in the horizontal direction and $5\pm5$\,{\textmu}rad in the vertical direction for a total deflection angle of $11\pm7$\,{\textmu}rad. The deflection angle of the \ac{COB} optics was also measured using an autocollimator once the \ac{COB} was removed. In these measurements the autocollimator beam passed through the \ac{COB} optics, was then normally incident on a mirror, and then once again passed through the \ac{COB} optics, before being incident on the autocollimator detection system. This procedure was then repeated several times pushing the \ac{COB} in and out of the setup, but always being normally incident on the mirror. The difference of the angle when the \ac{COB} was in the setup versus when it was not was measured by the autocollimator to be $9\pm2$\,{\textmu}rad in the horizontal direction and $1\pm1$\,{\textmu}rad in the vertical direction. This would correspond to a total \ac{COB} deflection angle of $5\pm1$\,{\textmu}rad and is within the error of the measurements using the He-Ne laser. Considering that the \ac{COB} optics were mounted with a random azimuthal orientation, this is consistent with the wedge angles measured by the manufacturer. If the system were ideally aligned with the \ac{HPL} perfectly coupled to the \ac{RC} during the open shutter periods, such a deflection angle would lead to an additional loss of the spatial power overlap of the regenerated field to the \ac{RC} eigenmode of 2\%. If there were an additional 10\,{\textmu}rad offset in the alignment of the \ac{HPL} with respect to the \ac{RC} eigenmode along the same axis as the wedge angle (corresponding to a loss in the spatial overlap of 8\%), this would correspond to an additional uncertainty in the spatial power overlap of between 6 and 10\%. Therefore, an additional 10\% uncertainty has been incorporated into the closed to open shutter power ratio because such misalignments were typical during the first science campaign.

Another effect which could reduce the measured signal power is the loss in phase coherence between the regenerated field and the heterodyne detection system due to optical path length changes on the \ac{COB}. One cause of this could be temperature changes in the substrates on the \ac{COB} that the various lasers pass through on their way to forming interference beatnotes for the \ac{PLL}s. The \ac{COB} was originally designed to reduce the influence of this effect as the faces of the optics are oriented such that, after being split, the \ac{RL} will pass through two optics on both paths before it interferes with the \ac{HPL} and the \ac{AL} in transmission of the \ac{RC}.\footnote{On the path to the point of interference with \ac{AL} one of these optics is oriented at $0^\circ$ as opposed to the others which are oriented at $35^\circ$ and a more rigorous calculation should include this additional difference in path length.} Thus, if these optics experience common changes in their temperature, their optical path lengths will all change by the same amount. \ac{PLL}1 will adjust the phase of \ac{RL} to compensate for the changes in the optical path-length through the mirrors LT2 and \ac{RC}2. This should be nearly canceled by the optical path-length changes in the other \ac{RL} path through mirrors MZ1 and MZ3.\footnote{The word nearly is used here again due the path length difference through the $0^\circ$ mirror.} The \ac{HPL} only passes through the mirror  before interfering with \ac{RL}. Therefore only its path-length changes are encoded on the phase of the \ac{HPL}, and hence the \ac{BSM} field, with respect to the heterodyne detection system. 

Although this may appear to be ideal, that is, suppressing effects that lead to any decoherence between the bosonic field and the heterodyne detection system, it must also be considered that the open shutter path passes through all on axis substrates on the \ac{COB}. Therefore, in this configuration, this effect will cause the \ac{PLL}s to detune the open shutter phase with respect to the free space phase determined by the path length between the reflective interfaces of mirrors MZ3 and RC2. If the open shutter signal is then used to compensate for the potential phase changes in the system this effect will lead to a loss in coherence of the regenerated signal with respect to the heterodyne detection system. While this effect should produce a measurable change on the phase of the open shutter signals during the run, currently no way has been found to disentangle this effect from other effects that the open shutter calibration technique should compensate for. These other effects could be phase changes induced by misalignments in \ac{RL} or the alignment of the \ac{RL}-\ac{HPL} beatnote to the \ac{CH}-\ac{NL} optical fiber. If these signals were present and large enough to cause a decoherence in the open shutter data, this would appear as changes in the phase of the open shutter data unless they were perfectly canceled by other effects. Because the phase changes in the open shutter data were not large enough to produce a significant decoherence in the data, it seems this was not actually a problem for the optical system. Still, it was important to verify that the optical path length changes induced by temperature changes in the on axis \ac{COB} optics are significantly less than a wavelength.

The optical substrates each have a thickness of 9.5\,mm. Assuming an index of refraction for fused silica at a wavelength of 1064\,nm of 1.45, each of the 35$^\circ$ substrates have an optical path length of 15.0\,mm,\footnote{While the angle of incidence is 35$^\circ$, the angle the on-axis light propagates through the substrate is only 23.3$^\circ$ due to Snell's law.} while the flat cavity mirror has an optical path length of 13.8\,mm. Assuming a thermo-optic coefficient of 8.5\,ppm/K for fused silica at room temperature \cite{leviton2006temperature}, and a coefficient of thermal expansion of 0.5\,ppm/K \cite{hahn1972thermal}, the optical path length through the 35$^\circ$ mirrors will change by 95\,nm/K, while for the cavity mirror it will change by 87\,nm/K. Together, this will produce a 0.35\,cycle/K phase change in the open shutter signal with respect to the free space optical path length. 



Measurements of the temperature in the central cleanroom showed the RMS deviations from the mean were less than 0.1\,K over the duration of the closed shutter periods of $\rm S_{_\perp}$ and less than 0.07\,K during the closed shutter periods of $\rm S_{_\parallel}$. Therefore, the effects of path length changes in the substrates are expected to be less than 0.035\,cycles for $\rm S_{_\perp}$ and 0.025 for $\rm S_{_\parallel}$. In terms of the losses in the measured regenerated power, this would correspond to 4.8\% for $\rm S_{_\perp}$ and 2.4\% for $\rm S_{_\parallel}$. Each of these values have been included as an additional uncertainty in the closed to open shutter ratio in their respective runs.

Other effects that could change the phase of the open shutter signal were also investigated. They included misalignments of \ac{RL} with respect to the \ac{HPL} as well as alignment errors between these lasers and the long fiber used to transfer the \ac{RL}-\ac{HPL} beatnote to \ac{NL}. Both of these effects were shown to have an influence on the open shutter phase of the \ac{HPL} measured at the veto detector. Based on the misalignments experienced during the run and tests performed afterward, the alignment noise in these paths could have been responsible for phase changes as high as 0.1\,cycles during the run, although the unexplained phase changes in the run were not that large. Nevertheless, when misalignments of the optical system cause changes in the phase of the open shutter data, the open shutter calibration will compensate for this and therefore these effects were not considered when calculating the uncertainty in the open shutter calibration. A more detailed discussion can be found in Section~\ref{Sec:Op_cl_cal}. 

The calculation of the error bars on the open-to-closed shutter ratio also does not include effect from the wedge angles and path length changes of the half-wave plates on the \ac{COB} as they are specified to have a parallelism better than 0.5\,{\textmu}rad, with a thickness of less than 3\,mm. Also, differential changes in free space path length of the arms of the \ac{MZ} are not believed to be significant with respect to the other effects mentioned in this section and are therefore also not considered in the calculation of the error.

\subsection{System Locked Flags}

\label{Sedc:flags}

For the closed or open shutter data to be considered valid, the magnets must be operating at full current and all control loops must be engaged and stable at the correct frequencies. The loops include the \ac{PDH}, \ac{PLL}1, \ac{PLL}2, and \ac{PLL}3 control loops. The status of the system was recorded in the form of a number of `flags'. In addition to being stably locked, i.e. the control system engaged with a stable error signal, it was also important for the frequencies of all \ac{PLL}s to have the correct sign such that the \ac{HPL} would be resonant in the \ac{RC} and the beatnote between the \ac{HPL} and \ac{LO} would be at the proper frequency. The sign of the frequency difference between the two fields generating the beatnote was checked before each time the system was locked by varying the temperature of one of the laser crystals with the \ac{PLL} disengaged and examining the change in the beatnote frequency. Because the sign of the temperature to frequency calibration of the laser is known, it was possible to tell which laser had the higher frequency based on whether or not the beatnote frequency increased or decreased.

The flag for \ac{PLL}1 was based on monitoring the frequency of the PLL1 beatnote. If it deviated from its nominal value the flag would mark the data as invalid. We should note though that for $\rm S_{_\perp}$ this system was not yet implemented and the \ac{PLL}1 flags are only used for $\rm S_{_\parallel}$. Based on the flag data from $\rm S_{_\parallel}$, it does not appear that there were any sections of closed shutter data during $\rm S_{_\perp}$ in which \ac{PLL}1 was unlocked, but that the data during that stretch was otherwise flagged as valid. The reason for this is that when the \ac{PLL}1 lock is broken, this almost always breaks the \ac{PLL}2 lock as well. Furthermore, examining the stray-light signal over the duration of the run indicates that the system flags for $\rm S_{_\perp}$ are representative of the system status, as if the system is unlocked the stray-light signal disappears and the stray-light signal is ever present in the closed shutter data of $\rm S_{_\perp}$. 

It was possible to distinguish if the \ac{PLL}2 and \ac{PLL}3 loops were unlocked by observing the control signals sent to the  Piezoelectric and temperature actuators mounted to the laser crystal. The flags for these loops were therefore based on their control signals. 
Furthermore, the control signals were also inspected for times when they may have unlocked and then relocked on their own and the data from the time periods after the initial unlock were disregarded.

The flag for the magnet current was set to be valid if the magnet current was larger than 5.6\,kA. The magnet current, however did not vary over the duration of the run and was always within 0.0003\,kA of 5.7118\,kA during the closed shutter periods.

The open and closed shutter status flags were based on the recorded position of the shutter wheel such that it had to be oriented within 3 degrees of the center position of the open hole for the open status, or one of the closed holes for the closed status.

There was also a flag for the case where the operator was performing `maintenance' on the system. Maintenance involved taking measurements used for monitoring purposes as well as realigning the optics. In this case the flag was manually set to be invalid if maintenance was being performed as all data acquired during this period are disregarded. 

The Boolean product of each set of flags was used to form the complete set of flags for the shutter open and shutter closed periods. Out of caution, the complete flags were cut by 60\,s such the data from the first and last minute of every open and closed shutter section were disregarded.



Over the course of $\rm S_{_\perp}$ more than 580,000\,s of closed shutter data with valid flags was collected, while for $\rm S_{_\parallel}$ the total data from the closed shutter periods consisted of over 1,060,000\,s of valid data.

\subsection{Normalization by HPL Power}

\label{Sec:HPL_norm}
Recalling Equations~\ref{Eq:P_phi_intro} and \ref{Eq:P_open}, the \ac{HPL} power exiting the \ac{RC} in \ac{NR} during the closed ($P_\phi$) and open ($P_{\rm open}$)  shutter periods is
\begin{equation}
    P_\phi = \eta^2 \beta_{\!_{\rm RC}}  P_{\rm i } \,\mathcal{P}_{\gamma\leftrightarrow \phi}^2  
    \end{equation}
    and
    \begin{equation}
    P_{\rm open} = \eta^2 \beta_{\!_{\rm RC}} T_{\!_{\rm COB}}  T_{\!_{\rm RC2}} P_{\rm i} .
\end{equation}
As stated earlier, $T_{\!_{\rm COB}}$ and $T_{\!_{\rm RC2}}$ are static parameters that did not need to be measured constantly during the run, while the electromagnetic-\ac{BSM} field conversion rate $\mathcal{P}_{\gamma\leftrightarrow \phi}$ is the measurable for the experiment. The total field overlap $\eta$ and the resonant enhancement $\beta_{\!_{\rm RC}}$ are not static parameters and depend on attributes of the optical system, such as the alignment of the \ac{HPL} and storage time of the \ac{RC}, that could change throughout the run. 

\begin{figure}
    \centering
\begin{subfigure}[b]{0.49\textwidth}
    \centering
    \includegraphics[width=\textwidth]{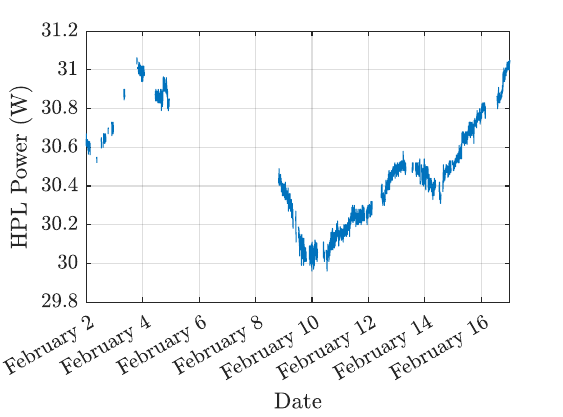}
    \caption{\ac{HPL} power ($\rm S_{_\perp}$)
    \label{fig:P_hpl_SC}}
    \end{subfigure}
\begin{subfigure}[b]{0.49\textwidth}
    \centering
    \includegraphics[width=\textwidth]{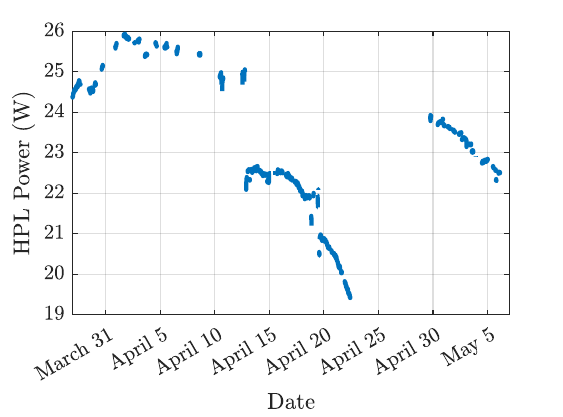}
    \caption{\ac{HPL} power ($\rm S_{_\parallel}$)
    \label{fig:P_hpl_PS}}
    \end{subfigure}
        \caption{The \ac{HPL} power measured exiting the production magnet string in CH during the open and closed shutter measurement periods of $\rm S_{_\perp}$ and $\rm S_{_\parallel}$.
    \label{fig:P_hpl}}
\end{figure}

$P_{\rm i}$ is also a dynamic variable, but in this case it was constantly measured during the entirety of the runs. By normalizing the $H$-function by the time series of the \ac{HPL} power, the effect of the changes in the power of \ac{HPL} on the closed/open shutter ratio can be removed, especially over long closed shutter periods. For clarity, consider the following example. Suppose the \ac{HPL} power was measured during a maintenance period that was immediately followed by an open shutter period that is then used to calibrate a long closed shutter period. During the closed shutter period the \ac{HPL} power fluctuates while all other variables remain the same. If the closed shutter data are not normalized to the \ac{HPL} power, the ratio between any point in the closed shutter time series and the open shutter period measured earlier will be affected by these fluctuations. Likewise, these fluctuations would also cause the apparent conversion rate, $\mathcal{P}_{\gamma\leftrightarrow \phi}$, to fluctuate, when in reality it should remain constant. If the entire data set including the open and closed shutter data is normalized to the measured \ac{HPL} power, the changes in the power will not affect the ratio or apparent coupling of the \ac{BSM} fields. The signal to noise ratio of the experiment will fluctuate due to changes in the \ac{HPL} power, however that would be the case with or without this normalization. 

As the amplitude of the heterodyne beatnote scales with $\sqrt{P_{\rm i}}$; dividing the $H$-function by $\sqrt{P_{\rm i}}$ removes the effect of fluctuations in the injected power, 
\begin{equation}
    H_{\rm p}[n] = \frac{H[n]}{\sqrt{P_{\rm i}[n]}}.
\end{equation}
Here, $H_{\rm p}[n]$ refers to the normalized $H$-function, while $P_{\rm i}[n]$ is the \ac{HPL} power measured at sample `$n$' by the out-of-vacuum power meter in \ac{CH}. The closed and open shutter powers will both be proportional to the product of the absolute square of the expectation value of $H_{\rm p}$ and the injected power, as shown in the  expressions 
\begin{equation}
    P_\phi \propto |\langle H_{\rm p,closed}[n]\rangle|^2 \,P_i
    \end{equation}
    and
    \begin{equation}
    P_{\rm open}\propto |\langle H_{\rm p,open}[n]\rangle|^2 \,P_i,
\label{Eq:P_phi_P_open}
\end{equation}
with the only difference between $H_{\rm p,\,closed}$ and $H_{\rm p,\,open}$ being the status of the shutter. By this definition $H_{\rm p}$ is a complex number with units of $\rm V/\sqrt W$ for either the open or closed shutter case.

The measured power injected into the production magnet string is shown in Figures \ref{fig:P_hpl_SC} and \ref{fig:P_hpl_PS} for $\rm S_{_\perp}$ and $\rm S_{_\parallel}$, respectively. These measurements were used to calculate $H_{\rm p}$. The power ranged from 29.9\,W to 31.1\,W for $\rm S_{_\perp}$ with a mean value of 30.4\,W during the closed shutter periods. For $\rm S_{_\parallel}$, the range was 19.4\,W to 25.9\,W with a mean value of 23.1\,W during the closed shutter periods. The fluctuations in power during the runs are related to changes in the alignment of the \ac{HPL} to the mode cleaning cavity on the \ac{NL} table either due to environmental noise or a manual realignment of the system.

There was also a drop in power between $\rm S_{_\perp}$, which saw a maximum power of 31\,W, and $\rm S_{_\parallel}$, with a maximum power of 26\,W. This was related to the lowering of the injection current sent to pump diodes for $\rm S_{_\parallel}$ because the \ac{HPL} began to overheat in the period between $\rm S_{_\perp}$ and $\rm S_{_\parallel}$ due a heatsink malfunctioning.

\subsection{Closed to Open Shutter Ratio}
\label{Sec:Op_cl_cal}

\begin{figure}
    \centering
\begin{subfigure}[b]{0.49\textwidth}
    \centering
    \includegraphics[width=\textwidth]{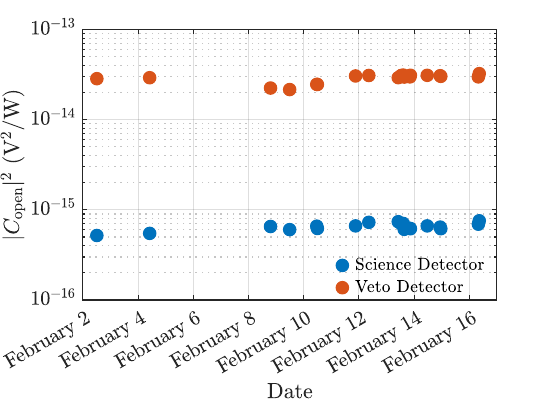}
    \caption{Open shutter magnitude squared ($\rm S_{_\perp}$) 
    \label{fig:Zopen_norm_SC}}
    \end{subfigure}
    \begin{subfigure}[b]{0.49\textwidth}
    \centering
    \includegraphics[width=\textwidth]{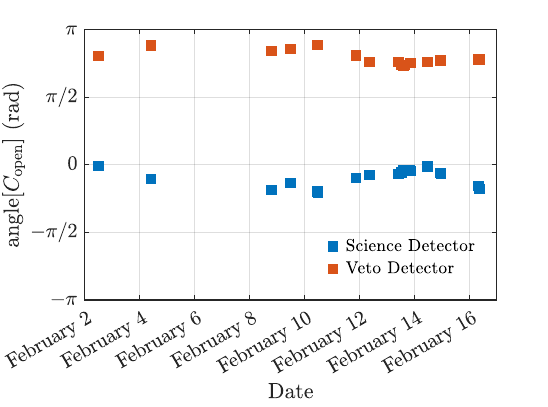}
    \caption{Open shutter phase ($\rm S_{_\perp}$) 
    \label{fig:Zopen_phase_SC}}  \end{subfigure}    \caption{Magnitude squared (a) and phase (b) of  $C_{{\rm{open,}}j}$ measured during $\rm S_{_\perp}$ for the science detector (blue) and veto detector (orange).
    \label{fig:Zopen_SC}}  
\end{figure}

In Section~\ref{Sec:Het}, the integration of the time series of the $H$-function to produce the `$Z$-function' was discussed. In order to remove systematic drifts and uncertainty from the coherent sum, the $H$-function must first be expressed in terms of the closed/open shutter ratio before the closed shutter data series is integrated, by calculating the $Z$-function for each open shutter section of data. Then each closed shutter section of data is divided by its nearest valid point in the set of the open shutter data denoted by $C_{{\rm{open,}}j}$, which can be expressed as
\begin{align}
    C_{{\rm{open,}}j} &=\langle H_{\rm p,open}\rangle_j \\
    &= \frac{1}{m[j]-k[j]}\sum_{n = k[j]}^{m[j]}H_{\rm p,\,open}[n].
\end{align}
Here, the index $j$ is assigned to each specific period of continuous open shutter section of data such that $j=1$ refers to the first period of open shutter data. The series $k[j]$ gives the sample number corresponding to the start of each open shutter section according to the flags discussed in the previous section, while $m[j]$ is the end of each open shutter period of data. The sum is normalized by $1/(m[j]-k[j])$ such that $C_{{\rm{open,}}j}$ is proportional to the average value of $H_{\rm p}$ in that section of data. 
Like $H_{\rm p,open}$, $C_{{\rm{open,}}j}$ is a complex number with units of $\rm V/\sqrt{W}$ as $m[j]$ and $k[j]$ are unitless.

The phase of $C_{{\rm{open,}}j}$ represents the phase of the mean value of each of the open shutter periods while Equation~\ref{eq:h_func} (with $P_{_{\rm S}}$ referring to $P_{\rm open}$) can be used to show that the magnitude squared of $C_{{\rm{open,}}j}$ is proportional to the ratio
\begin{equation}
    |C_{{\rm{open,}}j}|^2 \propto \frac{ P_{\rm open}}{ P_{\rm i} }  =  \eta^2 \beta_{\!_{\rm RC}} T_{\!_{\rm COB}} T_{\!_{\rm RC2}}.
    \label{Z_op_P}
\end{equation}
This demonstrates why periodically measuring $C_{{\rm{open,}}j}$ is important as it represents a direct measurement of the dynamic variables $\eta$, the field overlap, and $\beta_{\!_{\rm RC}}$, the resonant enhancement. The transmissivity of the \ac{COB} $T_{\!_{\rm COB}}$ and the flat \ac{RC} mirror $T_{\!_{\rm RC2}}$ are, on the other hand, static variables that were measured independently from the open shutter periods (See Section~\ref{Sec:COB} and Ref.~\cite{kozlowski2024designperformancealpsii} for more details on these measurements). 

\begin{figure}
    \centering
\begin{subfigure}[b]{0.49\textwidth}
    \centering
    \includegraphics[width=\textwidth]{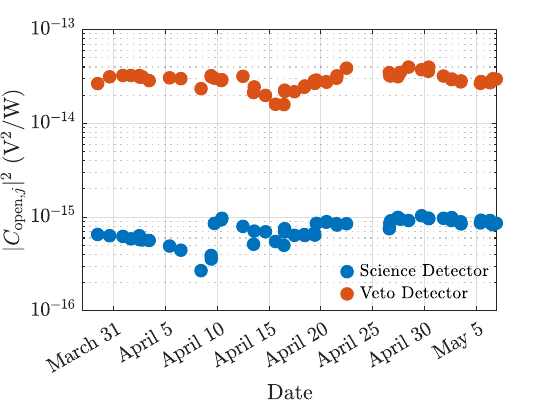}
    \caption{Open shutter magnitude squared ($\rm S_{_\parallel}$) 
    \label{fig:Zopen_norm}}
\end{subfigure}
\begin{subfigure}[b]{0.49\textwidth}
    \includegraphics[width=\textwidth]{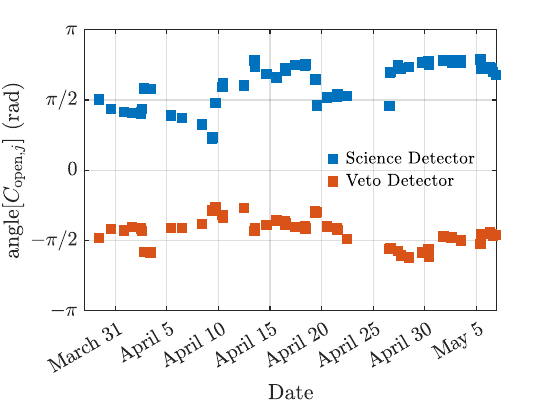}
    \caption{Open shutter phase ($\rm S_{_\parallel}$)
    \label{fig:Zopen_phase}}
    \end{subfigure}
        \caption{Magnitude squared (a) and phase (b) of  $C_{{\rm{open,}}j}$ measured during $\rm S_{_\parallel}$ for the science detector (blue) and veto detector (orange).
    \label{fig:Zopen_PS}}
\end{figure}

The magnitude squared and the phase of the open shutter measurements of both science and veto detector are shown in Figures~\ref{fig:Zopen_norm_SC}, \ref{fig:Zopen_phase_SC}, \ref{fig:Zopen_norm}, and \ref{fig:Zopen_phase}, for $\rm S_{_\perp}$ and $\rm S_{_\parallel}$. Each data point represents the 19 and 57 distinct open shutter periods in $\rm S_{_\perp}$ and $\rm S_{_\parallel}$ respectively. These results show that $C_{{\rm{open,}}j}$ was extremely stable during  $\rm S_{_\perp}$, especially the science detector, where the changes in $|C_{{\rm{open,}}j}|^2$ were less than 30\% through the duration of the run and phase changes of roughly a tenth of a cycle. $C_{{\rm{open,}}j}$ appears to be less stable during  $\rm S_{_\parallel}$.\footnote{This is believed to be due to changes in the weather over the duration of the run.} Here the main driver behind the changes in the magnitude and phase of $C_{{\rm{open,}}j}$ during $\rm S_{_\parallel}$ were the length changes of the \ac{RC} detuning the \ac{HPL} from the cavity resonance. For more details please refer to Section~\ref{Sec:Long_over}.

The time stamp associated with the midpoint of each open shutter period $C_{{\rm{open,}}j}$ is used to assign each point in the closed shutter data to its nearest point in $C_{{\rm{open,}}j}$, provide no realignment of the system was performed between the closed shutter data point and $C_{{\rm{open,}}j}$. The calibration series $C[n]$, is the series composed of these assigned values of $C_{{\rm{open,}}j}$ for each value of $H_{\rm p, closed}[n]$. For the specific case in which no valid open shutter period was measured between two different times in which the system was aligned, the closed shutter data between these times is disregarded.

The summation of the closed shutter $H$-function is then performed:\footnote{In this equation, the $H$-function has been named $H_{\rm p, closed}[n]$ as it is assumed that the $H$-function has been truncated to remove the data which are invalid either because of the status of the system flags or the lack of an open shutter period between alignments.}
\begin{equation}
    Z_{\rm closed} = \frac{1}{N}\sum_{n = 1}^N \frac{H_{\rm p, closed}[n]}{C[n]}.
    \label{Eq:Zn_cl}
\end{equation}
Here, the series corresponds to the product of the full normalized $H$-function $H_{\rm p, closed}[n]$ and one over the calibration series, or $1/C[n]$, while $N$ is the total number of samples in the series. It should be noted here that the $Z_{\rm closed}$ is a unitless complex number. The lack of units of $Z_{\rm closed}$ is a consequence of the fact that the calibration procedure not only compensates for the parameters $\eta$ and $\beta_{\!_{\rm RC}}$, but also for attributes of the heterodyne detection system, such as the conversion gain of the photodetector or the optical losses between RC1 and $\rm PD_{\rm sci}$. This is because they are common to both the open and closed shutter signals and is an important benefit of the calibration procedure. 

Working under the assumption that there are no uncontrolled effects that can change the relative amplitude and phase of the regenerated field with respect to the signal during the open shutter periods, this method provides an automatic correction for the coupling and phase of the \ac{HPL} with respect to the \ac{RC} and \ac{LO}. If, for example, the spatial coupling of the \ac{LO} to the \ac{RC} drops, this will affect both $C_{{\rm{open,}}j}$ and $H_{\rm p}[n]$, provided it does not change significantly between sample $n$ and sample $(m[j]-k[j])/2$.\footnote{Later in Section~\ref{Sec:C_n_uncert} and Appendix~\ref{APP:Del_C_j} it is shown that on time scales of roughly two days or less the system is stable and these changes are small compared to other sources of uncertainty in the measurement.}

Likewise, because $H_{\rm p}[n]$ and $C_{{\rm{open,}}j}$ are left as complex numbers, this effect also compensates for any slow phase drifts that are common mode to both the open and closed shutter periods. An example of this compensation occurs if there is an offset between the frequency difference of the \ac{HPL} and \ac{LO} and the frequencies used in the double demodulation scheme. This is only possible if such detunings lead to fractions of a cycle in phase drift or less between the open shutter periods. Therefore, this method can only compensate for this effect up to the order of tens of {\textmu}Hz. (It should be clear that the unexplained phase effects seen in the open shutter periods in Figures~\ref{fig:Zopen_phase_SC} and \ref{fig:Zopen_phase} were much less than a full cycle and frequencies less than {\textmu}Hz.)

Another dynamic effect that this method compensates for is the slow changes in the cavity length that can detune the \ac{HPL} from the resonance of the \ac{RC}, this could lead to the cavity inducing a phase offset into the science signal. These changes in phase would be common mode to the phase changes introduced to the open shutter signal as it transmits through the cavity. Therefore, as long as the cavity length changes only induce phase changes on the order of fractions of a cycle between the open shutter periods, this method can compensate for these drifts.

In either case, the optical system was actively monitored and re-aligned during maintenance periods during the science campaign to mitigate the phase drifts and changes in the coupling as much as possible in order to reduce the systematic uncertainty in the measurement. An analysis of the results of this effort is discussed in Section~\ref{Sec:Perf}.

The power ratio is equivalent to the squared magnitude of $Z_{\rm closed}$:
\begin{equation}
    \frac{P_\phi}{P_{\rm open}}\equiv|Z_{\rm closed}|^2 
    \label{Eq:Zn_cl_P}.
\end{equation}
Together with Equation~\ref{Eq:g_a_rat}, $\mathcal{P}_{\gamma\leftrightarrow \phi}$ can be expressed as the product of $|Z_{\rm closed}|$ with the square root of $T_{\!_{\rm COB}}$ and $T_{\!_{\rm RC2}}$: 
\begin{equation}
\mathcal{P}_{\gamma\leftrightarrow \phi} = \sqrt{T_{\!_{\rm COB}}T_{\!_{\rm RC2}}}\left|Z_{\rm closed}\right|.
\label{eq:P_g_a_Z}
\end{equation}
This equation represents the primary achievement of the calibration procedure as it provides an elegant method to directly translate the results measured by the heterodyne detection system in terms of the  conversion rate between electromagnetic and \ac{BSM} fields $\mathcal{P}_{\gamma\leftrightarrow \phi}$. This requires the use of only one other dynamic variable, the open shutter measurements,  as well as the measured combined transmissivities of the optics between the two magnet strings along the optical axis of the system. The results presented in Section~\ref{Sec:results}, in terms of $\mathcal{P}_{\gamma\leftrightarrow \phi}$, are then interpreted in Ref.~\cite{ALPSII_science} as limits on $g_{\phi\gamma\gamma}$.


\subsection{Veto Detection System}

\label{Sec:veto_det}
A veto detection system was also operated during the first science campaign to check the validity of signals measured at the science detection system. The veto photodetector ($\rm PD_{\rm veto}$) is positioned in the \ac{CH} cleanroom such that it would detect the light reconverted from the \ac{BSM} signal that transmits RC2. As RC2 is nearly a factor of 20 less transmissive than RC1, for a given electromagnetic to \ac{BSM} field conversion rate, the reconverted signal power on $\rm PD_{\rm veto}$ would be significantly lower than the signal power present on the science photodetector. This factor is seen in the expression for the close shutter signal power transmitted by RC2 in the presence of a \ac{BSM} signal:
\begin{equation}
    P_{\phi,{\rm veto}} =\eta^2 \beta_{\!_{\rm RC}} \frac{T_{\rm RC2}}{T_{\rm RC1}} P_{\rm i} \,\mathcal{P}_{\gamma\leftrightarrow \phi}^2 
    \label{Eq:P_phi_veto}.
\end{equation}
Although it is still capable of measuring the reconverted field, the true purpose of the veto detector is to observe the stray light, due to its proximity to the shutter system and \ac{COB}, while also providing independent means of monitoring the performance of the optical system. Stray light incident on the veto detector does not need to pass through the beam tube or \ac{RC} like the stray light present at the science detector. Furthermore, stray light on the optical axis of the system that directly couples to the \ac{RC} should have a power roughly a factor of $R_{_{\rm RC,CH}}/T_{_{\rm RC}}\sim 30 $ higher at the veto detector. Here $R_{_{\rm RC,CH}}\sim0.95$ is the reflectivity of the \ac{RC} for a laser that is perfectly spatially mode matched to the cavity eigenmode (see Section~\ref{Sec:RC_R} and Equation~\ref{Eq:R_CH} for more details). $T_{_{\rm RC}}\sim 3\%$ is the transmissivity of the \ac{RC} for a laser sharing the spatial eigenmode of the cavity.

The analysis for the data from the veto detection system is nearly the same as for the science detector. The signal from the veto photodetector is amplified and then digitized by an \ac{ADC} and the digital signal is demodulated at the veto heterodyne frequency $f_{\rm veto}=33f_0$, where $f_0$ is the \ac{FSR} of the \ac{RC}. As the diagram of the laser frequencies in Figure~\ref{fig:freq_plan} shows, the veto heterodyne frequency is different from the science heterodyne frequency because the interference beatnote measured at the veto detector is between the signal field and the \ac{AL} light in transmission of the \ac{RC}. The demodulated data are used to form the open and closed shutter arrays $H^{\rm veto}_p$ and the veto systems version of the calibration series $C_{\rm veto}[n]$ is constructed with the open shutter veto series $C^{\rm veto}_{{\rm open,}j}$.  Just like the calibration procedure for the science detector data, the results at the veto detector are formulated in terms of the normalized closed shutter power $|Z^{\rm veto}_{\rm closed}|^2$.

When the shutter is open, the \ac{HPL} light reflected by RC2 that is in the spatial mode of the \ac{AL}\footnote{The \ac{HPL} light must be in the spatial mode of the \ac{AL} to generate a beatnote measurable by the heterodyne detection system.} can be expressed by 
\begin{equation}
    P_{\rm open}^{\rm veto} =\frac{\eta^2}{\eta_{\hat z}^2} R_{\!_{\rm RC,CH}}  T_{\!_{\rm COB}}  P_{\rm i} .
    \label{Eq:P_open_veto_intro}
\end{equation}
The spectral component of the field overlap is canceled from the total overlap as the term $\eta^2$ is divided by $\eta_{\hat z}^2$. Again, $R_{_{\rm RC,CH}}$ (Equation~\ref{Eq:R_CH} in Section~\ref{Sec:RC_R}) is the reflectivity of the \ac{RC} for a laser occupying the spatial mode of the cavity. Here, the longitudinal overlap $\eta_{\hat z}^2$ will play a small role in $R_{_{\rm RC,CH}}$. From this equation it is apparent that $|Z^{\rm veto}_{\rm closed}|^2$ has a very different meaning than its counterpart calculated from the science detector as the dynamic variables $\beta_{_{\rm RC}}$ and $\eta_{xy}$ do not perfectly cancel here.

\subsection{Calibration Uncertainty}

\label{Sec:C_n_uncert}

The optical system is not static and parameters such as the cavity length and \ac{HPL} coupling to the \ac{RC} did change over time, producing changes in the expected signal for a given \ac{BSM} coupling during the run. These changes are encoded on the open shutter signal which is why open shutter data was taken as frequently as possible. The uncertainty in the projection of this data to the calibration series $C[n]$ for nearby closed-shutter periods is one of the leading contributors to the systematic uncertainty in the calibration of the regenerated power in terms of $\mathcal{P}_{\gamma\leftrightarrow \phi}$. 

To set error bars on the open shutter calibration, the changes in the open shutter results were examined for points between which no realignment of the system was performed. This was done by compiling sets of data expressing the changes between the magnitude squared and phase of different open shutter sections as a function of the time between them.

The normalized absolute value of the difference in the magnitude squared for two points in the open shutter data $C_{{\rm{open,}}j}$ and $C_{{\rm{open,}}m}$, separated in time by $t(m)-t(j)$ (here $t(m)$ and $t(j)$ are the time at the midpoint of their corresponding open shutter periods), is
\begin{equation}
    \Delta |C_{{\rm open,}j\rightarrow m}|^2 = \frac{\left||C_{{\rm{open,}}j}|^2-|C_{{\rm{open,}}m}|^2 \right|}{|C_{{\rm{open,}}j}|^2 }.
\end{equation}
A first order polynomial was then fit to $\Delta |C_{{\rm open,}j\rightarrow m}|^2$ as a function of $t(m)-t(j)$ to find the average normalized changes as a function of the time between the open shutter periods. This process gave a linear trend of $1.5\% {\rm\,/day}$ for $\rm S_{_\perp}$ in the science detector with a static offset of $7.3\%$. That is, open shutter periods measured only minutes apart would vary on average by $7.3\%$ in terms of $|C_{{\rm open,}j}|^2$, but if these measurements were separated by a day, they would vary on average by $8.8\%$. From these data it can be seen directly that it would take nearly 5 days for the uncertainty due to the long term drifts in the system to have a larger impact than the uncertainty measured over shorter times between the open shutter periods. 
The linear trend was $3.1\%{\rm\,/day}$ at the science detector for $\rm S_{_\parallel}$, with a static offset of $6.0\%$. The positive slope in all of the fits indicates that the changes between open shutter periods tend to be larger for larger periods of time as expected. 
Figures \ref{fig:diff_open_abs_sq_SC} and \ref{fig:diff_open_abs_sq_PS} in Appendix~\ref{APP:Del_C_j} show scatter plots for $\rm S_{_\perp}$ and $\rm S_{_\parallel}$ of the relative absolute change in $|C_{{\rm open},j}|^2$ for different open shutter periods as a function of the time between them.

These trends were then used to estimate the uncertainty in $|C[n]|^2$. Here, each point in $C[n]$ is assigned error bars based on how far in time it is from the nearest open shutter period with no realignment of the system performed in between. 
The average of the error bars of $|C(n)|^2$ for all closed shutter points found with this process was then used to calculate contribution of the changes in the open shutter data to the relative error on $|Z_{\rm closed}|^2$. This process gave an average relative uncertainty of 7.8\% on $|Z_{\rm closed}|^2$ with a range from 7.3\% to 8.8\% for $\rm S_{_\perp}$. For $\rm S_{_\parallel}$ the average uncertainty due to the changes in the open shutter periods was 7.0\% with a range from 6.0\% to 11.5\%.


The change in the phase of the open shutter periods was also calculated. The first-order polynomial fit to the data gave $0.123{\rm \,rad/day}+0.020$\,rad for the science detector and $0.041{\rm \,rad/day}+0.033$\,rad for the veto detector during $\rm S_{_\perp}$. In the  case of $\rm S_{_\parallel}$, the first-order polynomial fit gave $-0.010{\rm \,rad/day}+0.152$\,rad at the science detector and $0.035{\rm \,rad/day}+0.059$\,rad for the veto detector. These trends shows that longer times between the open shutter periods corresponded to larger changes in the phase of the open shutter measurements for $\rm S_{_\perp}$ for both the science and veto detectors, as well as $\rm S_{_\parallel}$ at the veto detector. The phase changes measured at different open shutter periods of $\rm S_{_\parallel}$ appear actually to decrease at the science detector as a function of the time between the points. This is believed to be due to several outliers biasing the data at earlier times. The average uncertainty on the angle of $C[n]$ was 0.062\,rad for $\rm S_{_\perp}$ with a range of 0.026\,rad to 0.1394\,rad. 
For $\rm S_{_\parallel}$ the average uncertainty on the angle of $C[n]$ was 0.148\,rad for with a range of 0.136\,rad to 0.151\,rad. This would correspond to an average uncertainty on the magnitude squared of $C[n]$ of 0.45\% for $\rm S_{_\perp}$ and 2.2\% for $\rm S_{_\parallel}$. As these errors were insignificant compared to other sources of error in $C[n]$, the contribution of the changes in the phase between open shutter periods was not considered when calculating the error of $C[n]$ and $|Z_{\rm closed}|^2$. Scatter plots of the absolute change in angle of the open shutter periods as a function of the time interval between them can be found in Appendix~\ref{APP:Del_C_j} with Figure~\ref{fig:diff_open_ang_SC} showing the results from $\rm S_{_\perp}$ and Figure \ref{fig:diff_open_ang_PS} $\rm S_{_\parallel}$.


It must be also noted that interpreting the changes in the open shutter data should be approached with a certain degree of caution. As the previous paragraph mentions, it is possible that the data have some bias related to the stability of the system. For example, when the system was unstable, it may have become unlocked more frequently during the open shutter periods. This would lead to more frequent open shutter periods during these times when the system was less stable because the operators were instructed to relock the system and restart the open shutter period if a certain integration time was not reached. Likewise, when the system was more stable, it would be more likely that no relock was required and therefore there may not even be any data at all for shorter durations between open shutter periods during the more stable periods. This may also explain why the shorter time periods appear to be less stable in some cases, than the longer time periods in the plots shown in Appendix~\ref{APP:Del_C_j}.

Later, in Section~\ref{Sec:het_mv}, the time evolution of the closed shutter data is examined on the complex plane. The effects in the open shutter data discussed in the previous paragraphs can also be examined on the complex plane. Figures~\ref{fig:stray_cpx_sci_open_close_SC} and \ref{fig:stray_cpx_veto_open_close_SC} show plots of the real and imaginary components of $C_{{\rm open},j}/C_{{\rm open},m}$  during $\rm S_{_\perp}$, for the science and veto detectors, respectively, in comparison to a moving average of the closed shutter data (for more information see Section~\ref{Sec:het_mv}). Here, only the changes between open shutter periods in which the system was free-running and no alignment took place between them were considered. This means that these are not necessarily all consecutive measurements of the open shutter periods, as often realignments of the system occur in between the open shutter periods. In these plots, if no changes occurred in the open shutter results, all points would be at $1+0i$. In the science detector, the mean value of the data shown in these plots was $0.98 - 0.08i$ with a standard deviation of 0.23. The mean of the absolute value of the real components of these data minus one was 0.04, while for the imaginary components the mean absolute value was 0.18. These changes can be used to calculate a rough estimation of the relative error in $|Z_{\rm closed}|^2$ of 8\%. The estimates of the errors from both methods are therefore equal with respect to their significant digits.

The reason that this result is slightly different from the estimate of the error compiled using the trends earlier in this section is that this is just an average of all of the changes in the complex open shutter data with no attempt to project these changes on to the closed shutter periods or account for the time between the open shutter periods. In either case, t

Data for $\rm S_{_\parallel}$ are shown in Figures~~\ref{fig:stray_cpx_sci_open_close_PS} and \ref{fig:stray_cpx_veto_open_close_PS} for the science and veto detector, respectively. In these plots, the mean value of the data at the science detector was $0.95+0.08i$ with a standard deviation of $0.26$. For these data the mean of the absolute value of the real component of the data minus one was 0.06, while the mean of the absolute value for the imaginary component was 0.14 leading to a rough estimate of the relative error of 12\% on $|Z_{\rm closed}|^2$. This shows relatively good agreement with the estimate of the error using the linear trend of the magnitude square of $C_{{\rm open},j}$ of 7.0\%. The discrepancy between the two methods is believed to be due to the fact that this method simply finds an average of all of the changes in the complex open shutter data with no attempt to project these changes on to the closed shutter periods or account for the time between the open shutter periods as the method using the linear trends in the data does.

The technical noise of the system (discussed in more detail in Section~\ref{Sec:Tech_noi_2}) also contributed to the uncertainty calculated here. As the technical noise is symmetric about the actual signal, over long integration times it will average to zero. This can be seen explicitly in Figures~\ref{fig:exp_op_SC} and \ref{fig:exp_op_PS} (discussed later in Section~\ref{Sec:Time_Ev}) as the 1-$\sigma$ statistical error on the data converges proportionally to a trend of $1/\sqrt{t'}$ with $t'$ being the integration time. Some of the open shutter periods were short enough that the impact of the technical noise was significant. The technical noise contribution to the uncertainty was calculated by performing a Fourier transform on the data from each of the open shutter periods and calculating the mean noise in the bins excluding the frequency bin where the actual signal resides.\footnote{Only frequency bins within the bandwidth of the filters applied to the data were considered.} For $\rm S_{_\perp}$, the mean value of the relative error in $|C_{{\rm open},j}|^2$ due to technical noise during the open shutter periods was only 0.15\% for the science detector however the shortest section of open shutter data at 493\,s gave an uncertainty of 3.99\%. For $\rm S_{_\parallel}$ the average uncertainty due to technical noise was only 0.30\% for the science detector, but the shortest open shutter section lasted only 59\,s and had an uncertainty of 12.69\% due to the technical noise. 

\begin{table*}[t]
    \centering
    \makebox[\textwidth][c]{
    \begin{tabular}{c|cccccccc}
& Observed result: & Estimated background: &
 Estimated background: \\
 & $|Z(f_k=0)_{\rm closed}|^2$  &  $\mathcal M_{_{\rm TH}}(f_k=0)$ &   $\mathcal M_{_{\rm SG}}(f_k=0)$ \\ \hline 
Science ($\rm S_{_\perp}$) & $(1.95\pm0.26)\times10^{-5}$ & $(2.54\pm0.34)\times10^{-5}$ & $(2.28\pm0.31)\times10^{-5}$ \\
Veto ($\rm S_{_\perp}$) & $(4.58\pm0.53)\times10^{-5}$ & $(2.59\pm0.30)\times10^{-5}$ & $(2.46\pm0.29)\times10^{-5}$ \\

Science ($\rm S_{_\parallel}$) & $(2.87\pm0.36)\times10^{-5}$ & (0.85$\pm0.11)\times10^{-5}$ & (1.07$\pm0.13)\times10^{-5}$  \\ 
Veto ($\rm S_{_\parallel}$) & $(4.90\pm0.57)\times10^{-5}$  & (3.87$\pm0.45)\times10^{-5}$ &  (4.09$\pm0.52)\times10^{-5}$

    \end{tabular}}
    \caption{Summary of the heterodyne data in the signal bin for both detectors and both science runs are shown in the column labeled $|Z(f_k=0)_{\rm closed}|^2$ . The errors are derived from the uncertainty in the calibration time series $C[n]$.  The columns labeled $\mathcal M_{_{\rm TH}}(f_k=0)$  and $\mathcal M_{_{\rm SG}}(f_k=0)$  give the background estimates discussed in Section~\ref{Sec:AFA_res} calculated using the top-hat and Savitzky-Golay filters respectively.}
    \label{tab:Het_Sig}
\end{table*}

The primary reason for the lower relative noise in the $\rm S_{_\perp}$ open shutter periods was that the average duration was their longer median duration of 2070\,s for $\rm S_{_\perp}$ compared to 960\,s for $\rm S_{_\parallel}$. In either case, technical noise appears to impact the uncertainty of the calibration only marginally. These effects are already accounted for in the analysis from this section as this will also play a role in the changes observed between different periods of open shutter data.\footnote{It is not entirely clear what the threshold is for when the drifts in the setup take over from the technical noise as the dominant source of noise in the open shutter data. From Figures~\ref{fig:exp_op_SC} and \ref{fig:exp_op_PS} this appears to be longer than 1000\,s for sections of open shutter data where the system was continuously locked based on the fact that the error bars never stop converging. In Appendix~\ref{APP:Del_C_j}, Figures~\ref{fig:diff_open_abs_sq_SC}-\ref{fig:diff_open_ang_PS}, the data compiled from periods in which the system was not continuously locked show that the drifts in the system are the dominant source of noise after only a few hours as the uncertainty on the shortest time scales shown in these plots is larger that the uncertainty shown on the longest time scales of Figures~\ref{fig:diff_open_abs_sq_SC}-\ref{fig:diff_open_ang_PS}. Nevertheless, as the text states, the results over short time scales in Figures~\ref{fig:diff_open_abs_sq_SC}-\ref{fig:diff_open_ang_PS} should be treated with caution, as there may be some bias in the data toward times when the system was less stable.}



\subsection{Uncertainty in the Results}

\label{Sec:results_uncert}

Three sources are considered in the  estimate of the uncertainty in $|Z_{\rm closed}|^2$. Two of these, the wedge angles of the \ac{COB} optics changing the spatial overlap of a \ac{BSM} signal with respect to the open shutter signal (introducing a relative uncertainty of 10\% on $|Z_{\rm closed}|^2$) and the loss of phase coherence of potential \ac{BSM} signals with respect to the open shutter data due to temperature changes on the \ac{COB} (contributing a relative uncertainty of 4.8\% to $|Z_{\rm closed}|^2$ for $\rm S_{_\perp}$ and 2.4\% for $\rm S_{_\parallel}$), were discussed in Section~\ref{Sec:Sys_eff_op}. The third source of error in $|Z_{\rm closed}|^2$ was the uncertainty of $C[n]$ discussed in the previous section. In this case the relative uncertainty introduced to $|Z_{\rm closed}|^2$ by $C[n]$ was estimated to be 7.8\% for $\rm S_{_\perp}$ and 7.0\% for $\rm S_{_\parallel}$. Summing these quadratically gives a total relative error on $|Z_{\rm closed}|^2$ of 13.3\% for $\rm S_{_\perp}$ and 12.4\% for $\rm S_{_\parallel}$.

The uncertainty of $\mathcal{P}_{\gamma\leftrightarrow \phi}$ will be affected by all of the sources discussed in the previous paragraph as well as the uncertainties on the transmissivities of the \ac{COB} optics along the optical axis of the system (discussed later in Section~\ref{Sec:COB_in}) and the transmissivity of RC2 (discussed in Ref.~\cite{kozlowski2024designperformancealpsii}). With a relative uncertainty in $T_{_{\rm COB}}$ of 7.1\% and a 9.8\% in $T_{_{\rm RC2}}$ the relative error bars on $\mathcal{P}_{\gamma\leftrightarrow \phi}$ come out to 9.1\% for $\rm S_{_\perp}$ and 8.7\% for $\rm S_{_\parallel}$.\footnote{It must be considered that because $\mathcal{P}_{\gamma\leftrightarrow \phi}$ is the square root of the product of the transmissivities of these optics and the $|Z_{\rm closed}|^2$, its relative error is roughly half the quadratic sum of the errors of these parameters.}

\section{Results and Backgrounds}
\label{Sec:Results_full}

The results acquired by the heterodyne detection system using the science and veto detectors are shown for both runs in terms of $|Z_{\rm closed}|^2$ in Table~\ref{tab:Het_Sig}. The measured value given by the science detector data was $(1.95\pm0.26)\times10^{-5}$ for $\rm S_{_\perp}$ and $(2.87\pm0.36)\times10^{-5}$ for $\rm S_{_\parallel}$. For the veto detector $|Z_{\rm closed}|^2$ was $(4.58\pm0.53)\times10^{-5}$ for $\rm S_{_\perp}$ and $(4.90\pm0.57)\times10^{-5}$ for $\rm S_{_\parallel}$. The error bars shown in Table~\ref{tab:Het_Sig} represent the estimated systematic uncertainty that was calculated with the methods discussed in the previous section.

The resulting values of $|Z_{\rm closed}|^2$ in Table~\ref{tab:Het_Sig} lie in what we will refer to as the signal frequency bin, because the data was demodulated at the heterodyne frequencies given in Section~\ref{Sec:Het}. As Section~\ref{Sec:Freq_alt} explains, the $Z$-function was also evaluated at alternative frequencies where no \ac{BSM} contribution was expected. This allowed the backgrounds to be estimated without requiring background measurements with the magnets off, under the assumption that the neighboring bins are indicative of the background in the signal bin. The estimated background in the signal bin is shown as the second and third columns of Table~\ref{tab:Het_Sig}and this process is summarized and compared with the measured data in Section~\ref{Sec:AFA_res}.

\begin{figure}[t]
    \centering
    \begin{subfigure}[b]{0.49\textwidth}
  \centering
    \includegraphics[width=\textwidth]{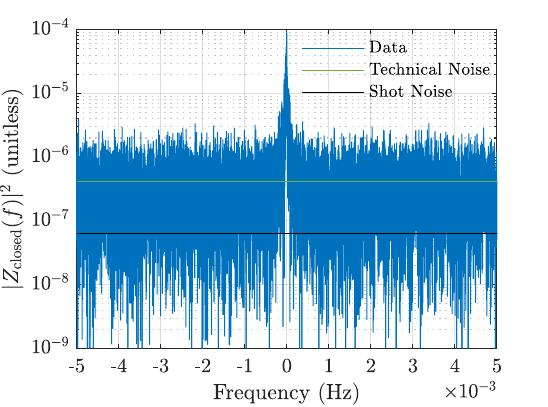}
    \caption{Science detector ($\rm S_{_\perp}$) \label{fig:alt_sci_SC}}
    \end{subfigure}
        \begin{subfigure}[b]{0.49\textwidth}
          \centering
    \includegraphics[width=\textwidth]{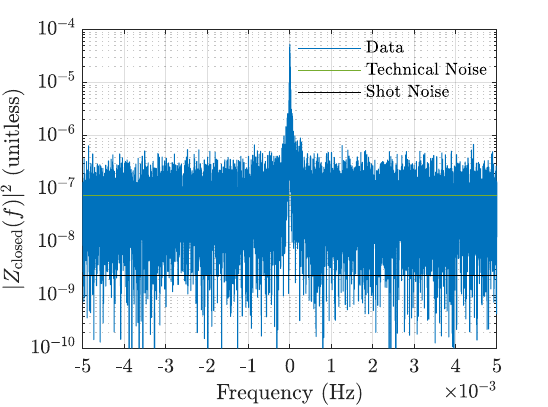}
    \caption{Veto detector ($\rm S_{_\perp}$) \label{fig:alt_veto_SC}}
    \end{subfigure}
      \caption{$|Z(f)_{\rm closed}|^2$ measured by the science (a) and veto (b) detectors during $\rm S_{_\perp}$ are shown in blue over the frequency range from $-5$\,mHz to $+5$\,mHz around the signal frequency bin. The mean technical noise is shown in green, while the shot noise limit is shown in black.    \label{fig:alt_SC}}
\end{figure}

Section~\ref{Sec:BG_stat_ex_lim} explains how the statistical distribution of the background was calculated by comparing the estimated background to the measured values at frequencies outside the signal bin. This distribution was then used to set detection thresholds that determined whether or not a signal was present at a given confidence level.

The results in the signal bin given in Table~\ref{tab:Het_Sig} are consistent with the estimated background levels during both runs. Thus neither run showed an excess statistically significant enough to claim the presence of a signal generated by new physics.  Section~\ref{Sec:BG_stat_ex_lim} therefore also examines how exclusion limits on the conversion rate between electromagnetic and \ac{BSM} fields were set based on the background distribution. The resulting exclusion limits are summarized in terms of $\mathcal{P}_{\gamma\leftrightarrow \phi}$ in Section~\ref{Sec:results}.

Section~\ref{Sec:Res_ver} discusses how the behavior and data measured at the science and veto detection systems indicate that the backgrounds could be due to stray light.
This is done by examining the time evolution of $|Z_{\rm closed}|^2$ as well as the ratio of the measured signals and backgrounds at the science and veto detectors. 

Measurements of the transmissivities of the \ac{COB} optics are then described in Section~\ref{Sec:COB_in}, as these parameters are critical to calculating $\mathcal{P}_{\gamma\leftrightarrow \phi}$ from the results of  $|Z_{\rm closed}|^2$.

\subsection{Alternative Frequency Analysis}
\label{Sec:Freq_alt}

As mentioned earlier, the signal strengths at frequencies other than the \ac{BSM} heterodyne frequency were also examined. The $Z$-function for a frequency $f$ away from the signal bin (in this equation the signal bin is at $f=0$) is
\begin{equation}
    Z(f)_{\rm closed} = \frac{1}{N}\sum_{n = 1}^N \frac{H_{\rm p}[n]e^{i2\pi \frac{f}{f_{\!_n}}n}}{C[n]}.
\end{equation}
Here $f_n$ is the sampling frequency (1\,Hz) of the down-sampled data series. To be clear, $H_{\rm p}$ is still the same $H$-function that has undergone double demodulation, down sampling, and normalization by the \ac{HPL} injected power. Also, it should be noted that for calculating the signal strength at other frequencies, there is no mathematical difference  between this method and changing one of the heterodyne demodulation frequencies given in Section~\ref{Sec:Het}.

The data were compiled with a bin size in the frequency series of one over the total time elapsed between the first and last samples of valid closed shutter data. With 1.2 million seconds elapsed between the first and last samples of closed shutter data for $\rm S_{_\perp}$, the frequency resolution for $\rm S_{_\perp}$ was 0.85\,{\textmu}Hz. $\rm S_{_\parallel}$ saw 3.3 million seconds elapse between the first and last valid closed shutter samples and therefore the frequency series is sampled every 0.30\,{\textmu}Hz for this run. Using these resolutions in the frequencies does mean that the Fourier series does not form a linearly independent basis set, as the gaps in the data during periods in which the system was unlocked cause the Fourier components to lose their linear independence over the data run. The consequences of this are discussed in more detail in Appendix~\ref{Sec:Freq_spr}.

\begin{figure}[t]
   \centering
            \begin{subfigure}[b]{0.49\textwidth}
            \centering
    \includegraphics[width=\textwidth]{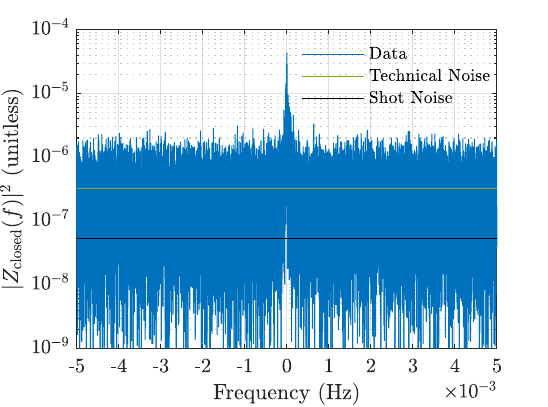}
    \caption{Science detector ($\rm S_{_\parallel}$)
    \label{fig:alt_sci_PS}}
        \end{subfigure}
        \begin{subfigure}[b]{0.49\textwidth}
          \centering
    \includegraphics[width=\textwidth]{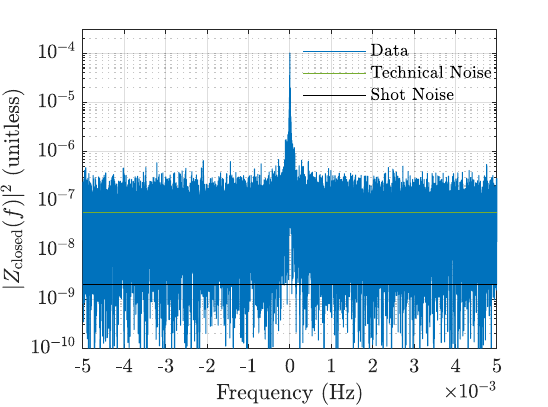}
    \caption{Veto detector ($\rm S_{_\parallel}$)
    \label{fig:alt_veto_PS}}
        \end{subfigure}
            \caption{$|Z(f)_{\rm closed}|^2$ measured by the science (a) and veto (b) detectors during $\rm S_{_\parallel}$ are shown in blue over the frequency range from -5\,mHz to +5\,mHz around the signal frequency bin. The mean technical noise is shown in green, while the shot noise limit is shown in black.
    \label{fig:alt_PS}}
\end{figure}

Because $H_{\rm p}/C$ is a complex number, $Z(f)_{\rm closed}$ is not necessarily symmetric about $f=0$ and it is important to examine both the positive and negative frequencies. The results of $\rm S_{_\perp}$ are shown from -\,5\,mHz to +\,5\,mHz in Figures \ref{fig:alt_sci_SC} and \ref{fig:alt_veto_SC} for the science and veto detectors, respectively. Likewise, the results for $\rm S_{_\parallel}$ over the same frequency range are shown in Figures \ref{fig:alt_sci_PS} and \ref{fig:alt_veto_PS}. All of these plots are centered on the heterodyne frequency such that $f=0$ is the signal frequency bin.

In all four plots, the measured data, shown in blue, is dominated by technical noise, whose mean is shown in green, for frequencies more than 100\,{\textmu}Hz away from the heterodyne frequency.\footnote{The mean technical noise is calculated for frequencies within the bandwidth of the filters applied to the data and more than 1\,mHz away from the heterodyne frequency to avoid the peak near the heterodyne frequency having and effect on the calculation.} In the case of $\rm S_{_\perp}$ the mean technical noise in terms of $\left|Z_{\rm closed}\right|^2$ was $4.2\times10^{-7}$, while for $\rm S_{_\parallel}$ it was $3.3\times10^{-7}$. 

The derivation of the shot noise limits, shown in black, and a comparison to the technical noise are discussed further in Section~\ref{Sec:Tech_noi}. A peak above the technical noise floor is apparent in the data at frequencies less than 100\,{\textmu}Hz away from the signal bin. The fact that these signals are spread over many frequency bins and not constrained to the signal bin indicates that this peak is the result of a background in the detection system. One possible source of background is stray light that scatters out of the \ac{HPL} and takes a random uncontrolled path through the optical system to the detection system. Because the length of this path will fluctuate in time due to environmental noise, the stray light can be Doppler shifted by {\textmu}Hz into the frequency bins neighboring the heterodyne frequency. This hypothesis is corroborated by the fact that similar peak amplitudes in terms of the closed to open shutter ratio are seen at the veto and science detectors. This point is discussed in more detail in Section~\ref{Sec:Res_ver}.
A statistical analysis of the background and plots showing the data from frequencies within 100\,{\textmu}Hz of the signal bin is discussed in Section~\ref{Sec:AFA_res}.

\subsubsection{Quantum Noise}
\label{Sec:Tech_noi}

The shot noise limit for the single sided \ac{PSD} of the photocurrent produced by a photodiode can be expressed using measured parameters by the equation 
\begin{equation}
    S_v(f) = \frac{2\bar P G^2 h\nu}{\epsilon},
\end{equation} 
in terms of $\rm V^2/Hz$.
Here $\bar P$ is the average laser power on the science \ac{PD}, $G$ is the gain in V/W of the detection system, $h$ is Planck's constant, $\nu$ is the laser frequency, and $\epsilon$ is the quantum efficiency of the photodetector.

Assuming an average power of 270\,{\textmu}W for the $\rm S_{_\parallel}$, a conversion gain at the heterodyne frequency of 3.00\,V/mW, a frequency of 282\,THz, and a quantum efficiency of 0.7, the voltage shot noise limited  \ac{PSD} would be $1.30\times10^{-15}\rm\,V^2/{\rm Hz}$. After 1.06\, million seconds of integration time this would produce a single sided flat power spectrum with an expectation value of $1.22\times10^{-21}\rm\,V^2$. Normalizing by the average \ac{HPL} power and open shutter calibration amplitude, and then also dividing by two as our measurement is equivalent to a double sided spectrum, gives the shot noise limit on $\left|Z_{\rm closed}\right|^2$ of $5.2\times10^{-8}$ for the science detector during $\rm S_{_\parallel}$. Calculating this noise limit using the parameters for the science detector during $\rm S_{_\perp}$ gives $6.3\times10^{-8}$. The mean technical noise, derived in Section~\ref{Sec:Tech_noi_2}, in terms of $\left|Z_{\rm closed}\right|^2$ was $3.3\times10^{-7}$ for $\rm S_{_\parallel}$, a factor of 6 above the shot noise limit, and $4.2\times10^{-7}$ for $\rm S_{_\perp}$, a factor of 7 above shot noise.

For the veto detector the expected single sided shot noise \ac{PSD} can be found to be $2.5\times10^{-15}\rm\,V^2/{\rm Hz}$ for  $\rm S_{_\parallel}$ and $2.2\times10^{-15}\rm\,V^2/{\rm Hz}$ for $\rm S_{_\perp}$. This leads to a shot noise limit of  $1.8\times10^{-9}$ on $\left|Z^{\rm veto}_{\rm closed}\right|^2$ for $\rm S_{_\parallel}$ and $2.2\times10^{-9}$ on $\left|Z^{\rm veto}_{\rm closed}\right|^2$ for $\rm S_{_\perp}$. In terms of $\left|Z_{\rm closed}\right|^2$, the shot noise limits for the veto detector are lower than at the science detector simply because the open shutter signal does not need to transmit the \ac{RC}. If both detectors reach the shot noise limit the veto detector will be more sensitive with respect to $\left|Z_{\rm closed}\right|^2$ by a factor of $1/T_{\rm RC}$ or roughly 30. This does not mean the veto detector is more sensitive to the coupling of \ac{BSM} fields. On the contrary, it is significantly less sensitive as the coupling of the \ac{BSM} signal to the detector in terms of the closed to open shutter power ratio is roughly a factor of 600 lower than at the science detector. This is explained in more detail in Section~\ref{Sec:Sci_Veto_rat}, however it is due to the fact that reflectivity of the \ac{RC}, and hence the open-shutter power of the \ac{HPL} in the \ac{RC}'s spatial eigenmode that is sent to the veto detector, is a factor of 30 larger than the transmissivity of the \ac{RC}, or the in-mode \ac{HPL} open-shutter power sent to the science detector. Likewise, the power reconverted from an actual \ac{BSM} signal would be roughly 20 times lower in transmission of RC2 (the cavity port directed to the veto detector) compared to RC1 (the cavity port directed to the science detector) due to the transmissivities of the mirrors.

\subsubsection{Technical Noise}
\label{Sec:Tech_noi_2}

Measurements of conventional sources of technical noise such as the noise of the \ac{ADC}s used in the data acquisition system and the noise equivalent power of the science \ac{PD} gave power spectral densities at the heterodyne frequency lower than the technical noise measured in the experiment or even the projected shot noise limit at that frequency. Therefore, a more thorough investigation was required to identify the source of this noise. This investigation included tests that revealed that the noise was related to the interaction of the \ac{LO} and the \ac{AL} when the \ac{AL} was frequency stabilized to a cavity resonance. One set of measurements was performed in reflection of the \ac{RC} with the circulating field of the cavity blocked, but the \ac{AL} reduced in power to the same power level as if it were on resonance. With the \ac{LO} at its normal power, the resulting \ac{PSD} of the voltage noise output by the photodetector was within 20\% of the predicted shot noise level at the heterodyne frequency. However, measurements performed with the system in its normal configuration (cavity unblocked, \ac{AL} resonant with the \ac{RC}) the \ac{PSD} of the voltage noise output by the science \ac{PD} was a factor of 6 above the predicted shot noise level, identical to the results achieved during $\rm S_{_\parallel}$. This was despite the fact that power incident on the science \ac{PD} was the same in both measurements. A third measurement with the \ac{AL} stabilized to the \ac{RC}, but the \ac{LO} blocked also produced a \ac{PSD} in the voltage noise output by the science \ac{PD} at the heterodyne frequency that was consistent with the predicted shot noise limit. It was not possible to measure the power noise at the science \ac{PD} with the \ac{AL} off resonance from the \ac{RC} and both it and the \ac{LO} at nominal power as this saturated the photodetector. While this points to the interaction between the \ac{LO}, \ac{AL}, and the \ac{RC} being the source of the technical noise, these effects are not entirely understood and therefore are still under investigation.

While it was not possible to increase the \ac{LO} power due to saturation effects in the science \ac{PD}, future upgrades plan for a new design of the science \ac{PD} that can accommodate higher powers to ensure that shot noise from the detection of the \ac{LO} is the dominant source of noise at the detector. 
This would lead to an apparent gain in sensitivity of a factor of roughly 2.5 with respect to $\mathcal{P}_{\gamma\leftrightarrow \phi}$ and 1.6 with respect to $g_{\phi\gamma\gamma}$ under the assumption the current science runs were limited by technical noise. However, both $S_\perp$ and $S_\parallel$ appear to be limited by an additional background discussed in the next section.

\begin{figure}
    \centering            
    \begin{subfigure}[b]{0.49\textwidth}
            \centering
    \includegraphics[width=\textwidth]{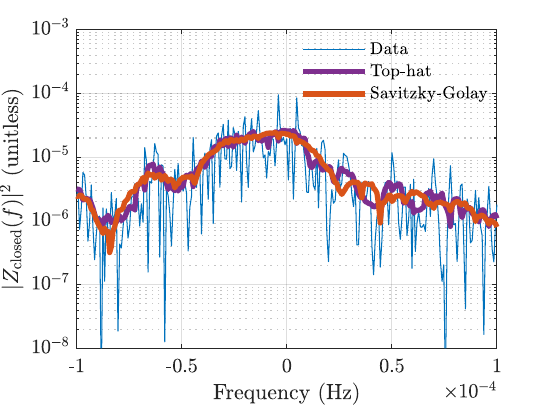}
    \caption{Science detector ($\rm S_{_\perp}$)
    \label{fig:alt_sci_nei_SC_100u}}
    \end{subfigure}
        \begin{subfigure}[b]{0.49\textwidth}
            \centering
    \includegraphics[width=\textwidth]{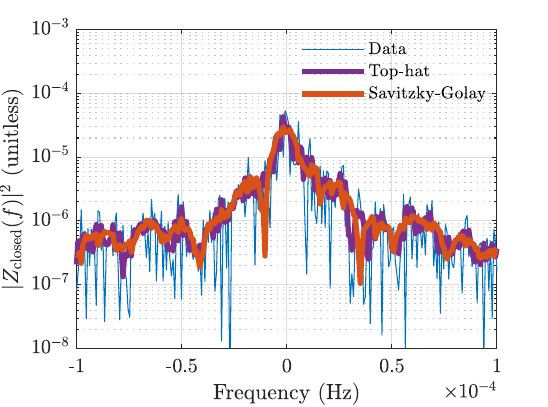}
    \caption{Veto detector ($\rm S_{_\perp}$)
    \label{fig:alt_veto_nei_SC_100u}}
    \end{subfigure}
        \caption{$|Z(f)_{\rm closed}|^2$ measured by the science (a) and veto (b) detectors during $\rm S_{_\perp}$ are shown in blue over the frequency range from -100\,{\textmu}Hz to +100\,{\textmu}Hz away from the signal frequency bin. The data point in the signal bin is not considered when calculating the filtered data to avoid the possibility that an actual signal would influence the estimation of the background (the estimated background at the signal frequency is still calculated).
    \label{fig:alt_nei_SC_100u}}
\end{figure}

\subsection{Background Estimation}
\label{Sec:AFA_res}

An estimation of the expected measured noise in the absence of a \ac{BSM} signal is performed by applying a filter to the data to calculate the expected value in the signal bin based on results in the frequency bins neighboring it. Two different filters were used in this process. One was a top-hat filter that produces a moving average of the eleven frequency bins centered on the bin in which the noise is being estimated for while the other was a Savitzky-Golay filter \cite{savitzky1964smoothing} that fits a 6-th order polynomial to the 101 points centered on that same bin.\footnote{The effects of spectral leakage discussed in Appendix~\ref{Sec:Freq_spr} are not considered.} In all cases, the signal bin was not included in the data considered by the filters to ensure that an actual signal would not influence the estimated background.\footnote{While the signal bin is excluded from consideration in each of the points of the filtered data, this should not be mistaken for the background in the signal bin not being estimated.}

\begin{figure}
    \centering            
    \begin{subfigure}[b]{0.49\textwidth}
            \centering
    \includegraphics[width=\textwidth]{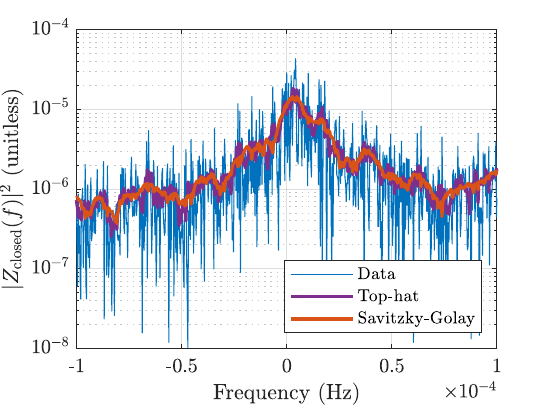}
    \caption{Science detector ($\rm S_{_\parallel}$)
    \label{fig:alt_sci_nei_PS_100u}}
    \end{subfigure}
        \begin{subfigure}[b]{0.49\textwidth}
            \centering
    \includegraphics[width=\textwidth]{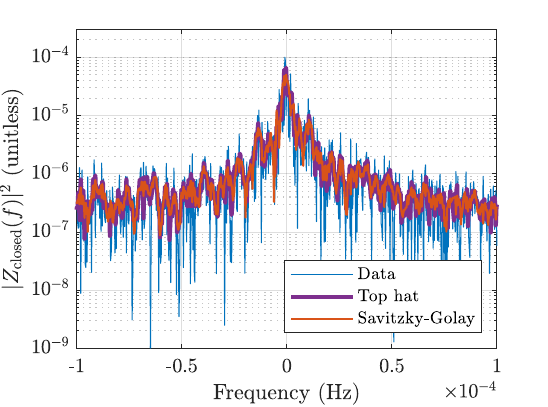}
    \caption{Veto detector ($\rm S_{_\parallel}$)
    \label{fig:alt_veto_nei_PS_100u}}
    \end{subfigure}
        \caption{$|Z(f)_{\rm closed}|^2$ measured by the science (a) and veto (b) detectors during $\rm S_{_\parallel}$ are shown in blue over the frequency range from -100\,{\textmu}Hz to +100\,{\textmu}Hz away from the signal frequency bin. The data point in the signal bin is not considered when calculating the filtered data to avoid the possibility that an actual signal would influence the estimation of the background (the estimated background at the signal frequency is still calculated). 
    \label{fig:alt_nei_PS_100u}}
\end{figure}

The results of applying the filters to $|Z(f)_{\rm closed}|^2$ at frequencies 100\,{\textmu}Hz above and below the signal bin are shown in Figures~\ref{fig:alt_sci_nei_SC_100u} and \ref{fig:alt_sci_nei_PS_100u}. 
Figures~\ref{fig:alt_veto_nei_SC_100u} and \ref{fig:alt_veto_nei_PS_100u} show the same process applied to the data from the veto detector. In all of these plots the data are shown in blue, while the results from the top-hat filter are represented by the purple lines with the orange lines giving the results of using the Savitzky-Golay filter. In the case of the veto detector results, a narrower Savitzky-Golay filter that considered the 41 bins and a top-hat filter that averaged 5 frequency bins were used to estimate the background as the data showed a narrower peak. 

The resulting expectation values of the background in the signal bin for $\rm S_{_\perp}$ and $\rm S_{_\parallel}$ estimated by the filters are shown in Table~\ref{tab:Het_Sig} in the two columns to the right of the results in the signal bin. In this table $\mathcal M_{_{\rm TH}}(f_k=0)$ refers to the background estimated by the top-hat filter and $\mathcal M_{_{\rm SG}}(f_k=0)$ is the background estimated by the Savitzky-Golay filter. The error bars here come from the uncertainty in the calibration array discussed earlier. From these results, it is apparent that $\rm S_{_\perp}$ did not show an excess in the signal bin with respect to the expected background, while the $\rm S_{_\parallel}$ data showed a slight excess. Significant background signals were also seen in the veto detectors. The ratio of backgrounds measured at the science and veto detectors along with the same ratio of the results measured in the signal bin is discussed more in Section~\ref{Sec:Sci_Veto_rat}, but is interpreted as evidence that these signals and background are the result of stray light scattering from the \ac{HPL} to the detection system and not the result of the coupling of the laser to \ac{BSM} fields.

The number of points considered in the filtering was selected based on Monte-Carlo simulations of randomly generated data with a probability distribution modeled after the general shape seen in the frequency spectrum of the measured data within 100\,{\textmu}Hz of the signal bin. For each iteration of the simulations, a Gaussian shape was used to model the background peak observed in the data. The width of the Gaussian, its frequency shift from 0\,Hz,\footnote{In Figures~\ref{fig:alt_sci_nei_SC_100u} and  \ref{fig:alt_sci_nei_PS_100u} it is apparent that the measured distributions are not centered on the signal bin.}  peak amplitude, and baseline amplitude were set to be similar to what was measured during the run, but with random noise giving the values a standard deviation of 20\% from their nominal values from iteration to  iteration. This was done to mimic the changes seen in the shape of the background spectra between the runs.\footnote{Because the background is believed to be due to stray light, this behavior could originate from {\textmu}Hz Doppler shifts in the stray light as its optical path length either expands or contracts. Over long enough integration times the Doppler shifts are expected to be symmetric about zero as the optical path length of the stray light to the detector should be just as likely to contract as it should be the expand. However, the measured background spectra only represent a snapshot of the background during the period in which the data runs were taking place and biases could exist due to the times of day when the system was in operation. The offsets seen in these measurements are therefore interpreted to be representative of the expectation value for magnitude of the shifts present in the center of background spectra over these time scales.}

While the measured data showed the broad features that were modeled in the Monte Carlo simulations by the process outlined in the previous paragraph, it also showed random behavior from bin to bin that adhered to a non-central $\chi^2$ distribution with two degrees of freedom. Therefore, each of the modeled spectra was then multiplied by a random series of numbers drawn from a $\chi^2$ distribution. After this, the filter being tested was applied to the modeled spectra and the deviation of the filtered value from the nominal value of the spectra before the noise was included was calculated. With this technique, it was possible to assess whether the filters produced an accurate representation of the underlying spectral features, and the filter parameters could be tuned for optimal performance based on the results.

\begin{figure}
   \centering
            \begin{subfigure}[b]{0.49\textwidth}
            \centering
    \includegraphics[width=\textwidth]{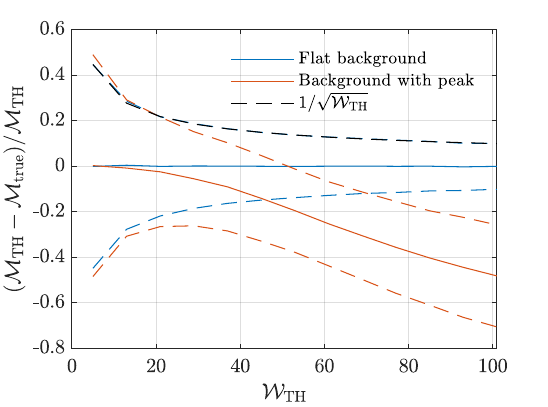}
    \caption{Top-hat Monte Carlo results
    \label{fig:alt_sci_mc_avg_100u}}
    \end{subfigure}
            \begin{subfigure}[b]{0.49\textwidth}
            \centering
    \includegraphics[width=\textwidth]{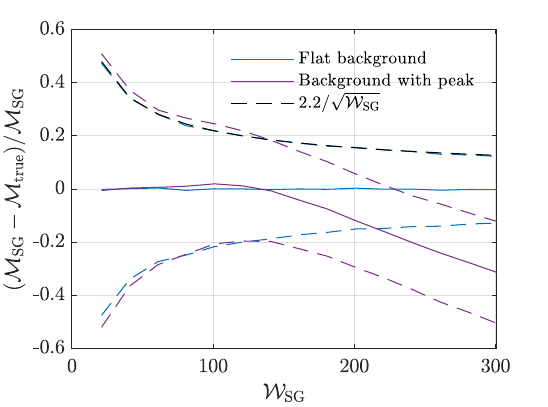}
    \caption{Savitzky-Golay Monte Carlo results
    \label{fig:alt_sci_mc_sg_100u}}
    \end{subfigure}
       \caption{ Results of Monte-Carlo simulations of applying a top-hat (a) and Savitsky-Golay (b) filters to data with a noise distribution modeled after data measured in $\rm S_{_\perp}$ and $\rm S_{_\parallel}$. The Gaussian background was randomly varied in its width, central frequency, peak amplitude and baseline amplitude with a standard deviation of 20\% of their observed values in the data, and noise from a non-central $\chi^2$ distribution was applied to the resulting spectra of each iteration.
    \label{fig:alt_sci_mc}}
\end{figure}

The results of the Monte-Carlo simulation can be seen in Figures~\ref{fig:alt_sci_mc_avg_100u} for the top-hat filter and \ref{fig:alt_sci_mc_sg_100u} for the Savitzky-Golay filter. Here, the number of frequency bins being filtered is represented by the horizontal axis. $\mathcal M_{_{\rm TH}}$ and $\mathcal M_{_{\rm SG}}$ are meant to represent the filtered values of the noisy spectra in central bin, while $\mathcal M_{\rm true}$ is the nominal `noiseless' value. The normalized deviation is therefore given by $(\mathcal M_{_{\rm TH/SG}}-\mathcal M_{\rm true})/\mathcal M_{\rm true}$ and is shown as the solid orange and purple lines for the top-hat and Savitzky-Golay filters, respectively. The dashed lines show the average standard deviations of the filtered data between iterations of the Monte Carlo simulations. The results of filtering the modeled background spectra are compared with the results of applying the filter to a flat background frequency spectrum with the same noise characteristics shown as the blue lines. Ideally, the filter parameters should be tuned such that data with the modeled spectrum match the data from the flat spectrum. If that is not the case, it can indicate that the filter is somehow biasing the data. The black dashed lines show the expected standard deviations for the filtered flat distribution as a function of the number frequency bins filtered. Both plots show that these traces agree well with the Monte Carlo results.

These plots show that the top-hat filter has a tendency to underestimate the value of the distribution in the signal bin after the number of points considered goes beyond 20, while the Savitzky-Golay filter was accurate out to filter sizes of 150 data points. It is also apparent that as the filter size increases, the variance in the results decreases for both filters. Therefore, it is advantageous to choose the largest filter size possible before the filter begins to skew the average of the data. With this in mind, the top-hat filter width was chosen to be ten points, while the 6th-order Savitzky-Golay filter was chosen to be 100 points wide to allow for sufficient margin from the filter parameters where the biases begin to appear in the simulations. It is also important to note that using the Savitzky-Golay filter is considered to be the more robust method due to the lower standard deviation of the results from the Monte-Carlo simulations at the selected filter size. For this reason, using the Savitzky-Golay with a filter width of 101 points (the central data point as well as the 100 neighboring frequency bins) was adopted as the formal method in the analysis. The top-hat filter is only included to allow in the following to show a comparison of the results with an alternative method.




It should be emphasized again that the individual data value in the signal bin is not considered. This is to avoid the possibility that an actual signal would influence the estimated background.

\subsection{Statistics of the Background}
\label{Sec:BG_stat_ex_lim}

Statistics on the data measured at frequencies outside the signal bin were also compiled. With these statistics, detection thresholds and exclusion limits on the signal rate were calculated using the probability distribution of this data. The calculation of the probability distribution of the data is discussed in Section~\ref{Sec:Lim_g} while the process of deriving the exclusion limits is discussed in Section~\ref{Sec:Sig_Ex}.

\subsubsection{Detection Thresholds}
\label{Sec:Lim_g}

Because the filtered spectrum of the closed to open power ratio gives an estimate of the expectation value of the background spectrum in terms of $|Z_{\rm closed}|^2$, it can be used to compile statistics on the noise in the data relative to that expectation value. In turn, these statistics can be used to assess the probability of measuring a given result in the signal bin and calculate a threshold on the signal strength for claiming a detection at a given level of confidence. To do this, the ratio of measured data to the filtered data was calculated over the frequency span of -100\,{\textmu}Hz to +100\,{\textmu}Hz. This ratio $W(f)$, is
\begin{equation}
W(f) = \frac{|Z(f)_{\rm closed}|^2}{\mathcal M(f)}.
\end{equation}
The further the value of $W$ is from one the less likely the result. 

\begin{figure}[t]
    \centering
    \begin{subfigure}[b]{0.49\textwidth}
            \centering
    \includegraphics[width=\textwidth]{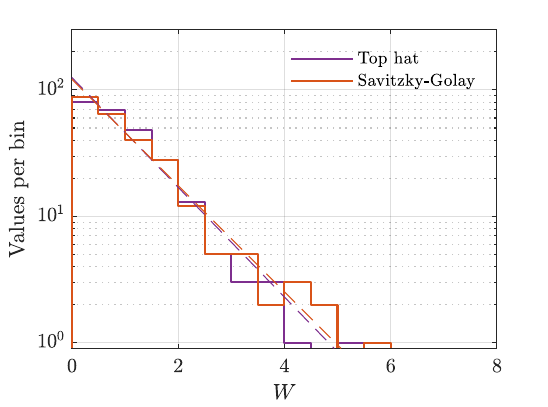}
    \caption{Histogram of the $W$ data ($\rm S_{_\perp}$)
    \label{fig:hist_SC}}
    \end{subfigure}
    \begin{subfigure}[b]{0.49\textwidth}
            \centering
    \includegraphics[width=\textwidth]{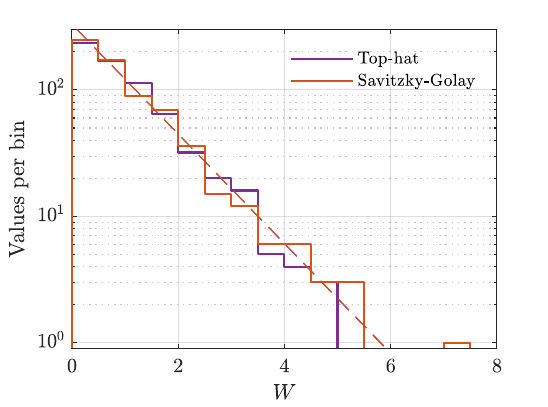}
    \caption{ Histogram of the $W$ data ($\rm S_{_\parallel}$) 
    \label{fig:hist_PS}}
    \end{subfigure}
        \caption{Histograms of the $W$ data from $\rm S_{_\perp}$ (a) and $\rm S_{_\parallel}$ (b) using the top-hat filter (purple) and the Savitzky-Golay filter (red-orange). The dashed lines show the resulting non-central $\chi^2$ distribution of two degrees of freedom by performing an unbinned maximum likelihood estimation of the non-centrality parameter for the $W$ data using the two filtering methods.
    \label{fig:hist_FSC}}
    \end{figure}

This helps reduce the impact of the shape of the frequency spectrum of the background near the signal bin. If only the average value of the data were used for normalization, the peak in the frequency spectrum would have a major influence on the resulting probability distribution. By filtering, the effects from the broad features of the sprectrum can be removed, leaving only the random variance of the data from point to point. The resulting histograms of $W$ from -100\,{\textmu}Hz to +100\,{\textmu}Hz are shown in Figure~\ref{fig:hist_SC} for $\rm S_{_\perp}$ and Figure~\ref{fig:hist_PS} for $\rm S_{_\parallel}$. In these plots, the purple line is the histogram of $W$ using the top-hat filter and the red-orange line is the histogram when using the Savitzky-Golay filter. At low values of $W$, the histograms show a similar shape that follows an exponential distribution. 

\begin{table*}[t]
    \centering
    \makebox[\textwidth][c]{
    \begin{tabular}{c|cc}    
    & \multicolumn{2}{c}{$\rm S_{_\perp}$ Run} \\   
    & $W_{\rm TH}$ & $W_{\rm SG}$  \\ \hline       
Observed result: $W(f_k=0)$      & 0.76 & 0.85  \\ 
Non-centrality parameter: $|\zeta|^2$ & $0.000^{+0.126}_{-0.000}$ & $0.067^{+0.114}_{-0.067}$ \\ 
Std. dev. of noise ($W$) & 1.00 & 1.03  \\
$5\sigma$ det. Thresh. $(W)$ & 14.37 & 14.81 \\ 
Significance of observed result & $0.7\,\sigma$ & $0.8\,\sigma$ \\
$5\sigma$ Det. Thresh. $\left(|Z_{\rm closed}|^2\right)$ &$(3.7\pm0.5)\times10^{-4}$ &$(3.4\pm0.5)\times10^{-4}$  \\
95\% C.L. $5\sigma$ Ex. Lim   $(W)$  & 23.97 & 24.54 \\ 
95\% C.L. $5\sigma$ Ex. Lim.  $\left(|Z_{\rm closed}|^2\right)$ & $(6.1\pm0.8)\times10^{-4}$ & $(5.6\pm0.8)\times10^{-4}$ \\ 
\multicolumn{3}{c}{ } \\  
  &  \multicolumn{2}{c}{$\rm S_{_\parallel}$ Run}\\   
    & $W_{\rm TH}$ & $W_{\rm SG}$   \\ \hline       
Observed result: $W(f_k=0)$      &     3.35 & 2.67 \\ 
Non-centrality parameter: $|\zeta|^2$ &  $0.000^{+0.080}_{-0.000}$ & $0.005^{+0.086}_{-0.005}$ \\
Std. dev. of noise ($W$) & 1.00 & 1.00  \\ 
Significance of observed result & $2.1\,\sigma$ & $1.8\,\sigma$ \\
$5\sigma$ det. Thresh. $(W)$ &   14.37 & 14.41 \\ 
$5\sigma$ det. Thresh. $\left(|Z_{\rm closed}|^2\right)$ &  $(1.2\pm0.2)\times10^{-4}$ & $(1.5\pm0.2)\times10^{-4}$ \\
95\% C.L. $5\sigma$ ex. lim   $(W)$ &  23.97 & 24.01 \\ 
95\% C.L. $5\sigma$ ex. lim.  $\left(|Z_{\rm closed}|^2\right)$ & $(2.0\pm0.3)\times10^{-4}$ & $(2.6\pm0.3)\times10^{-4}$ \\ 
    \end{tabular}  
}
    \caption{Summary of the results at the science detector of $W$ for the different analysis methods, including the unbinned maximum likelihood estimate of the non-centrality parameter, the $5\sigma$ detection thresholds in terms of $W$ and $|Z_{\rm closed}|^2$ that are calculated from the resulting probability distributions, as the 95\% confidence level $5\sigma$ exclusion limits.
    \label{tab:result_W}}
\end{table*}

A non-central $\chi^2$ distribution was fit to the histograms produced by the different filters because $|Z_{\rm closed}|^2$ is in terms of a squared magnitude and the noise on the signal is assumed to have the properties of a normal distribution on the complex plane \cite{NCXsq1}. The results of the fitting procedure is shown in Figures~\ref{fig:hist_SC} and \ref{fig:hist_PS} as the dashed lines.

The non-central $\chi^2$ distribution is a generalization of the $\chi^2$ distribution. Here we use a scaled version in which Gaussian noise on the complex plane with a variance in the real and imaginary components of 0.5 each, is offset from the origin by some value $\zeta$ giving rise to the non-centrality parameter $|\zeta|^2$. For two degrees of freedom, the \ac{PDF} of the non-central $\chi^2$ distribution is given by the equation
\begin{equation}
   f_{\rm nc\chi^2}(x|\zeta)  = e^{-\left(x+\frac{|\zeta|^2}{2}\right)}I_0\left(\sqrt{2|\zeta|^2 x}\right).
   \label{Eq:pdf_ncx}
\end{equation}
In this equation $I_0$ is the modified Bessel function of the first kind. In this formulation the non-central $\chi^2$ distribution has a mean value of $1+|\zeta|^2/2$ and a variance of $1+|\zeta|^2$, where $|\zeta|^2$ is the non-centrality parameter, .

Table~\ref{tab:result_W} gives a summary of the statistical analysis of the background data. In this table, the first row gives the ratio of the result in the signal bin versus the expectation value for the top-hat filter $W_{\rm TH}$ and the Savitzky-Golay filter $W_{\rm SG}$. The second row gives the non-centrality parameter resulting from an unbinned maximum likelihood estimation using the $W$ values from each of the filtering methods within 100\,{\textmu}Hz
of the signal bin with 95\% confidence intervals from this process. As the table shows, all of the fitted non-centrality parameters are extremely close to zero. 
The standard deviation of the noise in $W$ is calculated from the non-centrality parameter. This gives a measure of the statistical uncertainty of the results. Here, the sample standard deviation of the $W$ data is not shown  because it can be biased due to the correlations between bins that is introduced by dividing the alternative frequency data by the filtered data. Nevertheless, the Monte Carlo simulations also showed that these effects do not significantly affect our ability to calculate the probability distribution of the noise.

The $5\sigma$ detection threshold in terms of $W$ are displayed in the fourth row. This corresponds to the value of $W$ that a non-central $\chi^2$ distribution with the non-centrality parameter shown above in the table would produce with a probability of less than $5.73\times10^{-7}$.

The values of $W$ observed in the signal bin, shown in the first rows of each of the tables, are all far below than the $5\sigma$ detection thresholds derived in the sixth row. Therefore, there were no signals significant enough at the expected frequency to claim a discovery for either of the science runs. The significance of the result for $S_\perp$ was $0.7\,\sigma$ for the top-hat filter and $0.8\,\sigma$ for the Savitzky-Golay filter, while for $S_\parallel$ it was $2.1\,\sigma$ for the top-hat filter and $1.8\,\sigma$ for the Savitzky-Golay filter. Here, it should be emphasized that, despite the presence of a minor excess above the average background, the results of $S_\parallel$ are not interpreted as a hint of an interaction between electromagnetic and \ac{BSM} fields and is believed to be simply a statistical anomaly. 

The detection thresholds are also shown in terms of $|Z_{\rm closed}|^2$ in the sixth row of the table. This is calculated by multiplying the $5\sigma$ detection threshold in terms of $W$ by the expectation value of $|Z_{\rm closed}|^2$ calculated by applying the filters.

\subsubsection{Setting Exclusion Limits}
\label{Sec:Sig_Ex}

The non-central $\chi^2$ distribution can also be used to set exclusion limits $|Z_{\rm closed}|^2$ based on the $5\sigma$ detection thresholds described in the previous paragraph. These are found for a 95\% confidence level by calculating the non-centrality parameter, or signal strength, which would have a 95\% probability of producing a result above the $5\sigma$ detection threshold, given the measured probability distribution of the background. The resulting 95\% confidence level $5\sigma$ exclusion limits are shown in terms of the ratio of signal to estimated background $W$ in the seventh row of Table~\ref{tab:result_W}, and in terms of the closed to open shutter power ratio $|Z_{\rm closed}|^2$ in the eighth row. For $\rm S_{_\perp}$, a ratio of the signal to estimated background of 23.97 for the top-filter and 24.54 for the Savitzky-Golay filter can be excluded with 95\% confidence, while its possible to excluded at the same confidence level closed to open shutter power ratios of $(6.1\pm0.8) \times10^{-4}$ for the top-hat filter and $(5.6\pm 0.8)\times10^{-4}$ for the Savitzky-Golay filter. For $\rm S_{_\parallel}$, the 95\% confidence exclusion limits on the signal to background were 23.97 for the top-filter and 24.01 for the Savitzky-Golay filter, while the same exclusion limitis interpreted a closed to open shutter power ratio of $(2.0\pm0.3)\times10^{-4}$ can be excluded for the top-hat filter and $(2.6\pm0.3)\times10^{-4}$ can be excluded for the Savitzky-Golay filter.

\subsection{Summary of Results}
\label{Sec:results}

\begin{table*}
    \centering
    \begin{tabular}{c|cc}    
    & $\mathcal{P}_{\gamma\leftrightarrow \phi} ~(\rm S_{_\perp})$     & $ \mathcal{P}_{\gamma\leftrightarrow \phi}~ (\rm S_{_\parallel} )$ \\    \hline   
Observed Result  &$(4.4\pm0.4)\times10^{-14}$ & $(5.3\pm0.5)\times10^{-14}$ \\
Estimated Background &$(4.7\pm0.4)\times10^{-14}$&  $(3.2\pm0.3)\times10^{-14}$ \\ 
$5\sigma$ Det. Thresh. &$(18\pm2)\times10^{-14}$ & $(12\pm1)\times10^{-14}$ \\
95\% CL $5\sigma$ Ex. Lim. &$25
\times10^{-14}$&  $17\times10^{-14}$ \\ 
    \end{tabular}  
    \caption{Summary of the results in terms of $\mathcal{P}_{\gamma\leftrightarrow \phi} $.}
    \label{tab:results_P}
\end{table*}

Table~\ref{tab:results_P} gives a summary of the results $\rm S_{_\perp}$ and $\rm S_{_\parallel}$ in terms of $\mathcal{P}_{\gamma\leftrightarrow \phi}$. All results in this table are obtain using the Savitzky-Golay filter as this method is considered to be more robust than when the top-hat filter is used. Here the signal bin showed an electromagnetic power to \ac{BSM} power conversion rate of $(4.4\pm0.4)\times10^{-14}$ for $\rm S_{_\perp}$ and $(5.3\pm0.5)\times10^{-14}$ for $\rm S_{_\parallel}$. The average estimated background at the detection system for $\rm S_{_\perp}$ was $(4.7\pm0.4)\times10^{-14}$ and for $\rm S_{_\parallel}$ was $(3.2\pm0.3)\times10^{-14}$. 
The estimated backgrounds were then used to calculate a $5\sigma$ detection threshold of $(18\pm 2)\times10^{-14}$ for $\rm S_{_\perp}$ and $(12\pm 1)\times10^{-14}$ for $\rm S_{_\parallel}$. 
In $\rm S_\perp$ no excess was seen above the estimated background. Although a slight excess with a significance of $1.8\,\sigma$ was observed in $\rm S_\parallel$ (the significance of the result for $\rm S_\perp$ was $0.8\,\sigma$), this is well below the $5\sigma$ detection threshold. Therefore, the discovery of new physics cannot be claimed for either measurement run.  Signals with a $\mathcal{P}_{\gamma\leftrightarrow \phi}$ above $25
\times10^{-14}$ for $\rm S_{_\perp}$ and $17
\times10^{-14}$ for $\rm S_{_\parallel}$ can be excluded with a confidence level of 95\%. These limits were calculated using the upper limit of the systematic uncertainty interval such that for $\rm S_{_\perp}$ $(22\pm3)
\times10^{-14}\rightarrow25\times10^{-14}$ and for $\rm S_{_\parallel}$ $(15\pm2)
\times10^{-14}\rightarrow17\times10^{-14}$. 

\subsection{Verifying the Result}
\label{Sec:Res_ver}

For both $\rm S_{_\perp}$ and $\rm S_{_\parallel}$, background signals well above the technical noise were observed at 100\,{\textmu}Hz above and below the heterodyne frequency. This is not believed to be evidence of \ac{BSM} field, but rather due to stray light scattering out of the \ac{HPL} beam and eventually coupling to the detection system. One indication of this is that the background spectrum is spread over many frequencies rather than being concentrated within just the signal bin. The open shutter measurements over the course of both runs showed no indication for this to be the case for a true science signal. The reason why stray light may not remain at the signal frequency is that it takes a random path to the detection system and is Doppler shifted by {\textmu}Hz, due to uncontrolled environmental noise driving mechanical motion of reflecting surfaces and changing the path length to the detection system.

Other evidence that the signals measured during the science run are not the result of the conversion and reconversion of electromagnetic fields to \ac{BSM} fields are discussed in this section. This includes the ratio of the results at the veto detector system versus the science detector as well as the average time evolution of the variance of the closed shutter signal.
A discussion of how alternative calibration series $C[n]$ are used to check the results can be found in Appendix~\ref{Sec:alt_cn}.

\subsubsection{Science-to-Veto Ratio}
\label{Sec:Sci_Veto_rat}

As Table~\ref{tab:Het_Sig} shows, the results in the signal bin and the expected background are very similar for the science and veto detectors in terms of $|Z_{\rm closed}|^2$. This is not the expected result if the signals were due to the conversion of \ac{BSM} fields to electromagnetic fields inside the \ac{RC}. Based on Equations~\ref{Eq:P_phi_veto} and \ref{Eq:P_open_veto_intro}, a true science signal would produce a value of $|Z_{\rm closed}|^2$ over 600 times larger at the science detector. This is due to the fact that the cavity mirror at the science port RC1 is roughly 20 times more transmissive than the cavity mirror at the veto port RC2, and also the open shutter signal at the veto detector is roughly a factor of 30 larger due to the reflectivity of the \ac{RC} itself.

This discrepancy between the actual results and the expected results given the presence of \ac{BSM} fields could be explained by stray light coupling to both detectors. Scattered light from the \ac{HPL} still must pass through the optics that the open shutter beam passes in order to couple to the detectors. It is therefore reasonable to assume that the average background seen at the veto and science detectors would be similar if the cause is stray light.

If the background seen at the veto and science detectors is indeed because of stray light, the shape of the background spectrum, as well as the moving average of the closed shutter data, indicate that the stray light is not scattering into the \ac{RC} eigenmode before the flat cavity mirror on the \ac{COB}. Otherwise, there should be correlations present between the spectra measured at the science and veto detectors, as well as the time evolution of their data series demodulated at the heterodyne frequency. As the next section shows, such correlations do not appear to exist.

\subsubsection{Moving Averaged Data}
\label{Sec:het_mv}

\begin{figure*}[t]
    \centering
    \begin{subfigure}[b]{0.49\textwidth}
            \centering
    \includegraphics[width=\textwidth]{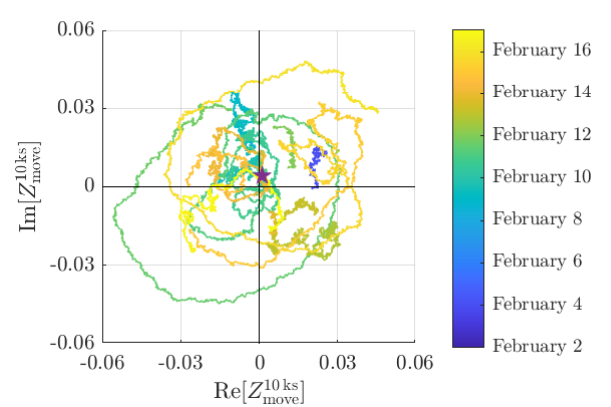}
    \caption{Science detector ($\rm S_{_\perp}$)
    \label{fig:stray_cpx_sci_close_SC}}
    \end{subfigure}
    \begin{subfigure}[b]{0.49\textwidth}
            \centering
    \includegraphics[width=\textwidth]{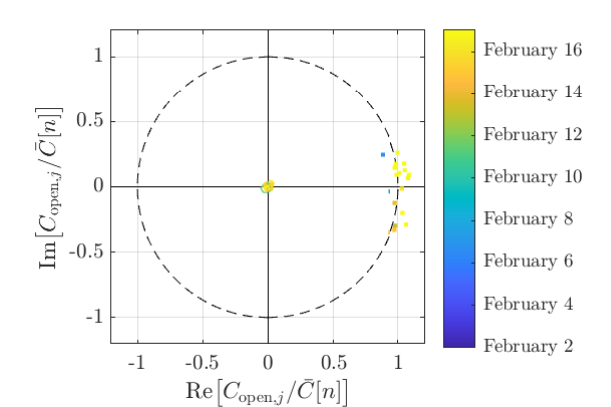}
    \caption{Science detector ($\rm S_{_\perp}$)
    \label{fig:stray_cpx_sci_open_close_SC}}
    \end{subfigure}
      \begin{subfigure}[b]{0.49\textwidth}
            \centering
    \includegraphics[width=\textwidth]{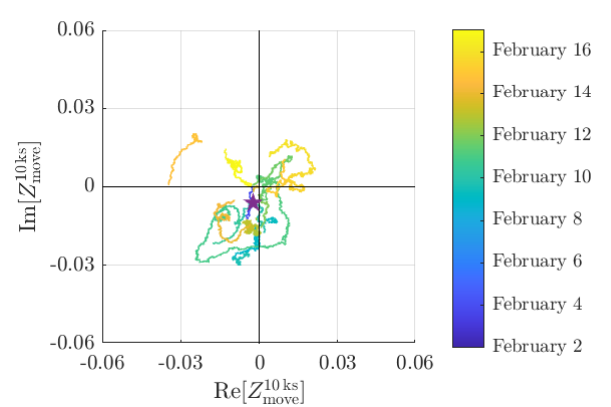}
    \caption{Veto detector ($\rm S_{_\perp}$)
    \label{fig:stray_cpx_veto_close_SC}}
    \end{subfigure}
    \begin{subfigure}[b]{0.49\textwidth}
            \centering
    \includegraphics[width=\textwidth]{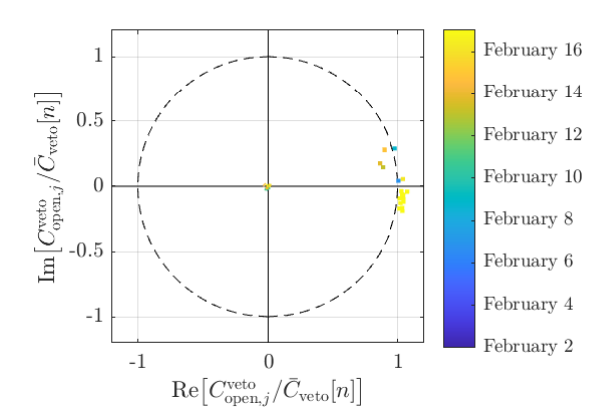}
    \caption{Veto detector ($\rm S_{_\perp}$)
    \label{fig:stray_cpx_veto_open_close_SC}}
    \end{subfigure}
       \caption{Moving average of the calibrated closed to open shutter ratio on the complex plane measured by the science detector during $\rm S_{_\perp}$. The color bar indicates the time since the beginning of data taking and the moving average filters over 10,000 seconds. The purple star indicates the mean value of the data.
    \label{fig:stray_cpx_sci_SC}}
\end{figure*}

\begin{figure*}[t]
    \centering
    \begin{subfigure}[b]{0.49\textwidth}
            \centering
    \includegraphics[width=\textwidth]{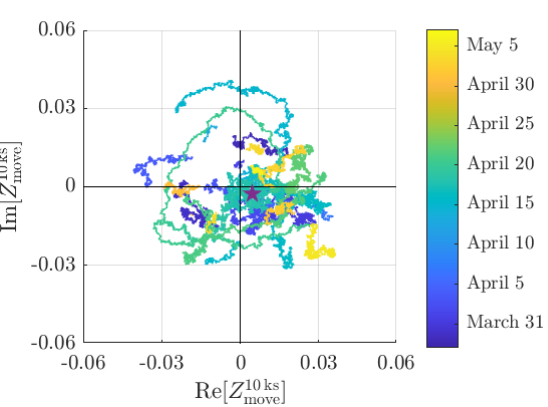}
    \caption{Science detector  ($\rm S_{_\parallel}$)
    \label{fig:stray_cpx_sci_close_PS}}
    \end{subfigure}
    \begin{subfigure}[b]{0.49\textwidth}
            \centering
    \includegraphics[width=\textwidth]{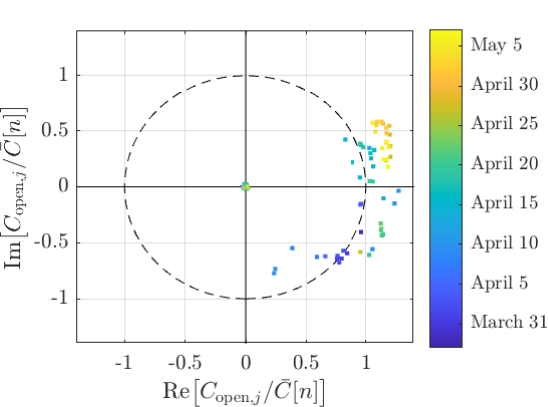}
    \caption{Science detector ($\rm S_{_\parallel}$)
    \label{fig:stray_cpx_sci_open_close_PS}}
    \end{subfigure}
       \begin{subfigure}[b]{0.49\textwidth}
            \centering
    \includegraphics[width=\textwidth]{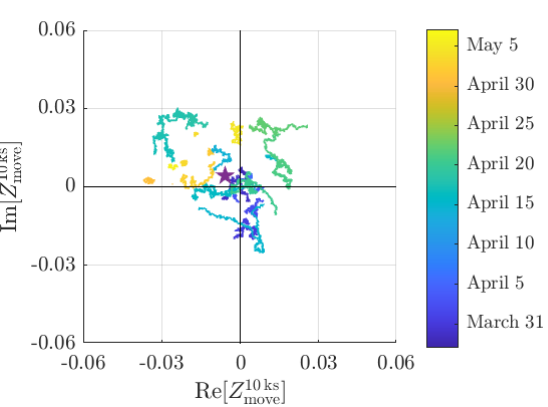}
    \caption{Veto detector ($\rm S_{_\parallel}$)
    \label{fig:stray_cpx_veto_close_PS}}
    \end{subfigure}
    \begin{subfigure}[b]{0.49\textwidth}
            \centering
    \includegraphics[width=\textwidth]{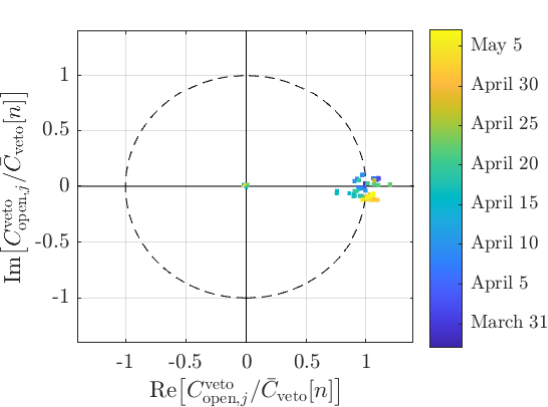}
    \caption{Veto detector ($\rm S_{_\parallel}$)
    \label{fig:stray_cpx_veto_open_close_PS}}
    \end{subfigure}
       \caption{Moving average of the calibrated closed to open shutter ratio on the complex plane measured by the veto detector during $\rm S_{_\parallel}$. The color bar indicates the time since the beginning of data taking and the moving average filters over 10,000 seconds. The purple star indicates the mean value of the data.
    \label{fig:stray_cpx_sci_PS}}
\end{figure*}

The time evolution of the data demodulated at the signal frequency was also examined by performing a moving average of the data. This time series can be expressed as
\begin{equation}
    Z_{\rm move}^{\rm 10\,ks}(t) = \frac{1}{N_{\rm 10\,ks}}\sum_{n = n_{t-5\,{\rm ks}}}^{n_{t+5\,{\rm ks}}} \frac{H_{\rm p}[n]}{C[n]},
\end{equation}
with $N_{\rm 10\,ks}$ representing the number of point in a 10\,ks period of data. With a moving average time constant of 10\,ks the summation on the heterodyne data takes place for the 5,000 seconds before and the 5,000 seconds after time $t$. The resulting moving average signal is shown on the complex plane for the science and veto detectors in Figures~\ref{fig:stray_cpx_sci_close_SC} and \ref{fig:stray_cpx_veto_close_SC}, respectively for $\rm S_{_\perp}$, and for $\rm S_{_\parallel}$ in Figures~\ref{fig:stray_cpx_sci_close_PS} and \ref{fig:stray_cpx_veto_close_PS}.

In these plots the colorbar to the right gives the dates associated with $Z_{\rm move}$. Breaks in the data indicate periods in time when the system was not locked. Here, only segments of data in which the system was locked for the entire duration of the 10,000\,s of averaging appear on the plot. Here, 10,000\,s is chosen because it is a long enough time that the stray light is well above the technical noise of the system, but short enough that the stray-light signal remains relatively coherent (this is explained in more detail in the next section). The axes on the plots are unitless, as the magnitude of the data represents the square root of the closed to open shutter ratio. The purple star gives the result of integrating $Z_{\rm closed}$ for each run, which is analogous to the mean value of the complex time series shown in the plots.


From these data, it is also interesting to consider that for time periods of over 10,000\,s, the signals measured in the heterodyne data appear to be more stable in phase at the veto detectors in both runs. This can be seen in Figures~\ref{fig:stray_cpx_sci_close_SC}, \ref{fig:stray_cpx_veto_close_SC}, \ref{fig:stray_cpx_sci_close_PS}, and \ref{fig:stray_cpx_veto_close_PS} as $Z_{\rm move}$ is much more dynamic on the complex plane for the science detectors in both runs. If one were to assume that stray light is responsible for the signals seen in these plots, this effect could be explained by the shorter path to the veto detector providing fewer opportunities for the phase of the stray-light signal to be modulated. Because of this, when $|Z_{\rm closed}|^2$ is calculated the constructive interference produces a larger value at the veto \ac{PD} relative to the science \ac{PD}.\footnote{The fluctuations in the phase of the signal measured at the heterodyne frequencies are believe to be the dominant source of this effect in comparison to fluctuations in the amplitude as the phase of the moving average times series appears to travel through several cycles over the duration of the run and in general more so than the time series moves in amplitude.} This can be seen directly in the ratio of the average of $|Z_{\rm move}|$ measured at the science and veto versus the same ratio for $|Z_{\rm closed}|^2$ and $\mathcal M$ in the signal bin. As Table~\ref{tab:Het_Sig} shows $|Z_{\rm closed}|^2$ is roughly two times larger at the veto detector compared to the science detector for both runs, corresponding to a roughly 40\% larger magnitude. For $\rm S_\perp$, $\mathcal M$ is very similar for both filters although the results are slightly larger at the veto detector. For $\rm S_\parallel$, $\mathcal M$ is nearly four times larger for both filters at the veto detector leading to the magnitude of $Z_{\rm closed}$ being twice as large at the veto detector.  In contrast, the results of $Z_{\rm move}$ discussed in the previous paragraph show a mean magnitude that was 30\% higher for the science detector during $\rm S_{_\perp}$ compared to the veto detector and a roughly equal average magnitude between the two detectors for $\rm S_{_\parallel}$. This indicates that there may be more instantaneous stray-light power relative to the open shutter power present at the science detector compared to the veto detector, but the coherent sum shows a lower closed-to-open-shutter power ratio over the duration of the run at the science detector due to the larger phase fluctuations in the stray-light there.

In either case, the dynamic nature of the signals on short time scales (10,000\,s) and their relatively large size compared to the expectations of the technical noise are an indication that these signals are due to the spurious coupling of \ac{HPL} power to the detection systems when the shutter is closed and not the reconversion of power from the \ac{BSM} fields.

Figures~\ref{fig:stray_cpx_sci_open_close_SC}, \ref{fig:stray_cpx_veto_open_close_SC}, \ref{fig:stray_cpx_sci_open_close_PS}, \ref{fig:stray_cpx_veto_open_close_PS} show a comparison of the moving average of the closed shutter data versus complex value of $C_{{\rm{open,}}j}$ normalized by the average value of $C[n]$ for each search. This is meant to convey what the phase evolution of an actual signal would look like compared to the data measured during the closed shutter periods. Here it is apparent that the values of $C_{{\rm{open,}}j}$ are well offset from origin clustered, for the most part, near $1+0\,i$. The difference between the $C_{{\rm{open,}}j}$ data measured at the veto detector, which in general has much smaller deviations from $1+0\,i$ than the data measured at the science detector, shows the effect that the detuning of the \ac{HPL} from the resonance of the \ac{RC} has on the complex phase. This effect is discussed in more detail in Section~\ref{Sec:Long_over}. Again, the open shutter calibration will compensate for these effects.


\begin{figure*}[t]
    \centering
           \begin{subfigure}[b]{0.49\textwidth}
            \centering
    \includegraphics[width=\textwidth]{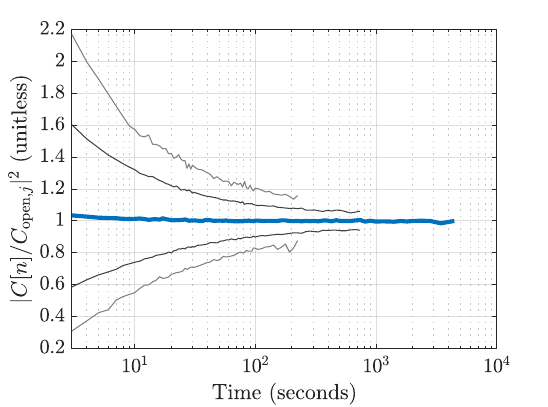}
    \caption{Open shutter ($\rm S_{_\perp}$)
    \label{fig:exp_op_SC}}
    \end{subfigure}
           \begin{subfigure}[b]{0.49\textwidth}
            \centering
    \includegraphics[width=\textwidth]{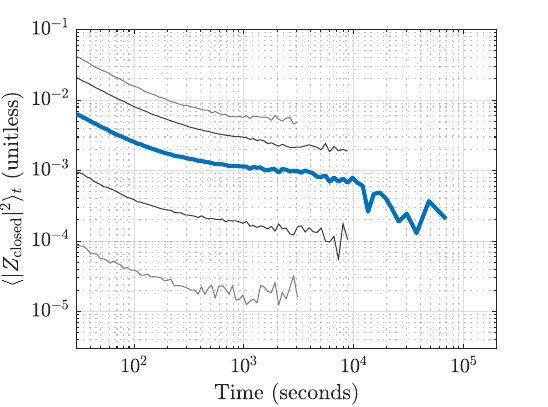}
    \caption{Closed shutter ($\rm S_{_\perp}$)
    \label{fig:exp_cl_SC}}
    \end{subfigure}
               \begin{subfigure}[b]{0.49\textwidth}
            \centering
    \includegraphics[width=\textwidth]{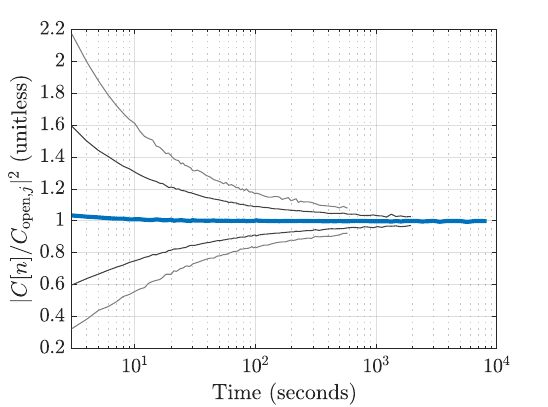}
    \caption{Open shutter ($\rm S_{_\parallel}$)
    \label{fig:exp_op_PS}}
    \end{subfigure}
           \begin{subfigure}[b]{0.49\textwidth}
            \centering
    \includegraphics[width=\textwidth]{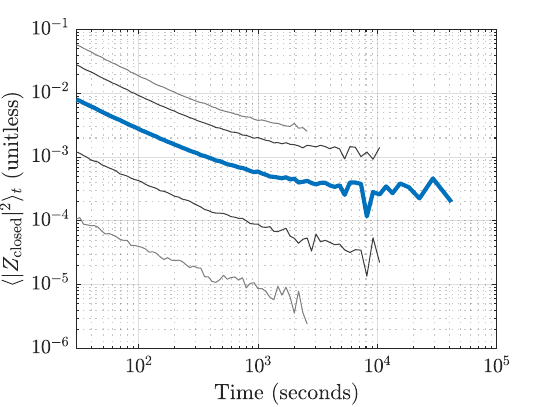}
    \caption{Closed shutter ($\rm S_{_\parallel}$)
    \label{fig:exp_cl_PS}}
    \end{subfigure}
        \caption{The mean open shutter signal (a) and closed shutter background (b)  for $\rm S_{_\perp}$ and open shutter signal (c) and closed shutter background (d)  for $\rm S_{_\parallel}$ is shown in blue as a function of the integration time. The black lines show the  1\,$\sigma$ uncertainty (between the 16\% and 84\%) and the gray lines show the 2\,$\sigma$ uncertainty interval (between 2.5\% and 97.5\%).
    \label{fig:exp_op_cl}}
\end{figure*}

\subsubsection{Time Evolution of Backgrounds}
\label{Sec:Time_Ev}

One potential risk in the experiment is that unknown systematic effects could alter the phase of an actual science signal and prevent it from coherently summing when $|Z_{\rm closed}|^2$ is integrated. This can be checked by examining stretches of data with shorter time scales where it is clear that the optical system is maintaining the coherence between the \ac{HPL} and the oscillators of the detection system, and keeping it on resonance with the \ac{RC}. Potential signals that may not integrate coherently on longer time scales would have a better chance to be measured in this case. The open shutter periods can show what these signals may look like. The blue traces in Figures~\ref{fig:exp_op_SC} and \ref{fig:exp_op_PS} show, for both runs, the average of every segment of open shutter data with a length given by the $x$-axis for $\rm S_{_\perp}$ and $\rm S_{_\parallel}$ respectively. In these plots, the open shutter periods were normalized by the calibration time series.
There are also lines for 1\,$\sigma$ uncertainty (between the 16th and 84th percentile) and 2\,$\sigma$ uncertainty interval (between  percentiles 2.5 and 97.5). It is important to note that these uncertainty intervals converge toward the average of the data as the time increases, indicating the presence of a stable signal in the data.

Figures~\ref{fig:exp_cl_SC} and \ref{fig:exp_cl_PS} show the same analysis, but instead performed on the closed shutter periods of both science runs. The range of integration times is much wider on these plots as there were more long stretches of data during the closed shutter periods. One notable difference between these plots and the plots of the open shutter periods is the total lack of convergence of the uncertainty interval in the closed shutter data. This could be interpreted as an indicator that the blue traces are not showing a coherent signal emerging from the noise in the system, but rather a random signal with mean magnitude squared that is equal to the blue line. This is strong evidence that the results in the signal bin are due to the presence of the background noise and not the conversion and reconversion of \ac{BSM} fields to an electromagnetic field.

It is also interesting to compare the mean backgrounds between $\rm S_{_\perp}$ and $\rm S_{_\parallel}$. For integration times of 3000\,s, the mean background is nearly a factor of two lower for $\rm S_{_\parallel}$, but by 10,000\,s integration times they are only on the order of 40\% lower.
These results appear to roughly agree with the heterodyne moving average analysis in Section~\ref{Sec:het_mv} and the estimate of the backgrounds from the stray-light models in Section~\ref{Sec:AFA_res}, that both appear to indicate a reduction of the stray light between $\rm S_{_\perp}$ and $\rm S_{_\parallel}$. Again, this is notable as the shutter material (see Section~\ref{Sec:COB}) between the runs was changed from the silver mirror to the plate coated with absorptive black foil. Additionally, all but one of the open slots in the shutter wheel were closed for $\rm S_{_\parallel}$, while for $\rm S_{_\perp}$, only the slot with the actual shutter was occupied.

\subsection{Central Optical Bench Transmissivity}

\begin{table*}
    \centering
    \makebox[\textwidth][c]{
    \begin{tabular}{c|c|c|c}    
     \multicolumn{2}{c|}{Ex-situ} &  \multicolumn{2}{c}{In-situ}\\ 
               Optic & Transmissivity & Optic & Transmissivity \\ \hline
LT1  & $(12.4\pm0.5)$\,ppm & LT1  & $(11\pm1)$\,ppm \\
LT2  &  $(12.4\pm0.5)$\,ppm   & LT2  & $(11\pm1)$\,ppm   \\
MZ3  & $(452\pm10)$\,ppm & $\rm MZ_3 \, MZ_4$  & $(0.137\pm0.007)$\,ppm \\
MZ4  & $(275\pm10)$\,ppm &  \\ \hline 
$T_{_{\rm COB}}$&$(1.9\pm0.1)\times10^{-17}$  & $T_{_{\rm COB}}$&$(1.7\pm0.2)\times10^{-17}$\\
$T_{_{\rm COB}}T_{_{\rm RC2}}$&$(9.7\pm1.2)\times10^{-23}$  & $T_{_{\rm COB}}T_{_{\rm RC2}}$&$(8.5\pm1.4)\times10^{-23}$
    \end{tabular}
    }
    \caption{The transmissivity of the on-axis \ac{COB} optics for $p$-polarized light. Here the combined transmissivity of the \ac{COB} on-axis optics, $T_{_{\rm COB}}$, is calculated from the product of the transmissivities of the individual components.}
    \label{tab:T_COB}
\end{table*}

\label{Sec:COB_in}
Precise measurements of combined transmissivity of the \ac{COB} mirrors mounted on the optical axis of the system (see Figure~\ref{fig:COB}),
\begin{equation}
T_{_{\rm COB}}=T_{_{\rm LT1}}T_{_{\rm MZ3}}T_{_{\rm MZ4}}T_{_{\rm LT2}},
\end{equation}
are essential to calibrate the ratio of the closed to open shutter power in terms of the conversion rate between electromagnetic and \ac{BSM} fields $\mathcal{P}_{\gamma\leftrightarrow \phi}$. This can seen in Equation~\ref{Eq:g_a_rat} with the critical parameter being $T_{\!_{\rm COB}}T_{\!_{\rm RC2}}$. $T_{\!_{\rm COB}}$ is also an important to  calculating the field overlap $\eta$ as Section~\ref{Sec:Sci_det_sys} will explain. 

Ex-situ measurements of these optics were performed after the completion of the first science campaign with the \ac{COB} removed from the vacuum system. As the transmissivities of these optics were strongly dependent on the angle of incidence ($\sim$\,1.5\,ppm per degree for the mirrors sealing the light-tight housings and $\sim$\,150\,ppm per degree for the \ac{MZ} mirrors) the flat cavity mirror of the \ac{RC} was used as an alignment reference to ensure that the laser used in the setup passed through the optics at the same angle as the \ac{HPL} light during the open shutter periods of the science runs. The measurement procedure involved shining a laser coupled to an optical fiber on the mirror being tested. The light reflected from the optic was then incident normal to the \ac{RC} flat mirror and reflected back through the fiber. This ensured that the incident beam was aligned to the angle of the \ac{RC} eigenmode (assuming no changes in the alignment of the mirror after the science campaign) to better than 1\,mrad. The results are shown in Table~\ref{tab:T_COB}. Similar results were expected for the pair of light tight housing mirrors and the pair of Mach-Zehnder mirrors, and the measurements of LT1 and LT2 both resulted in a transmissivity of $12.4\pm0.5\rm\,ppm$. MZ3 was measured to have a transmissivity of $452\pm10\rm\,ppm$ while the transmissivity of MZ4 was $275\pm10\rm\,ppm$. While MZ3 and MZ4 were expected to have the same transmissivity as their dielectric coatings were applied during the same coating run, the discrepancy between them is believed to be related to differences in their angles of incidence with respect to the optical axis of the system. The uncertainty in these measurements is due to a combination of the uncertainty in the calibration of the power meters used in the setup, the presence of stray light at the power meters measuring the weak transmitted signals, and the uncertainty of the angle of incidence of the laser on the mirrors. 

In-situ measurements of the on-axis \ac{COB} optics were also performed in between $\rm S_{_\perp}$ and $\rm S_{_\parallel}$. Here, the transmissivity of LT1 was observed via a measurement of the ratio of \ac{HPL} power directly at the input window to the vacuum system in \ac{NL} and then at the High-Power port of the \ac{CH} vacuum tank (see Figure~\ref{fig:COB}). This gave a value of $11\pm1$\,ppm. The transmissivity of LT2 was measured with a similar method, but instead measuring the \ac{RL} power just before the Injection Port at the \ac{CH} tank, also giving a value of $11\pm1$\,ppm. The combined transmissivity of MZ3 and MZ4 was measured using a power meter to measure the \ac{HPL} power at the \ac{PLL}2 port, and then at the Mach-Zehnder port by measuring the amplitude of the \ac{RL}-\ac{HPL} beatnote. The spatial overlap of the \ac{RL}-\ac{HPL} beatnote at optimal alignment could be directly measured at the \ac{PLL}2 port because it was possible to measure the static power of both lasers there. This value could then be used to compensate for these effects in the measurement of the beatnote amplitude at the Mach-Zehnder port.\footnote{It should be noted that to find the beatnote amplitude at each port for the `optimal' alignment, \ac{RL} was realigned to maximize the the beatnote amplitude in both cases. This was necessary to compensate for the alignment drift of the \ac{MZ} on the \ac{COB}.} The combined transmissivity of the \ac{MZ} mirrors found in this measurement was $0.137\pm0.007$\,ppm. 

The result of the combined transmissivities of the \ac{MZ} mirror from the ex-situ measurements gave $0.124\pm0.005$\,ppm. Thus, it is clear that the individual measurements of each of the on-axis \ac{COB} optics agree between the two methods. Even more relevant is the total transmissivity $T_{_{\rm COB}}$, in which a value of $(1.9\pm0.1)\times10^{-17}$ was found for the ex-situ measurement with $(1.7\pm0.2)\times10^{-17}$ measured for the in-situ measurements. In short, these measurements agree within their uncertainty. 

Finally, the most relevant parameter for the calibration of the results in terms of $\mathcal{P}_{\gamma\leftrightarrow \phi}$ is the product of the transmissivity of \ac{COB} optics and the cavity mirror RC2. For this, the ex-situ measurements gave $T_{_{\rm COB}}T_{_{\rm RC2}}=(9.7\pm1.2)\times10^{-23}$ while the in-situ results were $T_{_{\rm COB}}T_{_{\rm RC2}}=(8.5\pm1.4)\times10^{-23}$. Again, the measurements agree within their uncertainty. This gives us confidence in the accuracy of the overall calibration of the results. We consider the method for the ex-situ measurements to be more robust as we were able to directly access the optics. Therefore, the ex-situ measurements were used for the actual calibration of the results.  

\section{Performance of the Optical System}
\label{Sec:Perf}

Using the open shutter periods to calibrate the science data means that measuring the parameters of the optical system, such as the coupling of the \ac{HPL} to the \ac{RC}, is not strictly necessary. However, these measurements are important for verifying the performance of the optical system and also help identify the cause of changes in the calibration series.  Calculating the important metrics to assess the performance of the optical system, such as the resonant enhancement of the \ac{RC} and the total power overlap between the \ac{HPL} and the \ac{RC}, required measuring a number of other parameters. Some of these were dynamic parameters that were measured during maintenance periods during the run, while other static parameters were measured before and/or after the run. This section discusses how these measurements were made and gives a summary of the performance of the optical system during the science runs.

\subsection{Regeneration Cavity Parameters}
\label{Sec:RC}

The \ac{RC} is a critical component of the optical system. Its resonant enhancement factor and the power overlap between its eigenmode and the \ac{HPL} directly affect the sensitivity of the experiment. A more complete characterization of the \ac{RC} can be found in Ref.~\cite{kozlowski2024designperformancealpsii}. This section will focus on how the \ac{RC} performance was characterized during the first science campaign and the results of that effort.
Explanations of the resonant enhancement factor, transmissivity, and reflectivity of the \ac{RC} are given in Sections~\ref{Sec:RC_B}, \ref{Sec:RC_T}, and ~\ref{Sec:RC_R} respectively, along with a discussion of the measurements of these parameters during the science runs. 

\subsubsection{Resonant Enhancement}
\label{Sec:RC_B}

\begin{figure}
    \centering
           \begin{subfigure}[b]{0.49\textwidth}
            \centering
    \includegraphics[width=\textwidth]{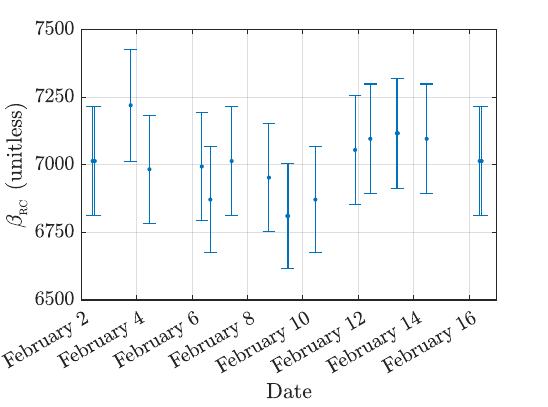}
    \caption{$\rm S_{_\perp}$
    \label{fig:beta_SC}}
    \end{subfigure}
           \begin{subfigure}[b]{0.49\textwidth}
            \centering
    \includegraphics[width=\textwidth]{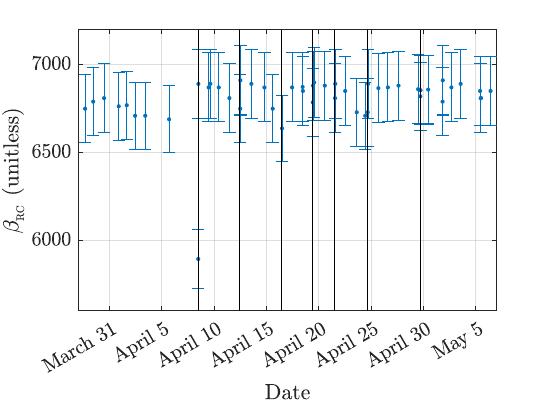}
    \caption{$\rm S_{_\parallel}$ 
    \label{fig:beta_PS}}
    \end{subfigure}
        \caption{The estimated resonant enhancement of the \ac{RC} is shown for $\rm S_{_\perp}$ (a) and  $\rm S_{_\parallel}$ (b) with error bars based on the uncertainty of the measured storage time. The black vertical lines in the plot of the $\rm S_{_\parallel}$ data give the times at which the \ac{RC} mirrors where realignmed to optimize its resonant enhancement. No vertical black lines are shown in the plot of the $\rm S_{_\perp}$ data as the \ac{RC} mirrors were never manually realigned during the course of this run. 
    \label{fig:beta}}
\end{figure}

The resonant enhancement factor of the \ac{RC}, $\beta_{\!_{\rm RC}}$, can be defined as the ratio of the power in the signal field that exits the cavity at the mirror coupled to the detection system, to the signal power at the same location if the cavity were not there at all, for a signal field that is on resonance and in the spatial mode of the \ac{RC},. 
$\beta_{\!_{\rm RC}}$ therefore gives a measure of the amplification of the science signal. In this case, the \ac{BSM} field reconverts to an electromagnetic field inside the cavity and is stored there while being reflected back and forth between the mirrors.
During this time the field experiences the cavity gain as it constructively interferes with the electromagnetic field that is newly reconverted from the \ac{BSM} field. In this way, the field is injected via the reconversion of the \ac{BSM} field to an electromagnetic field, rather than transmitting through an input coupling mirror. The electromagnetic field that has built up in the cavity is then transmitted by RC1 and directed to the science detector. 
The resonant enhancement, $\beta_{\!_{\rm RC}}$, can be expressed as \cite{ortiz2020design}
\begin{equation}
\beta_{\!_{\rm RC}} = \frac{T_{\!_{\rm RC1}}}{\left(1-\sqrt{1-A}\right)^2} \simeq \frac{4 T_{\!_{\rm RC1}}}{A^2}.
\label{Eq:B_rc}
\end{equation}
In this equation, $T_{\!_{\rm RC1}}$ is the power transmissivity of RC1 while $A$ is the total round-trip attenuation of the power circulating in the cavity. This is due to the combination of mirror transmissivities and other optical losses including scattering from the mirror surface roughness, absorption in the mirror coatings, and clipping losses on the combined free apertures of the magnet string and reflective surfaces of the mirrors. The approximation assumes $A\ll1$ and therefore $\sqrt{1-A} \simeq 1-A/2$, leading to an error in the resonant enhancement on the order of 0.01\%.  

Assessing the total attenuation $A$ is critical to knowing not only the resonant enhancement factor during the run, but a number of other parameters as well. Fortunately, $A$ can be measured  via a combination the cavity \ac{FSR} $f_0$ and storage time $\tau$. The relationship between these parameters is
\begin{equation}
    A = 1 - e^{-\frac{2}{f_0\tau}}\simeq \frac{2}{f_0\tau},
\end{equation}
resulting in
\begin{equation}
\beta_{\!_{\rm RC}} \simeq T_{\!_{\rm RC1}} f_0^2 \tau^2,
\end{equation}
allowing it to determined by also measuring the \ac{FSR} and storage time of the \ac{RC}. In this case, using the approximate equation leads to a relative error in the measured resonant enhancement factor on the order of 0.01\%. An estimation of the uncertainty in these measurements is discussed later in this section, but using this approximation leads to errors well below the other sources of error in the measurement.

The relationship between the finesse of the \ac{RC} and $\beta_{\!_{\rm RC}}$ can be approximated by
\begin{align}
\mathcal{F} = \frac{f_0}{2\nu_0} &= \frac{\pi\sqrt{1-\sqrt{A}}}{1-\sqrt{1-A}} \\
&\simeq \pi\frac{2}{A} \simeq \pi \sqrt{\frac{\beta_{\!_{\rm RC}} }{T_{\!_{\rm RC1}}}},
\end{align}
while the finesse can also be defined in terms of the storage time $\tau$ and \ac{FSR} $f_0$ of the cavity \cite{Siegman_1987,Isogai:13} as
\begin{equation}
\mathcal{F} = \pi f_0 \tau .
\end{equation}
Details on the technique for measuring the storage time and the values for $T_{\!_{\rm RC1}}$ and $T_{\!_{\rm RC2}}$ can be found in the Refs.~\cite{kozlowski2024designperformancealpsii} and \cite{Spector:24}. To measure the cavity \ac{FSR} $f_0$ during the run, the laser \ac{AL} was PDH locked to the cavity. Then the \ac{PLL}3 frequency was tuned near 45 \ac{FSR}s, until the power transmitted by the cavity was at a maximum. The value of the \ac{PLL}3 frequency at that point was then divided by 45 to give the \ac{FSR} of the \ac{RC}. The uncertainty in this measurement was 1\,Hz, leading to an uncertainty on the FSR of 0.02\,Hz (see Table\,\ref{tab:RC_param}).

The measurements of $\beta_{\!_{\rm RC}}$ during the maintenance periods are shown in Figures~\ref{fig:beta_SC} and \ref{fig:beta_PS} for $\rm S_{_\perp}$ and $\rm S_{_\parallel}$ respectively. For $\rm S_{_\parallel}$, the black vertical lines in the plot indicate times at which the \ac{RC} eigenmode position on the mirrors was aligned to improve the storage time and in turn the resonant enhancement factor (no such realignments were required for $\rm S_{_\perp}$). For $\rm S_{_\perp}$ the mean value of $\beta_{\!_{\rm RC}} $ associated with the closed shutter periods was 7150 with a range from 6810 to 7120 and \ac{RMS} fluctuations from its mean value of 90. The uncertainty on the measurements of $\beta_{\!_{\rm RC}} $ was 80, mostly due to the uncertainty on the transmissivity of RC1 and the uncertainty on the measurements of the storage time. For the closed shutter periods of $\rm S_{_\parallel}$ the mean value of $\beta_{\!_{\rm RC}} $ was estimated to be 6870 with a range from 6640 to 6910 with an \ac{RMS} of 70. These measurements had the same uncertainty as those performed during $\rm S_{_\perp}$. While a value below 5900 was measured during one of the periods of $\rm S_{_\parallel}$, and a value above 7200 was measured during one of the periods of $\rm S_{_\perp}$, there were no closed shutter periods during time periods associated with these measurements. For this reason this data point is not considered in the range of data mentioned above.

\subsubsection{Transmissivity}
\label{Sec:RC_T}
The transmissivity of the cavity for a laser occupying its spatial eigenmode (the spatial coupling will be examined later) can be expressed as \cite{Siegman_1987}
\begin{align}
T_{\rm RC} &=\frac{T_{\!_{\rm RC1}}T_{\!_{\rm RC2}}}{\left|1-\sqrt{1-A}\,e^{i2\pi \frac{\Delta\nu}{f_0}}\right|^2}\\
&\simeq \frac{4T_{\!_{\rm RC1}}T_{\!_{\rm RC2}}}{A^2}\frac{1}{1+\left(\frac{4\pi\Delta\nu}{A f_0}\right)^2},
\label{Eq:T_simp_rc}
\end{align}
with $\Delta\nu$ being the offset of the laser frequency from its nearest resonance. Assuming that the laser is very nearly on resonance with the cavity ($\Delta\nu\ll Af_0$), the maximum transmissivity of the cavity is 
\begin{equation}
T_{\rm RC,max} =\frac{T_{\!_{\rm RC1}}T_{\!_{\rm RC2}}}{\left(1-\sqrt{1-A}\right)^2} =   T_{\!_{\rm RC2}} \beta_{\!_{\rm RC}} .
\label{Eq:T_simp_rc_max}
\end{equation}
In this equation it is clear that due to the common factors, $T_{\rm RC,max}$ can be expressed as the product of the resonant enhancement and the transmissivity of the flat cavity mirror on the \ac{COB} (again ignoring the spatial or longitudinal field overlap).
It should also be noted that $T_{\rm RC,max}$ is independent of which side the input field is incident on.  The maximum transmissivity of the \ac{RC} on resonance can therefore be expressed as the product 
\begin{equation}
T_{\rm RC,max}\simeq T_{\!_{\rm RC1}} T_{\!_{\rm RC2}} f_0^2 \tau^2.
\end{equation}
\\ For $\rm S_{_\perp}$, during the closed shutter periods $T_{\rm RC,max}$ was estimated to have an average value of 3.6\% with a range from 3.5\% to 3.6\% and \ac{RMS} deviations from the mean of 0.05\%. The uncertainty of $T_{\!_{\rm RC2}}$ was the primary contributor to the absolute uncertainty of $\pm0.4\%$ on $T_{\rm RC,max}$. During the closed shutter periods of $\rm S_{_\parallel}$ the average of $T_{\rm RC,max}$ was estimated to be 3.5\%, with a range from 3.4\% to 3.5\%, an \ac{RMS} of 0.04\%, and an absolute uncertainty of $\pm0.3\%$.

As Equation~\ref{Eq:T_simp_rc} shows, detuning the incident laser frequency from the cavity resonance will reduce its transmissivity. Measurements of the detuning $\Delta\nu$ between the \ac{HPL} frequency and resonance of the \ac{RC} are discussed in Section~\ref{Sec:Long_over} and were used to infer the transmissivity of the \ac{RC} during the first science campaign. During the closed-shutter periods of $\rm S_{_\perp}$, $T_{\rm RC}$ had a mean value of 3.3\%, with \ac{RMS} fluctuations about the mean of 0.2\%, a maximum value of 3.6\%, and a minimum value of 2.6\%. The uncertainty of these measurements was 0.3\%. For $\rm S_{_\perp}$, the closed-shutter periods gave a mean value of 3.0\% for $T_{\rm RC}$, with \ac{RMS} fluctuations of 0.5\%, a maximum of 3.5\%, and a minimum of 1.4\%. The uncertainty was the same as the measurements during $\rm S_{_\perp}$ at 0.3\%. In both runs the uncertainty was almost entirely driven by the uncertainty in the measurement of the transmissivity of the flat \ac{RC} mirror mounted to the \ac{COB}, RC2.


\subsubsection{Reflectivity}
\label{Sec:RC_R}

The reflectivity of the \ac{RC} is used in the calibration of several critical parameters used to evaulate the performance of the optical system. Unlike the transmissivity, the reflectivity differs depending on the side of the cavity on which the laser is incident. The reflectivity experienced by the \ac{HPL} when incident on the flat \ac{RC} mirror in the \ac{CH} cleanroom ($\rm RC2$) will be referred to as $R_{\rm RC,CH}$ while the reflectivity for lasers incident on the curved cavity mirror in the \ac{NR} cleanroom ($\rm RC1$) will be referred to as $R_{\rm RC,NR}$. These quantities can be expressed as~\cite{Siegman_1987} 
\begin{widetext}
\begin{align}
R_{\rm RC,CH} & = \left|\frac{1}{R_{\!_{\rm RC2}}}\left(R_{\!_{\rm RC2}} - \frac{T_{\!_{\rm RC2}}} {1- \sqrt{1-A}\,e^{i2\pi \frac{\Delta\nu}{f_0}}} \right)\right|^2 \simeq  \left|
\left( 1 - \frac{2T_{\!_{\rm RC2}}}{A}\frac{1} {1- i\frac{\Delta \nu}{\nu_0}} \right)\right|^2  \label{Eq:R_CH}, {\rm \hspace{0.3cm} and}\\
R_{\rm RC,NR}& = \left|
\frac{1}{R_{\!_{\rm RC1}}}\left(R_{\!_{\rm RC1}} - \frac{T_{\!_{\rm RC1}}} {1- \sqrt{1-A}\,e^{i2\pi \frac{\Delta\nu}{f_0}}} \right)\right|^2 
\simeq  \left|
\left( 1 - \frac{2T_{\!_{\rm RC1}}}{A}\frac{1} {1- i\frac{\Delta \nu}{\nu_0}} \right)\right|^2 .\label{Eq:R_NR}
\end{align}
\end{widetext}
In both cases, the approximations assume the reflectivity of $\rm RC1$ and $\rm RC2$ are one. 
This can be expressed in terms of measured parameters as 
\begin{align}
R_{_{\rm RC,CH}} &\simeq \left|1 - \frac{T_{_{\rm RC2}} f_0\tau} {1-i2\pi\tau\Delta\nu} \right|^2,{\rm \hspace{0.3cm} and}
\label{Eq:R_RC_CH}\\
R_{_{\rm RC,NR}} &\simeq \left|1 - \frac{T_{_{\rm RC1}} f_0\tau} {1-i2\pi\tau\Delta\nu} \right|^2.
\label{Eq:R_RC_NR}
\end{align}
In Sections~\ref{Sec:eta_sci} and \ref{Sec:eta_total} $R_{_{\rm RC,CH}}$ is used to calibrate the spatial overlap and field overlap measured at the veto detection system respectively. During the closed shutter sections of $\rm S_{_\perp}$ average value of $R_{_{\rm RC,CH}} $ was 92\%, with a range from 92\% to 94\%, and \ac{RMS} fluctuations from the mean of 0.6\%. For $\rm S_{_\parallel}$ the \ac{RC} reflectivity gave an average value of 93\%, with values ranging between 92\% to 97\%, and an \ac{RMS} of 1\%. The uncertainty of these measurements was 1\%, largely driven by the uncertainty in measurements of the transmissivity of RC2.

The reflectivity of the \ac{RC} for the \ac{AL} is used in Section~ to calculate its spatial overlap with the cavity eigenmode. Since \ac{AL} is injected to the \ac{RC} via $\rm RC1$ and is frequency stabilized directly to a cavity resonance, Equation~\ref{Eq:R_RC_NR} for  $R_{_{\rm RC,NR}}$ can be used with $\Delta\nu=0$. In this case the reflectivity of the cavity had an average value of 2.6\% during the closed shutter periods of $\rm S_{_\perp}$ with a range from 2.4\% to 3.1\% and an \ac{RMS} of 0.2\%. During $\rm S_{_\parallel}$ the average reflectivity was 3.0\% while the range was from 2.8\% to 3.4\%. The \ac{RMS} of the reflectivity during these measurements was 0.2\%. For both runs the absolute uncertainty in the reflectivity was 1.3\% and was due to the uncertainty in the measurements of the cavity storage time and the transmissivity of $\rm RC1$.

\subsection{Field overlap}
\label{Sec:Sci_det_sys}

The field overlap between the \ac{HPL} and the \ac{RC}, $\eta$, is a critical metric for evaluating the performance of the optical system. As explained in Section~\ref{Sec:ALPSII_des}, it can be expressed as the product of spectral ($\eta_{\hat z}$), polarization ($\eta_{\rm pol}$), and spatial ($\eta_{xy}$) components.

The spectral component, also called the longitudinal overlap, is discussed in Section~\ref{Sec:Long_over} and can be calculated exclusively from measurements of the \ac{RC} storage time and the frequency detuning of \ac{HPL} from its resonance. The polarization overlap, discussed in Section~\ref{Sec:eta_pol}, is calculated from the orientation of polarization state of the \ac{HPL} with respect to the magnetic field.

The spatial overlap is measured by calibrating the results of the open shutter periods in terms of the \ac{HPL} power leaving the cavity $P_{{\rm open}}$, and dividing by the expected \ac{HPL} power at this point for a laser occupying the spatial eigenmode of the \ac{RC}. At the science detector, this can be expressed as 
\begin{equation}
    \eta_{{xy}}^2  = \frac{ P_{{\rm open}}}{T_{\!_{\rm COB}} T_{\!_{\rm RC}} P_{\rm i}}. 
\end{equation}
The expected power is therefore the product of the combined transmissivities of the \ac{COB} optics $T_{\!_{\rm COB}}$, the transmissivity of the \ac{RC},  $T_{\!_{\rm RC}}$, and the \ac{HPL} power traveling though the production magnet string $P_{\rm i}$. 

The veto detector can also be used to measure the field overlap between the \ac{HPL} and the \ac{RC}. There are even some advantages to using the veto detector as the \ac{HPL} does not need to pass through the cavity to reach it resulting in a \ac{HPL} power is higher at the veto \ac{PD}. Therefore, the following sections include a discussion of the field overlap measured at both the science and veto detectors. The process of measuring the ratio $P_{{\rm open}}/P_{{\rm i}}$ is explained in Section~\ref{Sec:Z_Pow_rat}. 
Projections of the spatial overlap during the first science campaign are discussed in Section~\ref{Sec:eta_sci}, while the total field overlap $\eta$ is discussed in Section~\ref{Sec:eta_total}. The methods used to optimize the spatial overlap are described in Section~\ref{Sec:eta_sci_opt}.


\subsubsection{Longitudinal Overlap}
\label{Sec:Long_over}
The longitudinal field overlap of the \ac{RC} $\eta_{\hat z}$, expresses the relative reduction in the resonant enhancement factor experienced by a light field which is detuned from the cavity resonance by some frequency $\Delta\nu$, the difference between the frequency of the \ac{HPL} and the nearest resonance of the \ac{RC}. It can be expressed as
\begin{equation}
\eta_{\hat z} = \frac{1-\sqrt{1-A}}{\left|1-\sqrt{1-A}\,e^{i2\pi \frac{\Delta\nu}{f_0}}\right|}
\simeq \frac{1}{\sqrt{1+\left(\frac{4\pi\Delta\nu}{A f_0}\right)^2}}.
\label{Eq:eta_z_approx_49}
\end{equation}
This can also be expressed in terms of the measured parameters $\tau$ and $\Delta\nu$ as
\begin{equation}
\eta_{\hat z} \simeq \frac{1}{\sqrt{1+(2\pi\tau\Delta\nu)^2}}.
\label{Eq:eta_z_meas}
\end{equation}
A laser that is detuned from the cavity resonance will also accumulate some additional phase offset during its transmission through the cavity. This phase offset $\theta_{\hat z}$, can be found using the equation 
\begin{align}
\theta_{\hat z} &= \arg\left[\frac{1-\sqrt{1-A}}{1-\sqrt{1-A}\,e^{i2\pi \frac{\Delta\nu}{f_0}}}\right]\\
&\simeq \arctan \left(2\pi\tau\Delta\nu\right).
\end{align}
This effect is important as it means a detuning of the \ac{HPL} from the resonance of the cavity can distort the open shutter phase. This is one of the reasons why the calibration array is expressed in terms of a complex variable as it will be able to compensate for these phase changes when the closed shutter data are normalized.

\begin{figure*}
    \centering
               \begin{subfigure}[b]{0.49\textwidth}
            \centering
    \includegraphics[width=\textwidth]{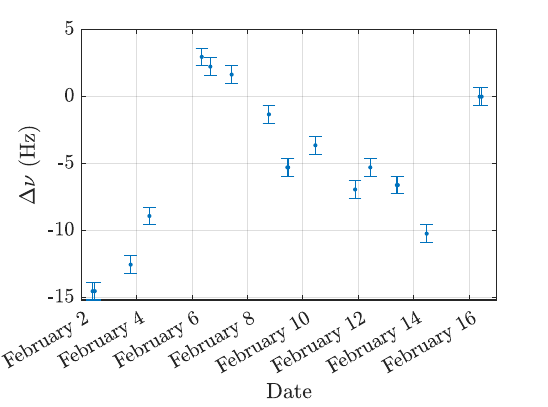}
    \caption{$\rm S_{_\perp}$
    \label{fig:nu_SC}}
    \end{subfigure}
           \begin{subfigure}[b]{0.49\textwidth}
            \centering
    \includegraphics[width=\textwidth]{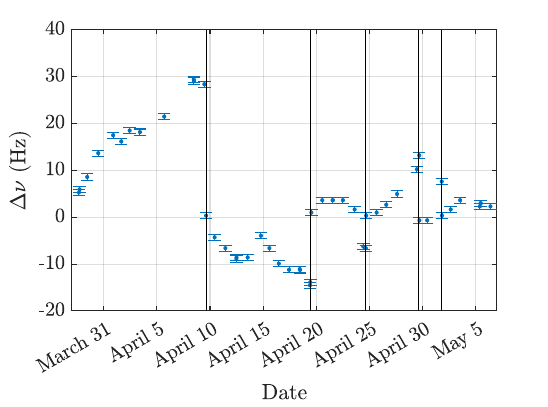}
    \caption{$\rm S_{_\parallel}$ 
    \label{fig:nu_PS}}
    \end{subfigure}
           \begin{subfigure}[b]{0.49\textwidth}
            \centering
    \includegraphics[width=\textwidth]{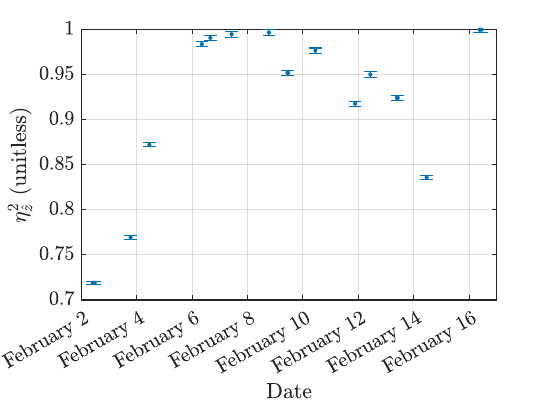}
    \caption{$\rm S_{_\perp}$
    \label{fig:eta_long_SC}}
    \end{subfigure}
           \begin{subfigure}[b]{0.49\textwidth}
            \centering
    \includegraphics[width=\textwidth]{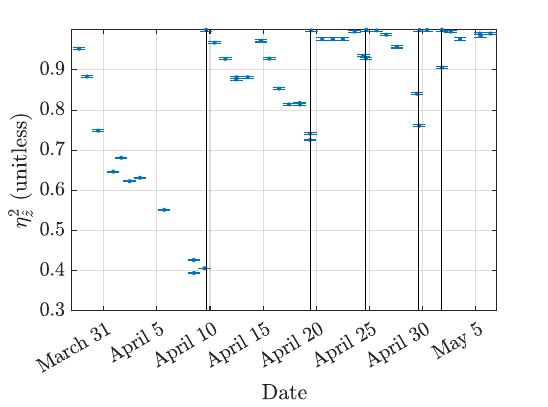}
    \caption{$\rm S_{_\parallel}$ 
    \label{fig:eta_long_PS}}
    \end{subfigure}
               \begin{subfigure}[b]{0.49\textwidth}
            \centering
    \includegraphics[width=\textwidth]{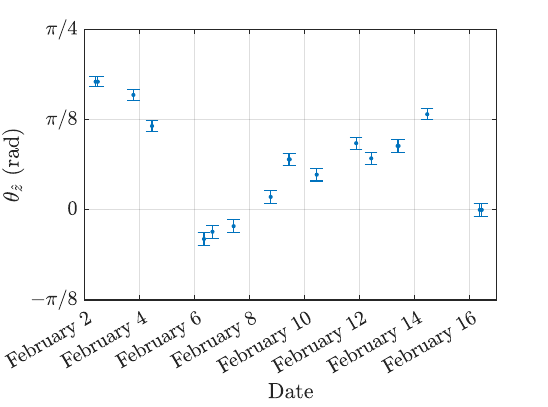}
    \caption{$\rm S_{_\perp}$
    \label{fig:phi_long_SC}}
    \end{subfigure}
           \begin{subfigure}[b]{0.49\textwidth}
            \centering
    \includegraphics[width=\textwidth]{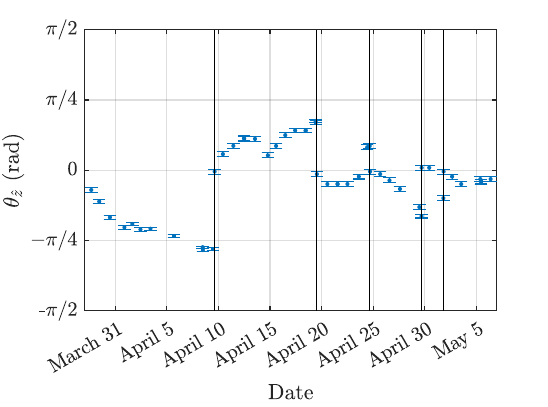}
    \caption{$\rm S_{_\parallel}$ 
    \label{fig:phi_long_PS}}
    \end{subfigure}
        \caption{The measured detuning $\Delta\nu$ of the \ac{HPL} with respect to the \ac{RC} resonance is shown for $\rm S_{_\perp}$ (a) and  $\rm S_{_\parallel}$ (b) with error bars based on the uncertainty of the measured \ac{FSR}. The corresponding longitudinal overlaps in terms of power $\eta_{\hat z}^2$ are shown in (c) and (d), while the phase shift induced by the detuning $\theta_{\hat z}$ is shown for each run in (e) and (f). The vertical black lines in (b), (d), and (f) show times during $\rm S_{_\parallel}$ in which the \ac{FSR} was manually reset to eliminate the detuning by adjusting the physical length of the cavity. No such adjustments were performed during  $\rm S_{_\perp}$.
    \label{fig:eta_long}}
\end{figure*}

The initial \ac{FSR} of the RC was used to establish the frequencies of the \acp{PLL} that control the frequency of the \ac{HPL} relative to the frequency of the \ac{AL} light transmitted by the \ac{RC}. At 33 \ac{FSR}s, this static frequency difference meant that the \ac{HPL} frequency was tuned to be on resonance with the \ac{RC} when the run was initially set up. However, as the length of the cavity drifted and the \ac{FSR} changed, the \ac{HPL} frequency was detuned from resonance and the longitudinal overlap $\eta_{\hat z}$, decreased. These effects also led to changes in $\theta_{\hat z}$, which was exhibited as a change in the phase of the open shutter measurements $C_{{\rm open}}$.

As Section~\ref{Sec:RC_B} described, it was possible to measure the \ac{FSR} with a precision a precision of the 0.02\,Hz. The frequency detuning $\Delta\nu$ of the \ac{HPL} with respect to the \ac{RC} could therefore be measured with a precision of 0.7\,Hz because the frequency spacing between \ac{AL} and the \ac{HPL} was 33~\ac{FSR}s.

Figures~\ref{fig:nu_SC} and \ref{fig:nu_PS} show $\Delta\nu$ for $\rm S_{_\perp}$ and $\rm S_{_\parallel}$, as a result of the cavity \ac{FSR} measurements. For $\rm S_{_\perp}$, the average absolute detuning of the \ac{HPL} from the \ac{RC} resonance (the mean value of $|\Delta\nu|$) estimated during the closed shutter periods was 5.4\,Hz with a minimum detuning of 0\,Hz, a maximum detuning of 14.5\,Hz, and  \ac{RMS} deviations from the mean of 4.0\,Hz. $\rm S_{_\parallel}$ showed an average absolute detuning of the \ac{HPL} frequency of 8.9\,Hz with a minimum detuning of 0.3\,Hz, a maximum detuning of 28.4\,Hz, an \ac{RMS} of 6.4\,Hz. The uncertainty in these measurements was 0.7\,Hz as explained in the previous paragraph.


Figures~\ref{fig:eta_long_SC} and \ref{fig:eta_long_PS} show the projected \ac{HPL}-\ac{RC} longitudinal power overlap $\eta_{\hat z}^2$, from measurements of the cavity \ac{FSR} and storage time. The black vertical lines in Figure~\ref{fig:eta_long_PS} show periods when the cavity length was adjusted to try to bring $\eta_{\hat z}^2$ back to one during $\rm S_{_\parallel}$. No such readjustments were performed on the length of the cavity during $\rm S_{_\perp}$. In these plots $\rm S_{_\perp}$ showed an estimated average value of $\eta_{\hat z}^2$ during the closed shutter periods of 92.8\% with a minimum of 71.9\%, a maximum value of 100\%, and \ac{RMS} fluctuations from the mean of 7.1\%. For $\rm S_{_\parallel}$ the average value of $\eta_{\hat z}^2$ was 85.5\%, while its minimum value was 40.6\%, its maximum value was 100\%, and its \ac{RMS} was 14.4\%. The uncertainty on these measurements were also primarily limited by the precision of the measurement of the \ac{RC} \ac{FSR} and was between 0.2\% and 0.3\%. 

The projections of $\theta_{\hat z}$ based on measurements of the cavity \ac{FSR} and storage time are shown in Figures~\ref{fig:phi_long_SC} and \ref{fig:phi_long_PS}. Again, the black lines show periods where the cavity length was readjusted. For $\rm S_{_\perp}$, the mean absolute deviation of the longitudinal phase offset from zero estimated during the closed shutter periods was 0.22\,rad and had a minimum value of 0\,rad with a maximum value of 0.56\,rad. The \ac{RMS} fluctuations from the mean in this case were 0.16\,rad. For $\rm S_{_\parallel}$ the average absolute deviation of $\theta_{\hat z}$ was 0.33\,rad with a minimum of 0.01\,rad, a maximum of 0.88\,rad and an \ac{RMS} of 0.22\,rad. Like the previous calculations, the uncertainty in these measurements was mainly due to the uncertainty of the \ac{FSR} measurement and was between 0.01 and 0.03\,rad.

It is interesting to compare the data shown in Figures~\ref{fig:eta_long_SC}, \ref{fig:eta_long_PS}, \ref{fig:phi_long_SC}, and \ref{fig:phi_long_PS}, to their corresponding plots of the open shutter data $C_{{\rm open}}$ measured at the science detectors in Figures~\ref{fig:Zopen_norm_SC}, \ref{fig:Zopen_phase_SC}, and \ref{fig:Zopen_norm} and \ref{fig:Zopen_phase}. Even though these two sets of plots were compiled from different sources of data, the relationship between $C_{{\rm open}}$, and $\eta_{\hat z}$ and $\theta_{\hat z}$ is obvious. This demonstrates how the detuning of the \ac{HPL} frequency from the \ac{RC} was perhaps the most significant source of drifts in $C_{{\rm open}}$.


\subsubsection{Polarization Overlap}
\label{Sec:eta_pol}

The polarization overlap $\eta_{\rm pol}$, quantifies the mismatch between the ideal polarization state of the \ac{HPL} for a given run ($\hat x$ for $\rm S_{_{\perp}}$ and $\hat y$ for $\rm S_{_\parallel}$) and the orientation of the magnetic field (defined to be in the $\hat y$ direction). 
The error in the orientation of the optical tables with respect to the magnetic fields is on the order of 100\,{\textmu}rad. The polarization state of the \ac{HPL} for $\rm S_{_\perp}$ was tuned for maximum transmission through LT1 (which was rigidly mounted to the \ac{COB}). For $\rm S_{_\parallel}$ an additional $\lambda/2$ wave-plate, whose orientation was precisely aligned, was also located on the \ac{COB} in the path to LT1. The polarization state of the \ac{AL} and \ac{LO} are set by a rotatable half wave-plate in \ac{NR}. The exact angle of the `roll' of the \ac{COB} with respect to the optical axis was not measured. 

A measurement of the polarization state of the \ac{HPL} before being injected to the magnet string via the vacuum window in \ac{NL} showed an azimuthal angle in the polarization state of  $7.9\pm0.1^\circ$ immediately following $\rm S_{_\perp}$. This would correspond to a power loss in the desired polarization state of less than 2\%. As this is well below the uncertainty from other sources, the polarization overlap was not considered in the analysis. For $\rm S_{_\parallel}$ the polarization overlap was also not considered as the azimuthal polarization angle was $89.6\pm0.1^\circ$ corresponding to a power loss in the desired polarization state of roughly 50\,ppm.

\subsubsection{Open Shutter Power Ratio}
\label{Sec:Z_Pow_rat}

The ratio of \ac{HPL} power leaving the cavity through $\rm RC1$ during the open shutter periods to the power injected to the production magnet string $P_{{\rm open}}/P_{{\rm i}}$, must be calculated to calibrate the field overlap measured at the science detector.  While the open shutter calibration array from Section~\ref{Sec:Op_cl_cal} $|C_{{\rm open}}|^2$, gives a measure of the ratio between the signal power and the injected power, it is still expressed in terms of the ratio of the squared voltage measured by the heterodyne detection system to \ac{HPL} power injected to the production magnet string (or units of $\rm V^2/W$). 
Nevertheless, $|C_{{\rm open}}|^2$ can be used with the expression
\begin{equation}
    \frac{P_{{\rm open}}}{P_{\rm i}} = 
    \frac{|C_{{\rm open}}|^2}{P_{\!_{\rm LO}} G_{\!_{\rm AC}}^2 T_{\!_{\rm sci}}  \eta_{\!_{xy,{\rm LO}}}^2},
\end{equation}
to calculate $P_{{\rm open}}/P_{{\rm i}}$ for a given open shutter period. Here $P_{\!_{\rm LO}}$ is the \ac{LO} (the laser used as the heterodyne local oscillator), $G_{\!_{\rm AC}}$ is the conversion gain of the science detection system including the full electronic chain at the first heterodyne demodulation frequency ($\sim$14\,MHz), $T_{\!_{\rm sci}}$ is the power transmissivity of the optical path from the \ac{RC} to the science \ac{PD}, and $\eta_{\!_{xy,{\rm LO}}}$ is the spatial overlap between the \ac{LO} and the \ac{RC} eigenmode.


The spatial overlap between the \ac{LO} and the \ac{RC} is analogous to the spatial overlap between the \ac{LO} and the \ac{HPL} in transmission of the cavity.  $\eta_{_{{xy,\rm LO}}}$ was measured by tuning the frequency of \ac{PLL}3 such that both the \ac{LO} and \ac{AL} were at a cavity resonance. Then the drop in the amplitude of the \ac{PLL}3 beatnote at the science PD in reflection of the cavity could be measured and compared to the amplitude of the \ac{PLL}3 beatnote in reflection of the cavity when the intracavity field was blocked (while also considering the reflectivity of the \ac{RC}).  We assume that the \ac{LO} and \ac{AL} share the same spatial eigenmode because they are initially interfered using a fiber beamsplitter (fused fiber coupler). The static reflected power from a cavity should adhere to the following equation.
\begin{equation}
    P_{\rm ref} = P_0 R_{\rm RC,NR}+P_1,
\end{equation}
Here, $P_{\rm ref}$ is the power reflected from the cavity, $P_0$ is the power in the spatial mode of the cavity, $R_{\rm RC,NR}$ is the cavity reflectivity in terms of power of a laser incident on the mirror in \ac{NR} (Equation~\ref{Eq:R_RC_NR}), and $P_1$ is the power outside the spatial mode of the cavity. With this, the spatial overlap can be defined using the expression
\begin{align}
    \frac{m_b}{m_a} & = \frac{P_0 R_{\rm RC,NR}+P_1}{P_0+P_1} \\
    & = \eta_{\!_{xy,{\rm LO}}}^2R_{\rm RC,NR} + 1 - \eta_{\!_{xy,{\rm LO}}}^2,
\end{align}
where $m_a$ represents the power measured when the cavity is blocked while $m_b$ is the value of the reflected power when the laser is on resonance. (This approximation assumes the reflectivity of RC1 is one for the measurements of $m_a$.) Knowledge of the \ac{PD} gain is unnecessary because the equation is expressed in terms of the ratio $m_b/m_a$, provided the same \ac{PD} is used to sense $m_a$ and $m_b$.
Thus, the spatial overlap in terms of power $\eta_{\!_{xy,{\rm LO}}}^2$, can  be calculated with the equation 
\begin{equation}
   \eta_{\!_{xy,{\rm LO}}}^2 =  \frac{\frac{m_b}{m_a} - 1}{R_{\rm RC,NR}-1}.
\end{equation}
The mean value of $\eta_{\!_{xy,{\rm LO}}}$ estimated during the closed shutter periods of $\rm S_{_\perp}$ was 94.5\%, with a range from 92.2\% to 95.7\% and \ac{RMS} from the mean of 0.9\%. During $\rm S_{_\parallel}$ the mean was 93.8\%, with a minimum value of 87.2\%, a maximum value of 96.4\% and an \ac{RMS} of 2.1\%. In both runs the uncertainty on the measurements of $\eta_{\!_{xy,{\rm LO}}}$ was between 0.2\% and 0.3\%, and was driven primarily by the uncertainty in the measurements of the beatnote amplitude.

$P_{\!_{\rm LO}}$ is also an important parameter as it directly affects the amplitude of the heterodyne signal at the science detector. It was measured by calibrating the static voltage output by the science detector by the conversion gain and subtracting the power of the \ac{AL} that was also present at this detector.  
The change in the power during the runs is believed to be related to changes in the alignment of each laser to its corresponding optical fiber, as well as changes in $T_{\rm sci}$. For $\rm S_{_\perp}$ the mean value of $P_{\!_{\rm LO}}$ during the closed shutter runs was 310\,{\textmu}W with a range of 296\,{\textmu}W to 322\,{\textmu}W and \ac{RMS} fluctuations from the mean of 10\,{\textmu}W. During $\rm S_{_\parallel}$ closed shutter periods $P_{\!_{\rm LO}}$ is estimated to have a mean value of 385\,{\textmu}W, with a minimum of 354\,{\textmu}W, a maximum of 408\,{\textmu}W, and an \ac{RMS} of 15\,{\textmu}W. The uncertainty in the measurements of $P_{\!_{\rm LO}}$ during both runs was between 10\,{\textmu}W and 12\,{\textmu}W. Although both the uncertainty in the measurements of the DC voltage from the science detector, as well as the uncertainty in the measurement of the reflectivity of the \ac{RC}, played a role in the overall uncertainty on $P_{\!_{\rm LO}}$, the dominant source of uncertainty in these measurements was the uncertainty in the measurements of the DC gain of the science detector electronic chain. This is discussed in more detail in the following section.


For the science detection system, both the DC conversion gain and the gain of the system at the heterodyne frequency, $G_{\rm AC}$, had to be measured to calculate $P_{{\rm open}}$. The DC conversion gain of the science \ac{PD} was measured by varying the power of a single laser incident on the science PD, measuring the power with a power meter and recording the voltage level output by the detector. The slope of a linear fit to the voltage as a function of the incident power gave an average value of 3.9$\pm$0.1\,V/mW over several measurements made at different times. The uncertainty is believed to be related to the variance in the measured gain as a function of the position of the beam on the active area of the \ac{PD}, as well as the uncertainty in the power meter calibration.

The conversion gain of the science detector electronic chain at the heterodyne frequency $G_{\rm AC}$, was also measured under conditions meant to resemble the actual conditions of the science campaign. The \ac{LO} and \ac{AL} were both incident on the curved \ac{RC} mirror in \ac{NR} with \ac{PLL}3 locked at its operating frequency. To make the measurements independent of the parameters of the \ac{RC} the circulating field of the cavity was blocked between \ac{NR} and \ac{CH} and the \ac{AL} power was reduced to 25.9\,{\textmu}W, a value near the mean of what was seen during the science campaign when the \ac{AL} was stabilized to a resonance of the \ac{RC}. A signal was sent to a fiber \ac{EOM} to generate phase modulation sidebands on \ac{AL} at the heterodyne frequency with roughly 0.01\% of its carrier power. Therefore, like $\rm S_{_\perp}$ and $\rm S_{_\parallel}$, the light on the science \ac{PD} had a large interference beatnote at the \ac{PLL}3 frequency and a low amplitude signal at the heterodyne frequency. With this setup, the conversion gain of the full detection chain at $14.603$\,MHz, $G_{\rm AC}$, could be measured. This was done using the ratio of the sideband to carrier beatnote amplitude measured at the \ac{PLL}3 PD (see Figure~\ref{fig:Opt_sys}) to estimate the relative power in the sideband. A previous measurement of the frequency-dependent gain of the \ac{PLL}3 PD at these two frequencies was used to correct for the difference in the gain of the \ac{PD} at 14.603\,MHz and 54.95\,MHz. The power of the \ac{LO} was then measured at the science \ac{PD} along with the amplitude of the signal at the heterodyne frequency. Thus the sideband power (with the sideband at the heterodyne frequency) could be calculated along with the conversion gain of the detection system, independent of the measurement of the DC conversion gain of the photodetector.  These measurements resulted in a conversion gain of  $2.9\pm0.1$\,V/mW. This is slightly lower than the DC conversion gain as a result of the electronic transfer function of the photodetector. Like the measurements of the DC conversion gain, the primary uncertainty in these measurements was related to the dependence of the gain on the position of the beam on the active area of the \ac{PD}, as well as the calibration of the power meter.

To check for saturation effects, the \ac{LO} power at the science PD was varied in power between 172\,{\textmu}W and 1\,mW and measured, holding the \ac{AL} power at 25.9\,{\textmu}W, while measuring the beatnote amplitudes at 54.95\,MHz and 14.603\,MHz at both the \ac{PLL}3 PD and the science PD. Between 172\,{\textmu}W and 680\,{\textmu}W no evidence of saturation was seen, however the saturation effects became clear once the \ac{LO} power reached 1\,mW. However, the power of \ac{LO} was at most 400\,{\textmu}W for both runs and therefore the science detection chain is not believed to have ever experienced saturation during the closed shutter periods.

$T_{\!_{\rm sci}}$ is the total percentage of power exiting the \ac{RC} that reaches the science photodetector. It was measured intermittently over the course of the science campaign by measuring the power at various points in the optical system in the \ac{NR} cleanroom. These points included before and after the Faraday isolator, at RC1, directly after the Faraday isolator at the detection port with the cavity circulating field blocked, and at the science \ac{PD}. 

To find $T_{\!_{\rm sci}}$, the ratio of power incident on the \ac{RC} cavity mirror versus the power incident on the science detector was calculated. This was done several times before, during, and after the first science campaign and gave values between 80\% and 90\%. 
The value used to calculate the field overlap was $85\%\pm5\%$. The losses were mostly due to optical pickoffs used to sample the light reflected from the cavity for the \ac{PDH} frequency stabilization system as well as other diagnostic purposes. Changes in the total reflectivity of the input optics that appear to be humidity dependent were also observed, along with clipping losses on the free aperture of the input optics after the beam expanding telescope in \ac{NR}. Both of these effects played a role in the uncertainty of the results of $T_{\!_{\rm sci}}$.

The open shutter calibration series for the veto detection system $C^{\rm veto}_{{\rm open}}$, is defined in the same way as for the science detector and is also in units of $\rm V^2/W$. The power ratio $P^{\rm veto}_{{\rm open}}/P_{\rm i}$ for the veto detector can be expressed as
\begin{equation}
   \frac{P^{\rm veto}_{{\rm open}}}{P_{\rm i}} = \frac{|C^{\rm veto}_{{\rm open}}|^2}{P_{\!_{\rm AL,veto}} G_{\!_{\rm AC,veto}}^2 T_{\!_{\rm veto}} }.
    \label{Eq:P_open_veto}
\end{equation}
Here $P_{\!_{\rm AL,veto}}$ is the \ac{AL} (the laser used as the heterodyne local oscillator for the veto detector) power measured at the veto detector in units of W, $G_{\!_{\rm AC,veto}}$ is the conversion gain of the full veto detection system electronic chain at the veto heterodyne demodulation frequency ($\simeq$40.35\,MHz) in V/W, $T_{\!_{\rm veto}}$ is the transmissivity of the optical path from mirror RC2 to the veto \ac{PD}. There is no need for a spatial overlap term between \ac{AL} and the \ac{RC} in this case because \ac{AL} is transmitted by the cavity and shares its spatial eigenmode.


The AC gain of the veto detector chain was measured by setting the \ac{PLL}3 frequency to the veto heterodyne frequency of $f_{\rm veto}=40.34920$\,MHz such that both \ac{AL} and \ac{LO} were resonant with the cavity. Then the power of \ac{LO} was varied and the peak to peak amplitude of the signal at $f_{\rm veto}$ output by the veto detection chain was measured. At each point the power at the veto detector was measured with a power meter with and without the \ac{LO} resonant (\ac{AL} was always resonant in these measurements). With this the beatnote amplitude in power could be projected because in transmission of the \ac{RC}, \ac{AL} and \ac{LO} are both entirely in the cavity eigenmode. Using this method a gain of $G_{\!_{\rm AC,veto}}=25\pm1$\,V/mW was measured. In the measurement the absolute accuracy of the power meter was the limiting factor in the uncertainty. 


The power measured at the veto photodetector $P_{\rm AL,veto}$ is an important parameter as it plays a direct role in the calibration of $C^{\rm veto}_{{\rm open}}$ in terms of optical power. It can also be used to estimate the optical losses in the path from the \ac{RC} output mirror to the veto \ac{PD} discussed in the following section. $P_{\rm AL,veto} $ can be calculated by dividing the DC voltage output by the veto \ac{PD} $V_{\rm DC,veto}$, by its DC conversion gain $G_{\rm DC, veto}$:
\begin{equation}
  P_{\rm AL,veto} = \frac{V_{\rm DC, veto}}{G_{\rm DC, veto}}  
  \label{Eq:P_phiL_veto}.
\end{equation}
The DC conversion gain $G_{\!_{\rm DC,veto}}$ of the veto detection system was measured by varying the power incident on the veto detector, and measuring the voltage at the monitoring point used during the maintenance periods. This gave a value of $G_{\!_{\rm DC,veto}}=86\pm1$\,V/mW. The error in this measurement is due to the variance in the results over multiple measurements. The primary reason for the difference between the AC and DC gain, was the presence of a three-way splitter in the AC path that allowed the veto \ac{PD} to also be used as the sensor for \ac{PLL}1.



One complication in the measurement of $P^{\rm veto}_{{\rm open}}/P_{\rm i}$ arises from the fact that the transmissivity of the optics on the path from the output mirror of the \ac{RC} to the veto \ac{PD}, $T_{\rm veto}$, can not be directly measured, as there are a significant number of optics in the portion of this path inside the central vacuum chamber. $T_{\rm veto}$ can be calculated using the ratio of the power of the \ac{AL} measured at the veto \ac{PD} to the expected power of \ac{AL} given its input power and coupling efficiency to the \ac{RC}, as well as the transmissivity of the \ac{RC}. 
This is measured by dividing the drop in the DC voltage output at the science \ac{PD} $\Delta V_{\rm AL,sci}$, when \ac{AL} is frequency stabilized to the cavity, by the DC gain of the science \ac{PD} detection chain ($G_{\rm DC, sci}$) and by $T_{\rm sci}$ as expressed by 
\begin{equation}
    \Delta P_{\rm AL,NR} =\frac{\Delta V_{\rm AL,sci}}{T_{\rm sci} G_{\rm DC, sci}}.
\label{Eq:Del_P_phiL}
\end{equation}
As the following shows, this definition of $\Delta P_{\rm AL,NR}$ is useful as it already takes the spatial overlap into account. Assuming minimal absorption and scattering losses in the substrate, the light incident on the cavity can either be reflected by the cavity, scattered out of the eigenmode or absorbed while circulating in the cavity, or transmitted by the cavity. Equation~\ref{Eq:Del_P_phiL} represents the power in the latter two categories, light leaving the cavity eigenmode due to excess optical losses inside the cavity, or light transmitted by the cavity. 

Because the total round trip attenuation $A$ is measured via the cavity storage time $\tau$ during the maintenance periods and the transmissivity of the cavity mirrors were measured before the science runs took place, the power directly in transmission of the \ac{RC} ($P_{\rm AL,CH}$) can be estimated from
\begin{align}
P_{\rm AL,CH} & = \frac{T_{\!_{\rm RC2}}}{A-T_{\!_{\rm RC1}}} \Delta P_{\rm AL,NR} \\
 & = \frac{T_{\!_{\rm RC2}}}{ \frac{2}{f_0 \tau} - T_{\!_{\rm RC1}}} \frac{\Delta V_{\rm AL}}{T_{\rm sci} G_{\rm DC, sci}}.
\label{Eq:P_phiLCH}
\end{align}
It should be noted that this equation can also be derived using the expressions for the spatial overlap between \ac{AL} the \ac{RC} along with the \ac{RC} transmissivity and that Equation~\ref{Eq:P_phiLCH} is the simplest expression for the \ac{AL} power in transmission of the \ac{RC}.

$T_{\rm veto}$ can then be derived by calculating the ratio between the power measured at the veto \ac{PD} and the power transmitted by the cavity in \ac{CH} that is predicted by the Equation~\ref{Eq:P_phiLCH}. This gives 
\begin{align}
 T_{\rm veto} & = \frac{P_{\rm AL,veto}}{P_{\rm AL,CH}} \\
  & = T_{\rm sci}\frac{V_{\rm DC, veto}}{ \Delta V_{\rm AL}} \frac{ G_{\rm DC, sci} }{G_{\rm DC, veto} }  \frac{\left( \frac{2}{f_0 \tau} - T_{\!_{\rm RC1}} \right)}{T_{\!_{\rm RC2}}}.
 \label{Eq:T_veto}
\end{align}
This equation shows that $T_{\rm veto}$ can be expressed as the product of $T_{\rm sci}$ and several ratios. The ratio of the DC voltage measured at the veto \ac{PD} when \ac{AL} is on resonance to the drop in DC voltage measured at the science \ac{PD}, the ratio of the science \ac{PD} DC conversion gain to the veto \ac{PD} conversion gain, and finally the ratio of the losses minus the transmissivity of the \ac{RC} input mirror to the transmissivity of the \ac{RC} output mirror.

For the closed shutter periods of $\rm S_{_\perp}$, the average value of $T_{\rm veto}$ was estimated to be 68\% with a minimum value of 65\%, a maximum of 71\%, \ac{RMS} fluctuations from the mean of 2\%, and an uncertainty of 9\%. During $\rm S_{_\parallel}$, the average value of $T_{\rm veto}$ estimated for the closed shutter sections was 65\% with a range from 55\% to 70\%, an \ac{RMS} of 4\%, and an uncertainty of 9\%. In both runs, the primary source of the uncertainty was the uncertainty in the measurement of $T_{\!_{\rm RC2}}$.

After the conclusion of the first science campaign the power of the \ac{AL} in transmission of the \ac{RC} was measured immediately after the vacuum flange in the veto detector path to be 9.0\,{\textmu}W, while the power at the veto \ac{PD} was measured to be 7.4\,{\textmu}W, indicating that 82\% of the power leaving the \ac{COB} at the veto port reaches the \ac{PD}. These losses are mostly due to a diagnostic pick-off that sends some light to a camera to examine the position and shape of the \ac{RC} eigenmode. The Brewster window at this port is also oriented such that it has an angle with respect to the vertical direction such that it has a transmissivity estimated to be 88\% for horizontally polarized light. In addition to this, the mirror before the quadrant \ac{PD} on the \ac{COB} was measured to have a reflectivity of 95\%. Therefore the path that the \ac{RC} transmitted light takes to the veto \ac{PD} is expected to have a transmissivity of 68\%. While this agrees well with the measured values from science runs, in principle it should not be possible to measure values $T_{\rm veto}$ higher than the measured transmissivity of the optics. This could indicate a source of systematic error in the measurements of the transmissivities of the optics or $T_{\rm veto}$.
Nevertheless, the transmissivity measured after the first science campaign is well within the error bars of the $T_{\rm veto}$ values measured during the science run.

\subsubsection{Spatial Overlap Results}
\label{Sec:eta_sci}

The spatial coupling of the \ac{HPL} to the \ac{RC} measured during the open shutter periods should be representative of the coupling of the \ac{BSM} field as long as there are no significant effects that can alter the amplitude or alignment of the \ac{HPL} as it travels between the production and regeneration areas. The spatial component of the field overlap between the \ac{HPL} and the \ac{RC} measured by the science detector was expressed earlier as Equation~\ref{eq:eta_xy} in Section~\ref{Sec:ALPSII_des}. The square of the spatial field overlap, $\eta_{{xy}}^2$, represents the spatial overlap in terms of power and can be expressed as a function of measured parameters as:
\begin{equation}
    \eta_{{xy}}^2 =
    \frac{|C_{{\rm open}}|^2}{T_{\!_{\rm COB}} T_{\!_{\rm RC}} P_{\!_{\rm LO}} G_{\!_{\rm AC}}^2 T_{\!_{\rm sci}}  \eta_{\!_{xy,{\rm LO}}}^2}.
    \label{eq:eta_spa}
\end{equation}
Here, $T_{\rm RC}$ is used instead of $T_{\rm RC,max}$ because $\eta_{{xy}}^2$ is defined as the spatial overlap rather than the total field overlap.

\begin{figure*}[t]
    \centering
               \begin{subfigure}[b]{0.49\textwidth}
            \centering
    \includegraphics[width=\textwidth]{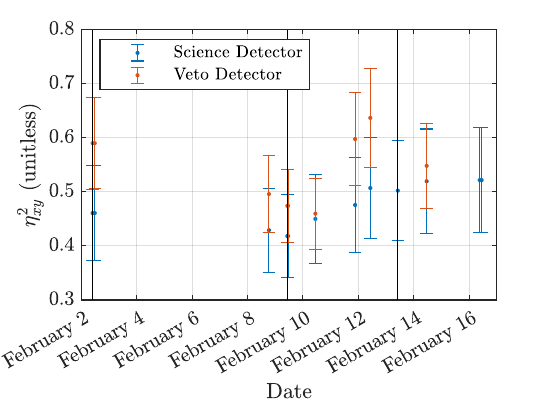}
    \caption{$\rm S_{_\perp}$
    \label{fig:eta_sp_LO_SC}}
    \end{subfigure}
               \begin{subfigure}[b]{0.49\textwidth}
            \centering
    \includegraphics[width=\textwidth]{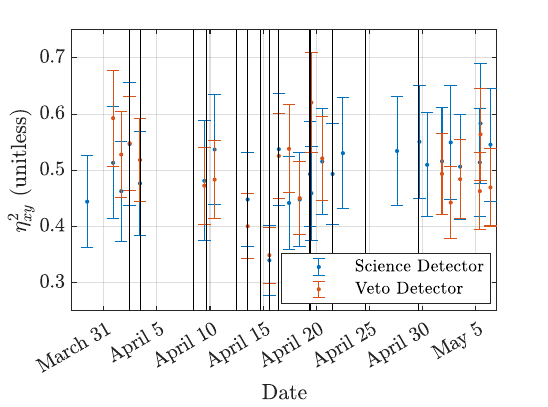}
    \caption{$\rm S_{_\parallel}$
    \label{fig:eta_sp_LO_PS}}
    \end{subfigure}
                   \begin{subfigure}[b]{0.49\textwidth}
            \centering
    \includegraphics[width=\textwidth]{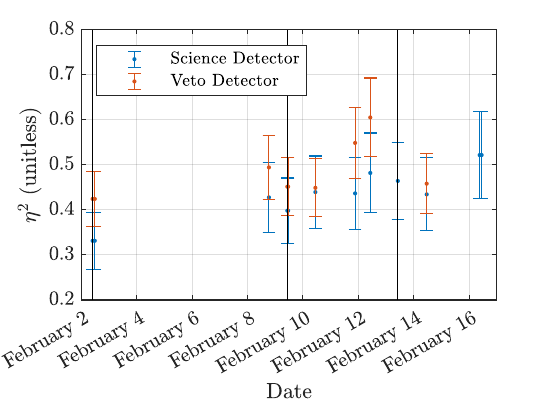}
    \caption{$\rm S_{_\perp}$
    \label{fig:eta_tot_LO_SC}}
    \end{subfigure}
               \begin{subfigure}[b]{0.49\textwidth}
            \centering
    \includegraphics[width=\textwidth]{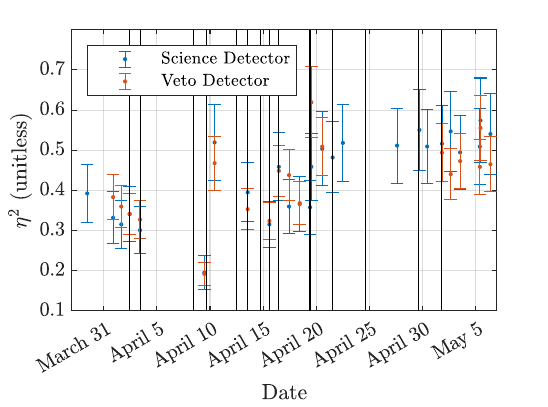}
    \caption{$\rm S_{_\parallel}$
    \label{fig:eta_tot_LO_PS}}
    \end{subfigure}
        \caption{The field overlap between the \ac{HPL} and \ac{RC} measured at the science detector and veto detectors respectively, from data taken during the maintenance periods of $\rm S_{_\perp}$ and $\rm S_{_\parallel}$. The squared spatial overlap $\eta_{xy}^2$ is show in (a) for $\rm S_{_\perp}$ and (b) for $\rm S_{_\parallel}$. The squared total field overlap $\eta^2$ is show in (c) for $\rm S_{_\perp}$ and (d) for $\rm S_{_\parallel}$. In all four plots the black lines indicate any times at which the cavity alignment was actively changed, either to optimize the spatial field overlap or the resonant enhancement, while in (d) they also indicate times in which the length of the cavity was change to optimize the spectral component of the field overlap. In the case of (c) the length of the cavity was never changed during $\rm S_{_\perp}$.
    \label{fig:eta_sp_t_LO_PS}}
\end{figure*}

The results of measuring $\eta_{{xy}}^2$ are shown in Figures~\ref{fig:eta_sp_LO_SC} and \ref{fig:eta_sp_LO_PS} respectively for $\rm S_{_\perp}$ and $\rm S_{_\parallel}$. For $\rm S_{_\perp}$, the average spatial overlap between the \ac{HPL} and the \ac{RC} during the closed shutter periods was estimated to be 47\% with a range of 41\% to 54\% and \ac{RMS} fluctuations from the average of 4\%. The uncertainty in these measurements varied from 7\% to 10\%, with an average value of 9\%, and was due mostly to the uncertainties in the measured transmissivities of the on-axis \ac{COB} mirrors and the flat cavity mirror mounted to the \ac{COB} (RC2), with some additional contribution due to the uncertainty on the calibration series. During the closed shutter periods of $\rm S_{_\parallel}$ the mean value of $\eta_{{xy}}^2$  was estimated to be 48\% with a minimum of 42\%, a maximum of 60\%, and an \ac{RMS} of 6\%. The uncertainty of these measurements was again driven by the uncertainties on the mirror transmissivities like in $\rm S_{_\perp}$, along with some additional uncertainty due to the calibration series. For $\rm S_{_\parallel}$ the uncertainty on $\eta_{xy}^2$ varied between 6\% and 11\% with an average value of 9\%.

To calculate the \ac{HPL}-\ac{RC} spatial power overlap measured by the veto detection system $\eta_{xy,{\rm veto}}^2$, the open shutter power defined in Section~\ref{Sec:Z_Pow_rat} is divided by the expected power of the \ac{HPL} passing through the production magnet string and the \ac{COB} optics, along with the reflectivity of the cavity from the direction of \ac{CH}:
\begin{equation}
    \eta_{xy,{\rm veto}}^2 = \frac{P^{\rm veto}_{{\rm open}}}{R_{\rm RC,CH} T_{_{\rm COB}}P_{\rm i}}.
\end{equation}
The spatial overlap measured by the veto detection system can be expressed in terms of measured parameters as: 
\begin{widetext}
\begin{equation}
     \eta_{xy,{\rm veto}}^2 \simeq  \frac{|C^{\rm veto}_{{\rm open}}|^2 \Delta V_{\rm AL} G^2_{\rm DC, veto} T_{\!_{\rm RC2}}   \left( 1  + (2\pi\tau\Delta\nu)^2 \right) }{G_{\!_{\rm AC,veto}}^2 T_{\rm sci}     \left( (1 - T_{\!_{\rm RC2}} f_0\tau)^2 + (2\pi\tau\Delta\nu)^2 \right) T_{_{\rm COB}} V^2_{\rm DC, veto}  G_{\rm DC, sci} \left( \frac{2}{f_0 \tau} - T_{\!_{\rm RC1}} \right) }.
\label{Eq:eta_xy_veto}
\end{equation}
\end{widetext}
Equations~\ref{Eq:R_RC_CH}, \ref{Eq:eta_z_meas}, \ref{Eq:P_open_veto}, \ref{Eq:P_phiL_veto}, and \ref{Eq:T_veto}  are used to derive these expressions. Figures~\ref{fig:eta_sp_LO_SC} and \ref{fig:eta_sp_LO_PS} show the \ac{HPL}-\ac{RC} spatial overlap measured at the veto detector during the maintenance periods of the science runs as the orange data points and the results of $\eta_{xy,{\rm veto}}^2$ agree very well witht he data acquire at the science detector. During the closed shutter periods of $\rm S_{_\perp}$, the estimated spatial overlap had an average value of 53\% with a range of 40\% to 63\%, \ac{RMS} fluctuations from the mean of 6\%, and an uncertainty of between 6\% and 10\% with an average uncertainty of 8\%. For $\rm S_{_\parallel}$ the average value was 49\% with a minimum of 34\%, a maximum value of 66\%, an \ac{RMS} of 8\%, and an uncertainty between 5\% and 10\% with an average value of 7\%. For both runs the uncertainty was primarily limited by the uncertainty on the measured transmissivities of the \ac{COB} optics and mirror RC2. While RC2 does not play a direct role in the \ac{HPL} power measured at the veto \ac{PD}, its uncertainty does effect the measurement of $T_{\rm veto}$.

\subsubsection{Field Overlap Results}
\label{Sec:eta_total}

The total power overlap between the \ac{HPL} and \ac{RC} is the product of the spatial overlap and the longitudinal overlap, or 
\begin{align}
\eta^2 & = \eta_{{xy}}^2 \eta_{\hat z}^2 \\
& = \frac{|C_{{\rm open}}|^2}{T_{_{\rm COB}} T_{\rm RC,max} P_{\!_{\rm LO}} G_{\!_{\rm AC}}^2 T_{\!_{\rm sci}}  \eta_{\!_{xy,{\rm LO}}}^2}.
    \label{eq:eta_T}
\end{align}
In this case $T_{\rm RC,max}$ is used because the effects of the longitudinal overlap must be included in this calculation. 

Measurements at the science detector of the total overlap $\eta^2$ are shown for $\rm S_{_\perp}$ and $\rm S_{_\parallel}$ in Figures~\ref{fig:eta_tot_LO_SC} and \ref{fig:eta_tot_LO_PS}. During the closed shutter periods of $\rm S_{_\perp}$, the mean value of the total overlap was estimated to be 43\% with a range from 33\% to 51\%, \ac{RMS} fluctuations from the mean  of 4\%, with an uncertainty that varied between 6\% and 10\% and had an average value of 8\%. These measurements were limited by the same sources of uncertainty as $\eta_{{xy}}^2$. For $\rm S_{_\parallel}$, the average estimated value of $\eta^2$ at the science detector during the closed shutter periods was 41\% with a range from 21\% to 58\%, an \ac{RMS} of 9\%, and an uncertainty between 4\% and 10\% with an average value of 8\%. The uncertainty on these measurements was also due to the same effects as were just mentioned for $\rm S_{_\perp}$.

Using Equation~\ref{Eq:eta_xy_veto} to express $\eta_{xy,{\rm veto}}^2$, the total field overlap between the \ac{HPL} and the \ac{RC} can be expressed in terms of measured parameters as
\begin{equation}
%
 \eta_{\rm veto}^2\simeq\frac{\eta_{xy,{\rm veto}}^2 }{ \eta_{\hat z}^2} \simeq \frac{\eta_{xy,{\rm veto}}^2 }{ 1  + (2\pi\tau\Delta\nu)^2 }.
\end{equation}
The total overlap between the \ac{HPL} and \ac{RC} measured at the veto detector is shown in Figures~\ref{fig:eta_tot_LO_SC} and \ref{fig:eta_tot_LO_PS}  for $\rm S_{_\perp}$ and $\rm S_{_\parallel}$ respectively. During the closed shutter periods of $\rm S_{_\perp}$, the average value estimated for $\eta_{\rm veto}^2$ was 48\%. The minimum value over this time was 39\%, while the maximum value was 60\%, the \ac{RMS} fluctuations from the mean  were 5\%, and the uncertainty varied between 6\% and 10\% with an average value of 8\%. For $\rm S_{_\parallel}$ the average value was 41\%, and the data varied over a range from 20\% to 66\% with an \ac{RMS} of 8\% and an uncertainty between 3\% and 10\% with an average value of 6\%. In general, the spatial overlap and total overlap measured at the veto and science detectors showed good agreement and the differences between the results were on the order of 10\% or less with the exception of a few outliers.

Finally, it should be noted that in both runs, like $\eta_{xy,{\rm veto}}^2$, $\eta_{{\rm veto}}^2$ agree very with the results of the spatial overlap and total overlap measured at the science detector and discussed in Section~\ref{Sec:eta_sci}.

\subsubsection{Optimizing the Spatial Overlap}

\label{Sec:eta_sci_opt}

The spatial coupling of the \ac{HPL} to the \ac{RC} can be affected by the relative position, alignment, and size of the \ac{HPL} and \ac{RC} eigenmodes. The initial mode-matching of the \ac{HPL} was performed by first building a 246\,m cavity with the curved mirrors of the \ac{PC} and \ac{RC} located on the optical tables in the \ac{NL} and \ac{NR} cleanrooms respectively (see Figure~\ref{fig:Opt_sys} for a diagram of the cleanrooms). Then the \ac{AL} was coupled to the cavity from \ac{NR} and the light in tranmission of the cavity in \ac{NL} served as a reference for the eigenmode of the cavity.
This light was then mode matched, using the \ac{HPL} input optics, to a triangular cavity which served as an input mode cleaner for the \ac{HPL} \cite{Willke:98}. The spatial mode of the 246\,m cavity should be representative of the \ac{RC} eigenmode because it had a nearly confocal geometry with twice the length of the \ac{RC}. 

A spatial overlap of roughly 90\% was achieved. The input mode cleaner was then left in place to serve as a reference for the mode of the input laser. The higher order mode content related to the waist size and position of the \ac{HPL} with respect to the \ac{RC} should have therefore remained optimized, although drifts in the alignment of the mirrors after the input mode cleaner could still cause a decoupling of the \ac{HPL} to the \ac{RC}. Therefore during the maintenance periods the \ac{HPL} alignment was corrected using the following procedure. After the eigenmode position of the \ac{RC} was optimized using the storage time, the \ac{RL} was manually realigned to maximize the \ac{PLL}1 beatnote amplitude. After this  the \ac{HPL} input optics were tuned to optimize the beatnote amplitude between \ac{HPL} and \ac{RL} at the \ac{MZ} port. This method was limited in its accuracy as the maximum power spatial overlap achieved between \ac{RL} and the \ac{AL} the light transmitted by the RC was on the order of 50\% due to the astigmatism of the \ac{RL} beam generated by a pair of curved mirrors in its mode matching telescope. Mirrors were used instead of lenses to avoid creating a direct path for \ac{HPL} stray light to couple to the \ac{RC} via backreflections from these lenses. 






\section{Conclusion}
\label{Sec:Conc}

Optical light-shining-through-a-wall experiments are a powerful tool to search for pseudo-Goldstone bosons that could reveal a new energy scale between the electroweak and Planck scales. With optical cavities before and after the wall these experiments offer an unprecedented photon flux along with a long interaction length and boosted signal strength. The \ac{ALPS}\,II first science campaign took place from February to May of 2024 with runs using the light-shining-through-a-wall technique to search for beyond the standard model bosonic fields. These searches were performed with a regeneration cavity resonantly enhancing the electromagnetic field potentially regenerated behind the wall, the first time this has been done with optical cavities. This campaign consisted of two science runs, one with the polarization states of the lasers perpendicular to the direction of the magnetic field ($\rm S_{_\perp}$) and one with the polarization states aligned parallel to the magnetic field ($\rm S_{_\parallel}$). 
The signals measured by the heterodyne detection system over the duration of the runs were coherently summed, allowing the experiment to have a frequency resolution on the order of {\textmu}Hz. Therefore, background signals that are not at the exact expected frequency can be distinguished from signals that are the result of the coupling between electromagnetic and beyond the standard model fields.

The calibration of the data was performed using measurements of the transmission of the high-power input laser through the system when a shutter in the wall was open. This allowed the laser light to directly couple to the regeneration cavity after passing through the optics on the central optical bench. The signal rates measured when the shutter was closed were then normalized by the measured open shutter transmission. 
The use of this ratio cancels most experimental parameters and systematic uncertainties except the transmissivity of the central optical bench optics.   
The relevant optics showed a combined transmissivity of  $(9.7\pm1.2)\times10^{-23}$, 
resulting in less than ten attowatts reaching the detection system when the shutter is open. Such low powers highlight the challenges faced when using this technique and make it clear that the operation and calibration of such a system is in itself an accomplishment. Via this procedure, it is possible to express the results in terms of $\mathcal{P}_{\gamma\leftrightarrow \phi}$, the rate at which energy converts between electromagnetic and beyond the standard model fields after a single pass through one of the magnet strings.

The results at the expected signal frequency did show an excess relative to the technical noise of the system, however, when this ratio was evaluated at alternative frequencies, a broad peak in the spectrum was evident. This indicated the presence of a background believed to be caused by stray light coupling from the high-power laser to the detection system. This was corroborated by an analysis of the time evolution of the closed shutter signals.

For each run, a statistical analysis was performed on the background signals within 100\,{\textmu}Hz of the signal frequency giving an expectation value for the background and its probability distribution. Based on this, the values measured at the signal frequency were well below $5\,\sigma$ detection thresholds. 
We therefore set limits that exclude conversion rates above a significance of  $5\,\sigma$ with a confidence level of 95\%. The resulting exclusion limits are $\mathcal{P}_{\gamma\leftrightarrow \phi}<2.5\times10^{-13}$  for $\rm S_{_\perp}$ and  $\mathcal{P}_{\gamma\leftrightarrow \phi}<1.7\times10^{-13}$ for $\rm S_{_\parallel}$. 

These results compare favorably to previous generations of light-shining-through-a-wall experiments 
as the limits on 
$\mathcal{P}_{\gamma\leftrightarrow \phi}$ for
scalar and pseudoscalar bosons surpass OSQAR~\cite{PhysRevD.92.092002} by a factor of $18$ and $31$ respectively, 
despite the fact that much stricter exclusion criteria were applied in this analysis.  
We would also like to emphasize that the sensitivity in terms of $\mathcal{P}_{\gamma\leftrightarrow \phi}$ is independent of the properties of the magnet string. It is therefore a useful metric to compare the optical systems of different light-shining-through-a-wall experiments. 
The exclusion limits on $\mathcal{P}_{\gamma\leftrightarrow \phi}$ reported here are interpreted in terms of the coupling between electromagnetic and pseudoscalar, scalar, vector, and tensor bosons in Ref.\,\cite{ALPSII_science}. Those results show that \ac{ALPS}\,II improved the limits on the electromagnetic coupling of scalar bosons by a factor of $18$ and  and pseudoscalar bosons by a factor of $23$ with respect to the OSQAR results. In both limits, the improvement in sensitivity due to the longer magnet strings used by \ac{ALPS}\,II and its more sophisticated optical system is roughly the same.

The optical system was constantly monitored during the first science campaign. Although parameters such as the resonant enhancement of the regeneration cavity were not directly used in the calibration of the science data, they provided useful information on the status of the optical system. With a world record storage time of 7.17\,ms \cite{kozlowski2024designperformancealpsii}, the regeneration cavity sustained a resonant enhancement factor of roughly 7000 during the science campaign. 



In the future, we plan to implement an optical cavity before the wall, further reduce backgrounds due to stray light and technical noise, boost the resonant enhancement factor of the regeneration cavity and increase its overlap with the beyond the standard model field. With these upgrades, the signal to noise ratio with respect to the power measured at the detection system will increase by seven orders of magnitude allowing \ac{ALPS}\,II to probe $\mathcal{P}_{\gamma\leftrightarrow \phi}$ down to rates of $3\times10^{-17}$, an improvement of nearly four orders of magnitude.

\section*{Acknowledgments}
We would like to thank the ALPS\,II collaboration, especially David Reuther, Sandy Croatto, and Sven Karstensen, the FTX group, especially, Karsten Gadow and Maik Dinter, the MVS group, especially Antonio de Zubiaurre-Wagner, MKS, FEB, MPC, MPS, ZM1 and ZM5, and the MEA2 and MEA5 groups at DESY as well as the FH mechanical workshop, for their support during the commissioning and operation of the experiment. We are also grateful to Manuel Meyer for the many valuable discussions and his feedback regarding the manuscript.

We acknowledge the support of the National Science Foundation (Grant No. 1802006), of the Heising-Simons Foundation (Grant No. 2015-154 and 2020-1841), of the Deutsche Forschungsgemeinschaft through project grant WI 1643/2-1 and EXC 2121 “Quantum Universe” – 390833306, of the Science and Technology Facilities Council (UK), grants ST/T006331/1 and ST/Y004515/1, as well as support by the German Volkswagen Stiftung and the European Research Council (ERC) under the European Union’s Horizon 2020 research and innovation program Grant agreement No. 948689.
\bibliography{dichroic_noise.bib}

@online{ALPSII_science,
      title={Any Light Particle Searches with {ALPS II}: first science results}, 
      author={Daniel C. Brotherton and Sandy Croatto and Jacob Egge and Aldo Ejlli and Henry Frädrich and Joe Gleason and Hartmut Grote and Ayman Hallal and Michael T. Hartman and Harald Hollis and Katharina-Sophie Isleif and Friederike Januschek and Kanioar Karan and Sven Karstensen and Todd Kozlowski and Axel Lindner and Manuel Meyer and Guido Müller and Gulden Othman and Jan H. Põld and David Reuther and Andreas Ringwald and Elmeri Rivasto and José Alejandro Rubiera Gimeno and Jörn Schaffran and Uwe Schneekloth and Christina Schwemmbauer and Aaron D. Spector and David B. Tanner and Dieter Trines and Li-Wei Wei and Benno Willke and Rachel Wolf},
      collaboration = {ALPS Collaboration},
      year={2025},
      eprint={2512.14110},
      archivePrefix={arXiv},
      primaryClass={hep-ex},
      url={https://arxiv.org/abs/2512.14110}, 
}

@techreport{okun1982limits,
  title={Limits of electrodynamics: paraphotons},
  author={Okun, Lev Borisovich and others},
  year={1982},
  institution={Institute of Theoretical and Experimental Physics, Moscow},
url={https://inis.iaea.org/records/de9zs-5z760/files/14738843.pdf},
}

@article{Anselm:1985obz,
    author = "Anselm, A. A.",
    title = "{Arion $\leftrightarrow$ Photon Oscillations in a Steady Magnetic Field. (In Russian)}",
    journal = "Yad. Fiz.",
    volume = "42",
    pages = "1480--1483",
    year = "1985"
}

@article{VanBibber1987759,
	author = {Van Bibber, K. and Dagdeviren, N.R. and Koonin, S.E. and Kerman, A.K. and Nelson, H.N.},
	title = {Proposed experiment to produce and detect light pseudoscalars},
	year = {1987},
	journal = {Phys. Rev. Lett.},
	volume = {59},
	number = {7},
	pages = {759 – 762},
	doi = {10.1103/PhysRevLett.59.759},
	url = {https://www.scopus.com/inward/record.uri?eid=2-s2.0-33845356563&doi=10.1103%2fPhysRevLett.59.759&partnerID=40&md5=fff7a7d5134023688ba588a73a26645d},
	type = {Article},
	publication_stage = {Final},
	source = {Scopus}
}

@article{Hoogeveen19913,
	author = {Hoogeveen, F and Ziegenhagen, T},
	title = {Production and detection of light bosons using optical resonators},
	year = {1991},
	journal = {Nucl. Phys. B},
	volume = {358},
	number = {1},
	pages = {3 – 26},
	doi = {10.1016/0550-3213(91)90528-6},
	url = {https://www.scopus.com/inward/record.uri?eid=2-s2.0-0040782753&doi=10.1016%2f0550-3213%2891%2990528-6&partnerID=40&md5=b51ae9818be3a6c44cbf6b04b058b32b},
	type = {Article},
	publication_stage = {Final},
	source = {Scopus},
}

@article{PhysRevLett.98.172002,
  title = {Resonantly Enhanced Axion-Photon Regeneration},
  author = {Sikivie, P. and Tanner, D. B. and Van Bibber, Karl},
  journal = {Phys. Rev. Lett.},
  volume = {98},
  issue = {17},
  pages = {172002},
  numpages = {4},
  year = {2007},
  month = {Apr},
  publisher = {American Physical Society},
  doi = {10.1103/PhysRevLett.98.172002},
  url = {https://link.aps.org/doi/10.1103/PhysRevLett.98.172002}
}

@article{PhysRevD.80.072004,
  title = {Detailed design of a resonantly enhanced axion-photon regeneration experiment},
  author = {Mueller, Guido and Sikivie, Pierre and Tanner, D. B. and van Bibber, Karl},
  journal = {Phys. Rev. D},
  volume = {80},
  issue = {7},
  pages = {072004},
  numpages = {10},
  year = {2009},
  month = {Oct},
  publisher = {American Physical Society},
  doi = {10.1103/PhysRevD.80.072004},
  url = {https://link.aps.org/doi/10.1103/PhysRevD.80.072004}
}

@article{ehret2010new,
  title={New {ALPS} results on hidden-sector lightweights},
  author={Ehret, Klaus and Frede, Maik and Ghazaryan, Samvel and Hildebrandt, Matthias and Knabbe, Ernst-Axel and Kracht, Dietmar and Lindner, Axel and List, Jenny and Meier, Tobias and Meyer, Niels and others},
  journal={Phys. Lett. B},
  volume={689},
  number={4-5},
  pages={149--155},
  year={2010},
  publisher={Elsevier},
  url={https://www.sciencedirect.com/science/article/pii/S0370269310005526}
}

@article{PhysRevD.92.092002,
  title = {New exclusion limits on scalar and pseudoscalar axionlike particles from light shining through a wall},
  author = {Ballou, R. and Deferne, G. and Finger, M. and Finger, M. and Flekova, L. and Hosek, J. and Kunc, S. and Macuchova, K. and Meissner, K. A. and Pugnat, P. and Schott, M. and Siemko, A. and Slunecka, M. and Sulc, M. and Weinsheimer, C. and Zicha, J.},
  collaboration = {OSQAR Collaboration},
  journal = {Phys. Rev. D},
  volume = {92},
  issue = {9},
  pages = {092002},
  numpages = {6},
  year = {2015},
  month = {Nov},
  publisher = {American Physical Society},
  doi = {10.1103/PhysRevD.92.092002},
  url = {https://link.aps.org/doi/10.1103/PhysRevD.92.092002}
}

@article{PhysRevD.82.115018,
  title = {Optimizing light-shining-through-a-wall experiments for axion and other weakly interacting slim particle searches},
  author = {Arias, Paola and Jaeckel, Joerg and Redondo, Javier and Ringwald, Andreas},
  journal = {Phys. Rev. D},
  volume = {82},
  issue = {11},
  pages = {115018},
  numpages = {14},
  year = {2010},
  month = {Dec},
  publisher = {American Physical Society},
  doi = {10.1103/PhysRevD.82.115018},
  url = {https://link.aps.org/doi/10.1103/PhysRevD.82.115018}
}

@online{garciacely2025stellarboundslightspin2,
      title={Stellar Bounds on Light Spin-2 Particles in Bimetric Theories}, 
      author={Camilo Garc\'ia-Cely and Andreas Ringwald},
      year={2025},
      eprint={2511.03707},
      archivePrefix={arXiv},
      primaryClass={hep-ph},
      url={https://arxiv.org/abs/2511.03707}, 
}

@article{NCXsq1,
 ISSN = {00063444, 14643510},
 URL = {http://www.jstor.org/stable/2332542},
 author = {P. B. Patnaik},
 journal = {Biometrika},
 number = {1/2},
 pages = {202--232},
 publisher = {[Oxford University Press, Biometrika Trust]},
 title = {The Non-Central {$\chi^2$}- and {F}-Distribution and their Applications},
 urldate = {2025-08-06},
 volume = {36},
 year = {1949}
}

@article{savitzky1964smoothing,
  title={Smoothing and differentiation of data by simplified least squares procedures.},
  author={Savitzky, Abraham and Golay, Marcel JE},
  journal={Anal. Chem.},
  volume={36},
  number={8},
  pages={1627--1639},
  year={1964},
  publisher={ACS Publications},
  url={https://pubs.acs.org/doi/pdf/10.1021/ac60214a047?casa_token=QXSZ15jWCe0AAAAA:4eII8shmx_PDp761EECkMSoypxsxtQvWcJdKSgSD_fTZAjg1S_gXk2qdcYGPG6paD_gqEi4qIYyTtDg}
}

@article{albrecht2021straightening,
  title={Straightening of superconducting {HERA} dipoles for the any-light-particle-search experiment {ALPS II}},
  author={Albrecht, Clemens and Barbanotti, Serena and Hintz, Heiko and Jensch, Kai and Klos, Ronald and Maschmann, Wolfgang and Sawlanski, Olaf and Stolper, Matthias and Trines, Dieter},
  journal={Eur. Phys. J. Tech. Instrum.},
  volume={8},
  number={1},
  pages={5},
  year={2021},
  publisher={Springer Berlin Heidelberg},
  url={https://link.springer.com/content/pdf/10.1140/epjti/s40485-020-00060-5.pdf}
}

@article{meinke1991superconducting,
  title={Superconducting magnet system for HERA},
  author={Meinke, Rainer},  
  journal={IEEE transactions on magnetics},
  volume={27},
  number={2},
  pages={1728--1734},
  year={1991},
  publisher={IEEE},
  url={https://ieeexplore.ieee.org/stamp/stamp.jsp?arnumber=133525&casa_token=-vgly0upy6AAAAAA:5XlOnSc0UOmPyNfXlIgoxQjwUQbfs5A1PV8OZYYJiiswsdmcENwshR4G-3mcxK7wiwPLEM0URA&tag=1}
}

@article{bahre2013any,
  title={Any light particle search {II}—technical design report},
  author={B{\"a}hre, Robin and D{\"o}brich, Babette and Dreyling-Eschweiler, Jan and Ghazaryan, Samvel and Hodajerdi, Reza and Horns, Dieter and Januschek, Friederike and Knabbe, E-A and Lindner, Axel and Notz, Dieter and others},
  journal={J. Instrum.},
  volume={8},
  number={09},
  pages={T09001},
  year={2013},
  publisher={IOP Publishing},
  url={https://iopscience.iop.org/article/10.1088/1748-0221/8/09/T09001/meta?casa_token=lB1vAMILS7YAAAAA:WqoZUuy6xJNYNwj5s0lspWgpoPhxGWJzOuvPQJecg1_BVRhSGrBAmnBbIX2xbazyRb8q4m0YKYNz_mlWzKNGx5jX1Io}
}

@article{hallal2022heterodyne,
  title={The heterodyne sensing system for the {ALPS II} search for {sub-eV} weakly interacting particles},
  author={Hallal, Ayman and Messineo, Giuseppe and Ortiz, Mauricio Diaz and Gleason, Joseph and Hollis, Harold and Tanner, David B and Mueller, Guido and Spector, Aaron},
  journal={Phys. Dark Universe},
  volume={35},
  pages={100914},
  year={2022},
  publisher={Elsevier},
  url={https://www.sciencedirect.com/science/article/pii/S2212686421001382}
}

@article{PhysRevD.99.022001,
  title = {Coherent detection of ultraweak electromagnetic fields},
  author = {Bush, Zachary R. and Barke, Simon and Hollis, Harold and Spector, Aaron D. and Hallal, Ayman and Messineo, Giuseppe and Tanner, D. B. and Mueller, Guido},
  journal = {Phys. Rev. D},
  volume = {99},
  issue = {2},
  pages = {022001},
  numpages = {10},
  year = {2019},
  month = {Jan},
  publisher = {American Physical Society},
  doi = {10.1103/PhysRevD.99.022001},
  url = {https://link.aps.org/doi/10.1103/PhysRevD.99.022001}
}

@article{kozlowski2024designperformancealpsii,
 title={Design and performance of the {ALPS II} regeneration cavity},
  author={Kozlowski, Todd and Wei, Li-Wei and Spector, Aaron D and Hallal, Ayman and Fr{\"a}drich, Henry and Brotherton, Daniel C and Oceano, Isabella and Ejlli, Aldo and Grote, Hartmut and Hollis, Harold and others},
  journal={Opt. Express},
  volume={33},
  number={5},
  pages={11153--11166},
  year={2025},
  publisher={Optica Publishing Group},
  url={https://opg.optica.org/directpdfaccess/b42789fd-e4fc-47f2-ae7844131e1b77eb_568824/oe-33-5-11153.pdf?da=1&id=568824&seq=0&mobile=no}
}

@article{Spector:24,
author = {Aaron D. Spector and Todd Kozlowski},
journal = {Opt. Express},
number = {16},
pages = {27112--27124},
publisher = {Optica Publishing Group},
title = {Optical cavity characterization with a mode-matched heterodyne sensing scheme},
volume = {32},
month = {Jul},
year = {2024},
url = {https://opg.optica.org/oe/abstract.cfm?URI=oe-32-16-27112},
doi = {10.1364/OE.527344},
}

@article{ortiz2020design,
  title={Design of the {ALPS II} optical system},
  author={Ortiz, M Diaz and Gleason, J and Grote, H and Hallal, A and Hartman, MT and Hollis, H and Isleif, K-S and James, A and Karan, K and Kozlowski, T and others},
  journal={Phys. Dark Universe},
  volume={35},
  pages={100968},
  year={2022},
  publisher={Elsevier},
  url = {https://www.sciencedirect.com/science/article/pii/S2212686422000115},
}

@article{Frede:07,
author = {Maik Frede and Bastian Schulz and Ralf Wilhelm and Patrick Kwee and Frank Seifert and Benno Willke and Dietmar Kracht},
journal = {Opt. Express},
number = {2},
pages = {459--465},
publisher = {Optica Publishing Group},
title = {Fundamental mode, single-frequency laser amplifier for gravitational wave detectors},
volume = {15},
month = {Jan},
year = {2007},
url = {https://opg.optica.org/oe/abstract.cfm?URI=oe-15-2-459},
doi = {10.1364/OE.15.000459},
}

@article{Willke:98,
author = {B. Willke and N. Uehara and E. K. Gustafson and R. L. Byer and P. J. King and S. U. Seel and R. L. Savage},
journal = {Opt. Lett.},
number = {21},
pages = {1704--1706},
publisher = {Optica Publishing Group},
title = {Spatial and temporal filtering of a {10-W Nd:YAG laser with a Fabry--Perot} ring-cavity premode cleaner},
volume = {23},
month = {Nov},
year = {1998},
url = {https://opg.optica.org/ol/abstract.cfm?URI=ol-23-21-1704},
doi = {10.1364/OL.23.001704},
}

@book{Siegman_1987, 
 title={Lasers},
 author={Siegman, Anthony E.},
 publisher={Oxford University Press, University Science Books},
 location={Oxford, England}, 
 year={1987}
}

@article{black2001introduction,
  title={An introduction to {Pound--Drever--Hall} laser frequency stabilization},
  author={Black, Eric D},
  journal={Am. J. Phys.},
  volume={69},
  number={1},
  pages={79--87},
  year={2001},
  publisher={American Association of Physics Teachers},
  url={https://pubs.aip.org/aapt/ajp/article-abstract/69/1/79/1055569/An-introduction-to-Pound-Drever-Hall-laser?redirectedFrom=PDF}
}

@article{drever1983laser,
  title={Laser phase and frequency stabilization using an optical resonator},
  author={Drever, RWP and Hall, John L and Kowalski, FV and Hough, J\_ and Ford, GM and Munley, AJ and Ward, H},
  journal={Appl. Phys. B},
  volume={31},
  number={2},
  pages={97--105},
  year={1983},
  publisher={Springer},
  url={https://link.springer.com/content/pdf/10.1007/bf00702605.pdf}
}

@article{pound1946electronic,
  title={Electronic frequency stabilization of microwave oscillators},
  author={Pound, Robert V},
  journal={Rev. Sci. Instrum.},
  volume={17},
  number={11},
  pages={490--505},
  year={1946},
  publisher={American Institute of Physics},
  url={https://pubs.aip.org/aip/rsi/article-abstract/17/11/490/296311/Electronic-Frequency-Stabilization-of-Microwave?redirectedFrom=PDF}
}

@article{paschotta2007mod,
  title={Mode matching},
  author={Paschotta, R{\"u}diger},
  journal={RP Photonics Encyclopedia},
  year={2007},
  url={https://www.rp-photonics.com/mode_matching.html}
}

@article{Isogai:13,
author = {T. Isogai and J. Miller and P. Kwee and L. Barsotti and M. Evans},
journal = {Opt. Express},
number = {24},
pages = {30114--30125},
publisher = {Optica Publishing Group},
title = {Loss in long-storage-time optical cavities},
volume = {21},
month = {Dec},
year = {2013},
url = {https://opg.optica.org/oe/abstract.cfm?URI=oe-21-24-30114},
doi = {10.1364/OE.21.030114},
}

@inproceedings{leviton2006temperature,
  title={Temperature-dependent absolute refractive index measurements of synthetic fused silica},
  author={Leviton, Douglas B and Frey, Bradley J},
  booktitle={Optomechanical Technologies for Astronomy},
  volume={6273},
  pages={800--810},
  year={2006},
  organization={SpIE},
  url={https://www.spiedigitallibrary.org/conference-proceedings-of-spie/6273/1/Temperature-dependent-absolute-refractive-index-measurements-of-synthetic-fused-silica/10.1117/12.672853.full}
}

@inproceedings{hahn1972thermal,
  title={Thermal Expansion of Fused Silica from 80 to 1000 {K} - Standard Reference Material 739},
  author={Hahn, TA and Kirby, RK},
  booktitle={AIP Conference Proceedings},
  volume={3},
  number={1},
  pages={13--24},
  year={1972},
  organization={American Institute of Physics},
  url={https://pubs.aip.org/aip/acp/article/3/1/13/732646/Thermal-Expansion-of-Fused-Silica-from-80-to-1000}
}

\appendix


\section{Appendix}
\label{Appendix}

\begin{table*}
    \centering
    \makebox[\textwidth][c]{
    \begin{tabular}{c|cccccccc}
    & \multicolumn{2}{c}{$\rm S_{_\perp}$}\\
System     & Set frequency      & Digitization offset \\ \hline
\ac{PLL}1  & 12.2\,MHz          & -11.37\,nHz\\
\ac{PLL}2  & 52.5468728\,MHz   & 699.04\,nHz\\
\ac{PLL}3  & 54.95\,MHz         & 1608.67\,nHz\\
DM1 (Science) & 14.6031296\,MHz& 1096.83\,nHz\\
DM1 (Veto) & 40.3468752\,MHz   & 908.99\,nHz\\ \hline
Total (Science) & 2.4\,Hz       & 198.58\,nHz \\
Total (Veto) & 2.4\,Hz          & 198.58\,nHz   \\    \hline
&\\
& \multicolumn{2}{c}{$\rm S_{_\parallel}$}\\
System     & Set frequency      & Digitization offset \\ \hline
\ac{PLL}1  & 12.2\,MHz          & -11.37\,nHz\\
\ac{PLL}2  & 52.54686689\,MHz   & 1098.23\,nHz\\
\ac{PLL}3  & 54.95\,MHz         & 1608.67\,nHz\\
DM1 (Science) & 14.60313551\,MHz& 697.65\,nHz\\
DM1 (Veto) & 40.34686929\,MHz   & 1308.17\,nHz\\ \hline
Total (Science) & 2.4\,Hz       & 198.58\,nHz \\
Total (Veto) & 2.4\,Hz       & 198.58\,nHz\\          
    \end{tabular}}
    \caption{Summary of the frequencies and their digital offsets for  $\rm S_{_\perp}$ and  $\rm S_{_\parallel}$.}
    \label{tab:moku_f}
\end{table*}

\subsection{Digital Frequency Representation}
\label{APP:Sec_DFR}

Table~\ref{tab:moku_f} summarizes the frequencies as they are defined in the user interface of the devices generating them in the column labeled `Set frequency', while the offset from those frequencies due to their representation in terms of the clock frequencies is shown in the column `Digitization offset'. The row label `Total' for the science and veto detectors shows the result of the linear combination of frequencies that produces the signal that is demodulated at the second stage in post-processing. In this case the input frequencies of the \acp{PLL} are tuned to produce a 2.4\,Hz signal frequency after the first demodulation. The digital offset on that signal is a result of the same linear combination of the digital offsets. The second demodulation should therefore be performed at the total set frequency plus its digital offset, so in all cases 2.4\,Hz + 198.58\,nHz. This is the digital representation of a 2.4\,Hz signal using a clock with a rate of 1\,GHz and a precision of 48 bits.  In post-processing, the frequency of the second demodulation can be represented with double-precision (64-bit) floating-point because this process is no longer limited by the precision of the oscillators driving the \acp{PLL}. While the numbers in this table are only shown with a precision of tens of pHz, the actual second demodulation frequency was defined to tens of fHz.\footnote{This precision is only defined with respect to the clock used for the system and should not be mistaken in terms of absolute frequency.}

\subsection{Changes in Open Shutter Data}
\label{APP:Del_C_j}
\begin{figure*}[t]
    \centering
               \begin{subfigure}[b]{0.49\textwidth}
            \centering
    \includegraphics[width=\textwidth]{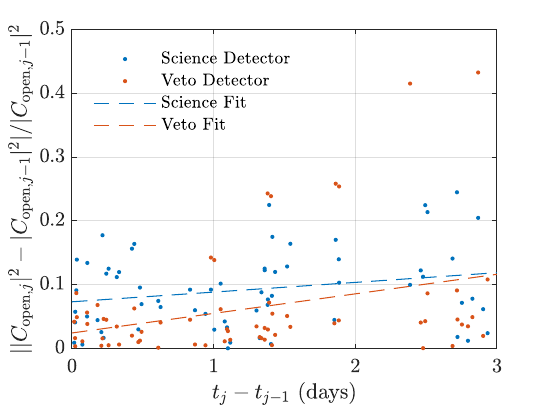}
    \caption{Magnitude squared changes open shutter ($\rm S_{_\perp}$)
    \label{fig:diff_open_abs_sq_SC}}
    \end{subfigure}               \begin{subfigure}[b]{0.49\textwidth}
            \centering
    \includegraphics[width=\textwidth]{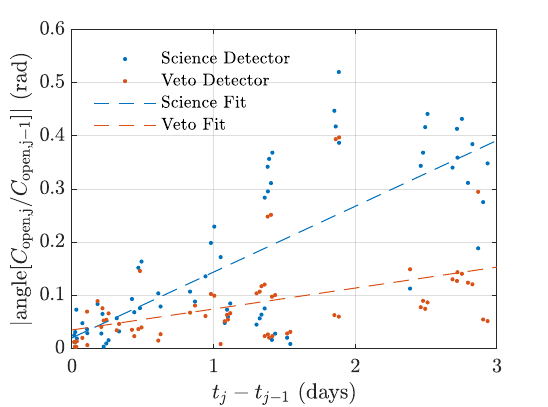}
    \caption{Angular changes open shutter ($\rm S_{_\perp}$)
    \label{fig:diff_open_ang_SC}}
    \end{subfigure}  
    
    \vspace{2mm}
    \begin{subfigure}[b]{0.49\textwidth}
            \centering
    \includegraphics[width=\textwidth]{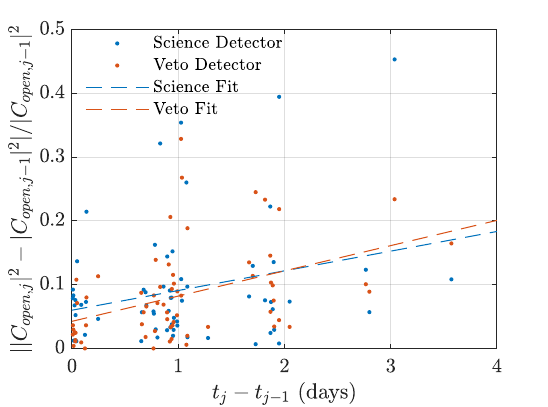}
    \caption{Magnitude squared changes open shutter ($\rm S_{_\parallel}$)
    \label{fig:diff_open_abs_sq_PS}}
    \end{subfigure}               \begin{subfigure}[b]{0.49\textwidth}
            \centering
    \includegraphics[width=\textwidth]{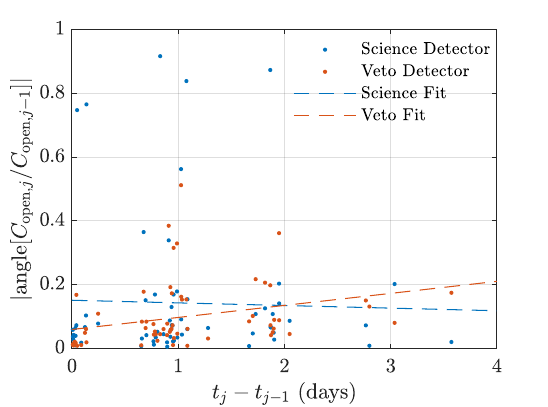}
    \caption{Angular changes open shutter ($\rm S_{_\parallel}$)
    \label{fig:diff_open_ang_PS}}
    \end{subfigure}
        \caption{Changes in the magnitude and angle of $C_{{\rm open},j}$ plotted as a function of the time between the points for $\rm S_{_\perp}$. Science detector data are shown in blue and the veto detector data are shown in orange. The dashed lines show the linear fits to the data.  The units of the $y$-axis in this plot are rad.
    \label{fig:diff_open_mag_ang}}
\end{figure*}

Scatter plots showing the changes between open shutter periods are shown in Figure~\ref{fig:diff_open_mag_ang}. From the plots in \ref{fig:diff_open_abs_sq_SC} and \ref{fig:diff_open_abs_sq_PS} it is apparent that the absolute value of the changes between the magnitude squared of two points in the open shutter calibration series $\left||C_{{\rm open},j}|^2-|C_{{\rm open},j-1}|^2\right|$ increased as the time between the points $j-1$ and $j$ increased. The data in these plots are also normalized to the value of $|C_{{\rm open},j-1}|^2$ to allow the relative error in the calibration series due to these effects to be calculated (discussed in Section~\ref{Sec:C_n_uncert}). The data for the science detector are shown as the blue points, while the veto detector data are shown in orange. The dashed lines give the fit of a first order polynomial to the data. The result of this fit is then used to calculate the relative uncertainty for each point in the calibration series $C[n]$, based on its length in time from the nearest open shutter measurement.

The changes in the angle of $C_{{\rm open},j}$ between open shutter points are also shown in Figures~\ref{fig:diff_open_ang_SC} and \ref{fig:diff_open_ang_PS}, again with blue represent the data measured at the science detector, orange the data measured and the veto detector, and the dashed lines giving the linear fit to the data. Interestingly, the fit of the data from $\rm S_{\parallel}$ shows a negative slope, indicating that changes in the angle decrease as the time between the points increases. We do not believe that it is actually the case that the uncertainty in the complex angle of the open shutter measurements decrease over time, but rather this is due to a bias in the data. This bias comes from the fact that the operation procedure required an open shutter period of at least 15 minutes before starting a closed shutter measurement. As the shorter periods are still used to assess the uncertainty, times in which the system was unstable would have produced more open shutter data. Partially because of this reason, the linear fit of the angular changes are not used in the calculation of the uncertainty on $C[n]$, and displayed here rather to illustrate the remarkable stability of the system.

\subsection{Signal Leakage}
\label{Sec:Freq_spr}

Irregular gaps in the closed shutter data exist due to the need for open shutter and maintenance periods, as well as the system coming unlocked during the closed shutter periods. These gaps lead to signals at a given frequency leaking into neighboring frequency bins as the integration of the product a static constant with these neighboring frequencies is no longer performed over a finite number of cycles and these components will not sum to zero. Figures~\ref{fig:freq_spread_a} and \ref{fig:freq_spread_b} show plots of this leakage into the different Fourier components. Here the spectrum of $Z(f)_{\rm closed}$ is calculated with $H_{\rm p}[n]$ set to a series of ones for both  $\rm S_{_\perp}$ (blue) and $\rm S_{_\parallel}$ (orange). Note that the data for $\rm S_{_\parallel}$ are offset downward by a decade to help show both sets of data on the same plots.

\begin{figure}[t]
    \centering
               \begin{subfigure}[b]{0.49\textwidth}
            \centering
    \includegraphics[width=\textwidth]{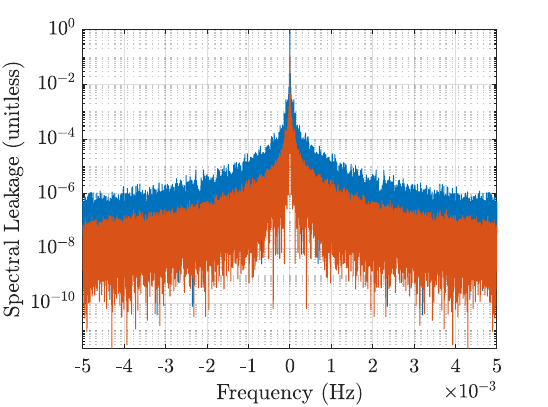}
    \caption{
    \label{fig:freq_spread_a}}
    \end{subfigure}
   \begin{subfigure}[b]{0.49\textwidth}
            \centering
    \includegraphics[width=\textwidth]{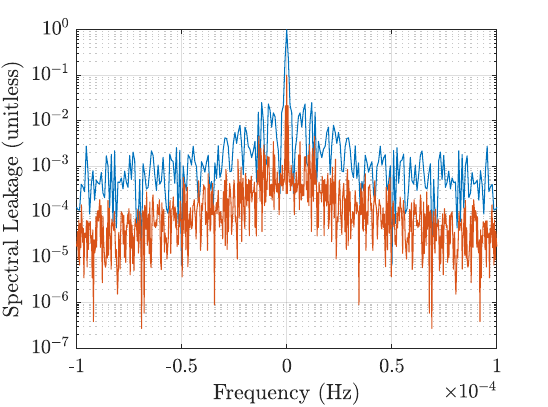}
    \caption{
    \label{fig:freq_spread_b}}
    \end{subfigure}
        \caption{Leakage from the signal bin into neighboring frequency bins for $\rm S_{_\perp}$ (blue) and $\rm S_{_\parallel}$ (orange) from -5\,mHz to +5\,mHz (a) and -100\,{\textmu}Hz to +100\,{\textmu}Hz (b). Note that the $\rm S_{_\parallel}$ data are offset a decade lower to show both data sets on the same plot.\label{fig:freq_spread}}
\end{figure}

In these plots it can be seen that the neighboring frequency bins at $f = \pm 0.85$\,{\textmu}Hz contain a value of $|Z|^2$ that is 0.093 times the value in the signal bin for $\rm S_{_\perp}$ and 0.214 at $f = \pm 0.60$\,{\textmu}Hz for $\rm S_{_\parallel}$. While in both cases none of these Fourier components are present in a continuous series of ones, they are present in a series of ones with gaps similar to our data sets. $\rm S_{_\parallel}$ has more of this cross coupling between the frequency bin at $f = \pm 0.60$\,{\textmu}Hz due to the fact that over 97\% of the valid closed shutter data came in three distinct periods each with nearly a week in between, from March 31 to April 6, from April 12 to April 22, and from April 29 to May 6. The closed shutter data of $\rm S_{_\perp}$, on the other hand, was acquired more regularly.

Gaps in the data are not expected to lead to excess coupling of white Gaussian noise into the signal bin or across other frequency bins with respect to the expected noise level if there were no gaps in the data (effects from the integration time notwithstanding).  It is therefore believed that any stochastic noise in the data should produce a spectrum in $Z(f)_{\rm closed}$ in accordance with its power spectral density. Deterministic signals, on the other hand, may couple into alternate frequency bins. Based on Figures~\ref{fig:freq_spread_a} and \ref{fig:freq_spread_b} this is not expected to reduce the power measured at the signal frequency as here the value at $f=0$ is 1 for both runs.

\begin{table}[b!]
    \centering
    \begin{tabular}{c|cccccccc}   
        & Scalar  & Pseudoscalar  \\  
        &  $|Z(f_j=0)_{\rm closed}|^2$ & $|Z(f_j=0)_{\rm closed}|^2$ \\ \hline  
$C[n]$ & $1.94\times10^{-5}$  & $2.87\times10^{-5}$ \\ 
$\langle C[n]\rangle$ &$2.61\times10^{-5}$  & $2.52\times10^{-5}$ \\ 
$|C[n]|$ &$2.66\times10^{-5}$  & $1.95\times10^{-5}$ \\ 
$C[n]_{\rm v}$ &$2.24\times10^{-5}$  & $1.80\times10^{-5}$ \\ 
    \end{tabular}  

    \caption{Summary of the heterodyne data in the signal bin for both detectors and both science runs.}
    \label{tab:alt_Cn}
\end{table}

\subsection{Double Checking the Calibration}
\label{Sec:alt_cn}

Three alternative versions of $C[n]$ were applied to the data to investigate different effects in the calibration array. For the first alternative series, the mean value of the calibration series, or $\langle C[n]\rangle$ was calculated over the full run, the coherent sum was performed independently, and  $\langle C[n]\rangle$ was used to calibrate the result of the summation. The second version of the calibration array simply used the absolute value of the calibration array and performed no correction to the phase of the heterodyne data and is referred to as $|C[n]|$. 
The phase of the veto calibration series is used in the last alternative calibration series. Here the $C_{\rm v}[n]$ refers to using the magnitude of the normal calibration series, but the phase of the veto calibration series. 

\begin{figure}[b]
    \centering
               \begin{subfigure}{0.49\textwidth}
            \centering
    \includegraphics[width=\textwidth]{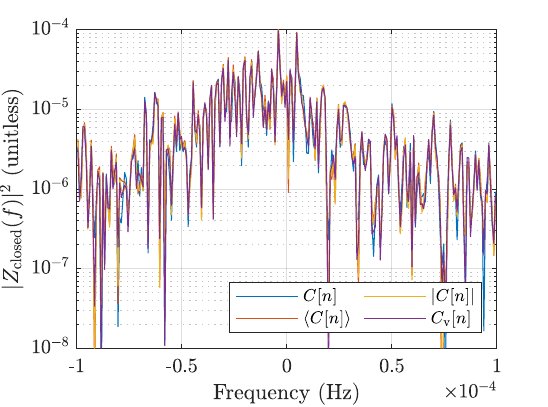}
    \caption{$\rm S_{_\perp}$ 
    \label{fig:alt_cn_SC}}
    \end{subfigure}
   \begin{subfigure}{0.49\textwidth}
            \centering
    \includegraphics[width=\textwidth]{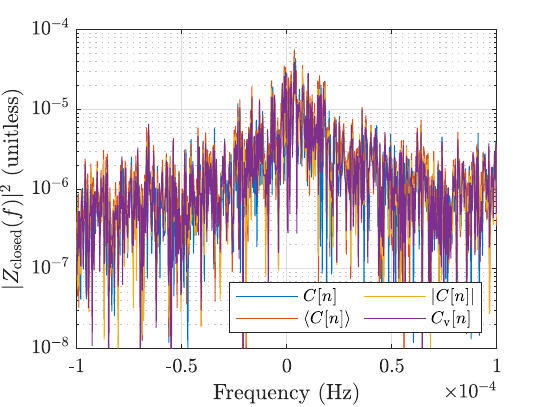}
    \caption{$\rm S_{_\parallel}$
    \label{fig:alt_cn_PS}}
    \end{subfigure}
        \caption{The stray-light signal from $\rm S_{_\perp}$ (a) and $\rm S_{_\parallel}$ (b) using different calibration functions.
    \label{fig:alt_cn}}
\end{figure}

The results of the alternative frequency analysis are shown in Figures~\ref{fig:alt_cn_SC} and \ref{fig:alt_cn_PS}, while Table~\ref{tab:alt_Cn} shows a summary of the results in the signal bin using the alternative calibration arrays. The data for the actual calibration array is shown as the blue trace in these plots. The results of $\langle C[n]\rangle$ and $|C[n]|$ demonstrate that the calibration method makes only a marginal difference in the results. This makes sense as the changes in the amplitude and phase of the open shutter data throughout the runs was not so large as to substantially effect the results. The results of using $C[n]_{\rm v}$ where also very similar to the results obtained with the normal calibration array.

For both runs it is apparent that the background levels do not show significant differences when one of these alternative calibration arrays was used. 

\end{document}